**Kurdistan Region Government**
**Ministry of Higher Education and Scientific Research**
**University of Sulaimani**
**College of Agricultural Engineering Sciences**

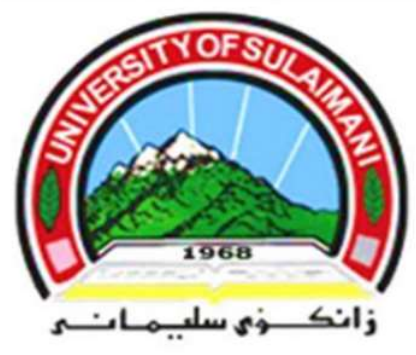

# MORPHO-PHYSIOLOGICAL AND GENETIC DIVERSITY OF *Crataegus* taxa (*ROSACEAE*) IN SELECTED LOCATIONS OF IRAQI KURDISTAN-REGION

**A Dissertation**

**Submitted to the Council of the College of Agricultural Engineering Sciences at the University of Sulaimani in Partial Fulfillment of the Requirements for the Degree of Doctor of Philosophy**

**in**
**Crop Sciences**
**Plant Biotechnology**

By

**Karzan Ezzalddin Mohammed**

B.Sc., Biotechnology and crop Sciences (2003), University of Sulaimani
M.Sc., Medicinal Plant (2012), University of Sulaimani

Supervisors

**Dr. Nariman Salih Ahmad** — Professor

**Dr. Saman Abdulrahman Ahmad** — Professor

**2724 K.** **2025 A.D.**

## Summary

One of the great phytogeography zones of semi-arid lands in the world is the Kurdistan region of Iraq which hosts many important fruit species due to its geographical location and ecology. Mountain Hawthorn (*Crataegus* spp.) is a vital wild edible deciduous fruit tree of the genus *Crataegus* for the region, which is highly beneficial for ornamental, economical, industrial and medicinal uses.

In the present study, morphological, phytochemical and molecular marker systems were applied on sixty-one Hawthorn accessions from different locations in the Iraqi Kurdistan region during April 2022 to September 2023.

Phenotypic markers have proven to be extremely useful in studies of genetic diversity in Hawthorn genotypes, the results of the present morphological study showed that there are seven taxa (five species, two hybrids) were observed including, *Crataegus azarolus*, *Crataegus meyrei, Crataegus monogyna, Crataegus orientalists, Crataegus pentagyna, Crataegus azarolus x Crataegus meyrei and Crataegus azarolus x Crataegus pentagyna.* There was significant variation among different ecotypes in terms of plant type, reproductive stage, and fruit morphology and production uses.

Fruit Physio-morphological data revealed a high level of significant variability ($P \leq 0.01$) among accessions based on the analysis of variance. The most important characteristics for explaining fruit morphological variability `were 11 varbales including fruit weight (FW), fruit length (FL), fruit width (FW), seed length (SL), seed width (SW), number of seeds per fruits (NSF), volume solution (VS), fruit fresh weight (WOF), seed weight (WS), Potentional of hydrogen (pH) and mositure content (MC). They all are significantly different for all the traits measured for the studied accessions.

Physio morphological characteristics were utilized in Hierarchical clustering analysis, producing seven clusters for fruit characteristics data from the *Crataegus* accessions using UPGMA analysis. The first cluster had six genotypes (G42, G24, G18, G44, G61, and G39) while the second cluster covered seven accessions of G29, G20, G26, G48, G15, G23, and 25. However G31, G60, G38, G8, G47, G35, G51, G59, and G32 were the nine genotypes in Cluster 3. G27, G36, G14, G52, G56, G54, G46, G40, G19, G16, G58, G57, G37, and G28 comprised the fourth cluster. G43, G34, G33, G17, G55, G45, G50, G41, G49, G12, and G11 were parts of cluster five. In the six-cluster had six genotypes such as (G1, G2, G4, G3, G30, and G6), while the final cluster G21, G22, G7, G9, G53, G10, G13, and G5 were parts of cluster seven.

Role of biochemical characterization for indicating genetic diversity of wild Hawthorn was also studied. through the utilization of biochemical markers such as Total soluble solid (TSS), Total phenolic content (TPC), Total flavonoid content (TFC), Antioxidant activity (AA), and Carotenoid content (CAC). The results, a considerable variation was revealed in the biochemical traits of Hawthorn species. The analysis of variance showed significant differences in the biochemical traits (TSS, TPC, TFC, AA and CAC) of the accessions examined ($P \leq 0.001$). The diversification of fruit components has identified to be referred to genetic structure and environmental factors.

Alongside morphological and biochemical characterizations, universal primer pairs ITS1 and ITS4 were utilized to study the genetic distance of Hawthorn accessions.

The nucleotide sequence alignments from online BLAST divided the genotypes into two groups. Frist group consists of one clade of only a genotype (G37), second group includes 60 accessions that are consent of nine sub-clades. These results revealed that there were discernible divisions in the molecular phylogeny of the various taxonomic forms of *Crataegus Taxa*.

To evaluate the genetic diversity among 61 accessions of Hawthorn in the different locations 11 SSR, 10 ISSR, and 10 SCoT markers were utilized. A total of 237 polymorphic bands were generated from the total primers used. Markers yielded 57 (5.18 per primer), 46 (4.6 per primer), and 134 (13.4 per primer) alleles for SSR, ISSR, and SCOT, respectively. According to the marker amplification SSR 58 recorded the maximum polymorphic bands of 13. SSR 40 had the best values of Na, Ne, I, He, uHe, and PIC ( 1.42, 1.35, 0.30, 0.20, 0.22, and 0.36) Respectivey. In case of SCoT markers, the highest value of 1.34, 1.37, 0.33, 0.22, 0.25, and 0.41 for the Na, Ne, I, He, uHe, and PIC, respectively were recorded for SCoT 12. The average values of Na, Ne, I, He, uHe, and PIC for the SSR primers were 0.98, 1.18, 0.17, 0.11, 0.12, and 0.19. Whereas the mean values of these indices for the ISSR markers were 1.12, 1.23, 0.20, 0.13, 0.15, and 0.23, and for the SCoT markers were 1.13, 1.28, 0.25, 0.17, 0.18, and 0.28, respectively. Unweighted pair group method with arithmetic mean (UPGMA) screened and clustered the Hawthorn accessions based on the all-primers data. All the Hawthorn accessions were clustered into seven main groups, relying of each marker set (SSR, ISSR, and SCoT markers). The genetic variation between and within the Hawthorn population was measured by analysis of molecular variance (AMOVA) using SSR, ISSR, and SCoT marker datasets. The variation amongst populations, described as 10 %, 19 % and 17 % of the total variance, relying on the SSR, ISSR, and SCoT marker data, respectively. Nevertheless, with the values of 90.00, 81.00, and 83.00%, the highest variation occurred within the population for the markers respectively.

# CHAPTER ONE

# INTRODUCTION

The Kurdistan region of Iraq is one of the great phytogeography zones of semi- arid lands in the world(Ahmad and Salih 2021; Qalatobzany et al., 2025). Mountain Hawthorn (Crataegus spp.) is a vital wild edible deciduous fruit tree of the genus Crataegus in the region, which is belonging to Rosaceae family, and globally represented in the northern hemisphere's temperate zones (Sheng et al. 2017). The genus (*Crataegus* spp.) belongs to Maloideae subfamily in the Rosaceae family, and widely distributed in regions located between 30° to 50°N latitude of as Asia, Europe, and America (Dickinson et al. 2021). The main Hawthorn species in central Europe are *C. monogyna* and *C. laevigata*, and other three species of *C. pentagyna, C. nigra*, and *C. azarolus* from southern and southeastern Europe. *Crataegus pinnatifida* and *C. scabrifoliaare* represent the main Hawthorn varieties in China (Sheng et al., 2017 and Du et al., 2019). Hawthorn is considered to be among the global important horticultural crop due to its considerable economic, ecological, and ornamental values (Brown et al., 2016 and Betancourt-Olvera et al., 2018). It is also highly contributed to providing a habitat for other organisms, being among the main sources of food and shelter for birds and other animals in a semi-arid ecosystem (Nazhand et al., 2020). In addition, these wild resources can be used as rootstocks and important breeding materials for cultivated Hawthorn (Sheng et al., 2017).

Morphological, phytochemical and molecular marker systems have been used for years to determine genetic diversity among plants, including Hawthorn (Swarup et al., 2021), however, applied researches on wild Hawthorn are limited. The available investigations are mainly focused on morphological characteristic description (Lin and Cui 2000 and Liao 2013), fruit pigment extraction (Lv et al., 2011) and flower organ research (Liu et al., 2014). To some extent, phenotypic indicators were utilized in selecting horticulturally favorable traits in Hawthorn such as vigor, some biotic and abiotic stress resistance. Hawthorn comprises both trees and shrubs mostly growing up to 5-15 m that considered to be recently among the global important edible wild horticultural fruits (Obideen et al., 2024). Phenological parameters are highly diverse including leaf morphology, seed traits, the number of seeds and the color of the fruits.

Hawthorn fruits containing different biochemical components, including polyphenols, flavonoids, antioxidants, vitamins, and organic acids. Carotenoid and anthocyanin are among the main pigments of Hawthorn fruit.  These chemicals are primary and secondary metabolites products of the plant, having important health benefits (Alirezalu et al. 2020). According to Zhang et al. (2022) Hawthorn nutrition values are varied according to the genetic construction and environmental effect on the plant growth; however, Hawthorn fruit contains in general 68.98 - 77% water, 3.03% protein, 1.22%, crude oil and 2.77% ash. It is with 4.08 pH and 56% acidity, consisting of 9.35 mg/g total phenolic content, 2 mg/g fat, 30–40 mg/g of pectin, 30-50 mg/g of organic acids, 3–8 mg/g of tannins, 0.5–1.5 mg/g of amino acids, 0.89 mg/g of vitamin C, 0.89 mg/g of vitamin D, and 0.65 mg/g of flavonoids. High polyphenolic content of Hawthorn is among the main advantage of the fruit, allows to be consumed as a food and medicinal treatment instead of synthetic medications. The plant leaves, are also rich in nutrients and beneficial bioactive compounds, with many health benefits, including medicinal and nutraceutical properties (Nazhand et al., 2020). Identifying the extend of genetic diversity in wild Hawthorn is crucial for the species conservation and uses, either directly or through the introgression of desire genes from wild to cultivated Hawthorn cultivars.

Genetically Hawthorn is assembled a genome size of 823.41 Mb (2n=2X=34). Even though the Kurdistan region is considered to be a diversity center for *Crataegus* spp. with a wide distribution across the Irano-Turanian region and Zagros mountains (having 17 Hawthorn species), still little information is available on the extent of the species genetic diversity of native Hawthorns (Saadatian et al., 2014). *Crataegus* is considered as a complex genus due to practicing natural hybridization, being the plant with both trees and shrubs (Kumar et al., 2012). The frequent occurrence of instinctive hybridization, apomixes and polyploidy in Hawthorn makes the diversity and biological evolution studies to be difficult based only on the phenotypic characteristics (Fineschi et al., 2005). The ability of this species to introgressive hybridization allowed Hawthorn to express with a wide range of phenotypic features (Depypere et al., 2006). In addition, birds and mammals are also contribute wide distributions and expanding the genetic variability through different geographical distances. Estimation of genetic variation in Hawthorn are mainly based on the limited markers of morphological traits with a low heritability until the new Millenium (Archak et al., 2003). During the last two decade several novel DNA-markers have been rapidly developed at the global level for the characterization of *Crataegus* genome and to study genetic diversity within and among wild landraces of this species (Güney et al., 2018). Such marker techniques include inter-simple sequence repeats (ISSRs) (Sheng et al., 2017 and Emami et al., 2018), and simple sequence repeats (SSRs) (Khiari et al., 2015). SSR, ISSR and SCoT markers are widely used in genetic research because they are robust, convenient codominant, and highly polymorphic (Amiteye, 2021).

All of these aspects are guiding pillars for the enhancement and development of the species and contribute to its competitiveness globally. Consequently, to determine genetic variation among Hawthorn genotypes with higher yield and better quality, it is essential to screen available Hawthorn genetic resources using different morphological, taxonomical and molecular approached to identify details about the extent of the species diversity with their characterization. These investigations facilitate the species conservation, uses and utilizing these resourced in the future improvement program of Hawthorn in Kurdistan Region of Iraq. Therefore, the objectives of this study were as follow:

- Floristic study, and the distribution of Hawthorn in the in representative spots of the Kurdistan of Iraq.
- Morphological characteristics for different collection sites in Kurdistan Iraq and determination of the wideness of genetic base of this studied taxa.
- Identifying fruit quality of Hawthorn accessions based on physiochemical, and pharmaceutical components, to be considered in the future medicinal uses of Hawthorn.
- Studying the genetic diversity among 61 Hawthorn accessions collected from different locations in the Kurdistan Region of Iraq, using morphological characteristics and molecular markers (SSR, ISSR and SCoT).

# CHAPTER TWO
# LITERATURE REVIEW

## 2.1 Hawthorn an overview; Availability and utilization

Hawthorn is a wild edible fruit tree a species member of the family Rosaceae and the genus *Crataegus* (Gundogdu, et al., 2014). The most recent comprehensive demographic study of the Asian and European *Crataegus* species was conducted by Christensen (Christensen 1992). The genus *Crataegus* L. is one of the most important genera of the subfamily Maloideae and it is currently considered to comprise between 150 and 1200 species depending on the species concept employed and on the insertion of many probable taxa hybrid origins (Attard and Attard 2019). Many morphological parameters represented polymorphism including leaves and seeds, the seed number and the color of the fruits, which is refer to the hybridization between *Crataegus* species therefore they have different synonyms, including; May blossom, May tree, or quick thorn, but, currently, it has been often used for "Hawthorn" in general.

Many species are polyploid which has a highly complex species in terms of systematic botany (Martinelli et al., 2021). The scientific name of Hawthorn comes from the Greek word "kràtaigos" which means "strength and robustness" due to its hard and durable wood. Natural habitats of Hawthorn are wooded and sunny areas on predominantly limestone soils up to 1500 m above sea level. This species is very rustic and is not very water demanding (Nazhand et al., 2020).

Hawthorn is widely available and mainly also distributed in the temperate regions of the Northern Hemisphere with many parts of the world, which is often found on forest margins or in scrub on mountain slopes, mostly growing to 15 m. *C. monogyna* and *C. laevigata* are the main Hawthorn species in Central Europe, *C. pentagyna*, *C. nigra*, and *C. azarolus* are the Hawthorn species of southern and southeastern Europe. *C. pinnatifida* and *C. scabrifolia* are the main Hawthorn varieties in China and most common *C. mexicana* and *C. douglasii*, grown in Europe, North Africa, West Asia, and North America (Du et al., 2019 and Zhang et al., 2022).

In addition, this genus has a wide geographic distribution across temperate Asia, Europe, and North America and also includes around 150 and 1200 species (Zeravan et al., 2007) from this number, around 50 species are found in the northern hemisphere (Gurlen et al., 2020) especially in Turkey and Iran, which contain a large number of species (Shahbaz and Sadeq 2003) and the mountains of Zagros are approximately home to 17 different species of Hawthorn (Saadatian et al., 2014). The Kurdistan region of Iraq (KRI) is restricted to the north, and the northeastern part of Iraq occupies 5% of the country's total area and is part of the Taurus–Zagros Mountain range, with altitude ranges between 1000 and 3000 m. at sea level (Al-Ansari 2021).

Furthermore, the taxonomy of *Crataegus* is complicated because of many factors have been affected including: Climatic, geological and biological factors are associated with the characteristics of polyploidy such as latitude, altitude, habitat variety, life cycle, reproduction system, high interspecific hybridization, cell size, chromosome size, chromosomal structure, sex chromosome mechanism and genotype diversity (Coughlan et al., 2017). Therefore, In Iraq, at present, studies about Hawthorn have only been conducted on the following scopes: for example, their chemical compositions (Saadatian et al., 2014) taxonomy and anatomy (Zeravan et al., 2007) and pharmacological activities (Mahmud et al., 2016). Others have investigated the genetic diversity of Hawthorn (Beigmohamadi et al., 2021) and morphological variety (Khadivi-Khub et al., 2015). In Egypt, Moustafa, et al. (2019) noted that the *Crataegus sinaica* environment has continuously worsened during the past few years. In Turkey, various studies explored the chemical characteristics of Hawthorn fruit (Özcan et al., 2005), molecu lar characterization (Gurlen et al., 2020), and morphology and chemical composition (Yanar et al., 2011). Moreover, globally, numerous studies have been published on diverse topics such as phylogeny, taxonomy, molecular analysis, and the complete chloroplast genome sequence in some Hawthorn species (Hu et al., 2021). There is still a gap in understanding their geographical distributions and their response to climate change conditions, research has examined these species' spatial dimensions concerning important environmental variables and their potential future statuses under scenarios of climate change in Iraq (Radha and Khwarahm, 2022).

From botanical point of view, the Hawthorn is a fruit. therefore, utilization of Hawthorn plays a multifaceted role in ecological balance, traditional, modern medicine, and economic activities. Its contributions to biodiversity, human health, and local economies underscore its importance as a valuable plant species in both natural and cultivated environments. As awareness of its benefits continues to grow, Hawthorn may increasingly be recognized and utilized in various sectors, from healthcare to sustainable agriculture and landscaping including; medicinal use, food, ornamental use, woodworking, fuel and fodder, cosmetic use, dye and traditional craft (Nazhand et al., 2020).

Kurdistan region of Iraq is one of the regions rich in wild plants and considered to be an essential national wealth in the region. Taxonomic studies of this wealth help to know and identify the species and their numbers (Al-Rawi, 1964). Taxonomic studies have taken good steps in different Iraqi areas that *Crataegus* spp play a significant ecological and economic role in fields such as biological diversity, human nutrition, wildlife, industrial wood materials, potential energy sources, erosion control, and urban afforestation (Özyurt et al., 2019).

## 2.2 Overview of the ecological, medicinal, and economic significance of Hawthorn

In biodiversity, represent important health and economic contents is medicinal plants, so it is necessary a complete inventory of the components that have a medicinal effect in the flora of plants for any country, to preserve it as well as for continued use (Khal, 2024). Hawthorn considered one of the most commercially significant families, including the woody plants, herb, fruit, nut, ornamental, and aromatic plants (Soundararajan et al., 2019). Ecological, medicinal and economic influence of Hawthorn to the national economy is quite high. Due to is widely utilized for human nutrition, nature wildlife, creating biological diversity, manufacturing wood materials, potential energy sources, pharmaceutical and beautifying materials and urban landscape design, farming, animal feeding and alternative medicine fields. Therefore, ecological, medicinal, and economic significance of Hawthorn can be described as following.

Ecological importance of this tree is commonly found in hedgerows, woodlands, grasslands and provide Hawthorn species is a crop with high biological important habitat, food for various wildlife, including birds and mammals. The Hawthorn, also utilized as good advantage for an ornamental or planting design that the flowers are also a crucial source of nectar and pollen for bees, butterflies, and other pollinators. In addition, they are used to create privacy in home gardens and town squares, thus they can easily adapt to urban climate and do not endanger electric lines. In some regions, it may grow along riverbanks and streams, benefiting from the moisture available in these environments (Hinsley and Bellamy 2000 and Özyurt et al., 2019). Because of Hawthorns have the strong root systems and wood bark. Consequently, to be used in industrial districts and have good effective in erosion control playing an important role in controlling surface flow in rocky slope areas. They can also be used in noise scarification facilities due to numerous branches and leaves, and planted for green belts against pollution. Overall, these key features highlight the adaptability, ecological significance, and ornamental value of Hawthorn species, making them important both in natural ecosystems and cultivated landscapes.

Hawthorn is one of the popular species among humans in arid and semiarid cold regions due to its fruitfulness that is also used for animal nutrition in terms of wildlife, as well as a source of income (Özyurt et al., 2019). Therefore, in the near future, the demand for this species should further increase (Özyurt et al., 2019) because of the main interest of this plant species is the high polyphenolic content and it has been consumed as a food and medicine source (Nazhand et al., 2020).

Regarding medicinal significance of Hawthorn which is one of the medicinal and edible plants, known as the “nutritious fruit” due to its richness in bioactive substances. Medicinal wild plants and herbs have recently received increased interest worldwide since they are rich sources of bioactive compounds and for their potential beneficial health properties, which have often been well known for centuries (Durazzo et al., 2020). Therefore, Hawthorn plants play an essential role in medical treatment traditional medicine, The World Health Organization (WHO) reported that about 80% of the world's population uses traditional drugs, including herbal medicine, for the treatment of diseases before considering conventional drugs when available some important modern drugs have been derived from natural sources on the basis of their use in traditional medicine (WHO, 2013). In addition, the Committee for Herbal Medicinal Products of the European Medicines Agency classified Hawthorn as a "traditional herbal medicinal product" (Nazhand et al., 2020). In addition, various parts of this plant in particular, the berries, flowers, and leaves are rich in nutrients, and have been traditionally associated with many health, medicinal or nutraceutical beneficial health effects (Attard, and Attard, 2019) for instance, anti-microbial, anti-inflammatory, antioxidant, anti-cancer, and anticoagulant properties. (Venskutonis, 2018). According to the holistic and traditional approach, Hawthorn leaves and flowers are used to prepare infusions that can be used to control palpitations, tachycardia, and nervousness (Zhang et al., 2022). Away from meals, Hawthorn has been used against hypertension and, before sleeping, for its relaxing and sedative actions (Nazhand et al., 2020).

Hawthorn is a good source of proteins, amino acids, sugars, minerals, vitamins and phytochemicals including terpenoids, phenols, and flavonoids, as a medical herb which is a highly marketable source of medicines worldwide. In addition, natural pigments of Hawthorn, and the development of new areas of Hawthorn utilization, such as: the utilization in special medical use formulae, will produce excellent economic and social benefits. Researcher confirmed that the beneficial effects of free sugar and organic acid for human health such as antioxidants and anticarcinogenic properties (Park and Kim, 2018).

The rich nutrients in Hawthorn seeds and leaves should not be overlooked. It is important to look at Hawthorn chemistry as a whole. In the future, it is important to continue to explore the nutritional and health benefits of Hawthorn and to develop value-added foods and supplements based on the functional components of Hawthorn, based on extracts and active substances from other parts of Hawthorn (Zhang et al., 2022).

From perspective of economic significance, the main requests for these products are: safety, minimum adverse unwanted effects, greater bioavailability, and lower cost when compared with synthetic medications available on the market (Nazhand et al., 2020). The berries of some Hawthorn species are edible and can be used in jams, jellies, wines, and herbal teas, contributing to local economies in regions where they are harvested. In addition, cultural significant value can enhance local tourism and heritage. Hawthorn's ability to thrive in various soil conditions and its low maintenance requirements make it an excellent choice for sustainable landscaping and erosion control projects. Traditionally, its fruits are harvested from the wild types, but currently, there is a growing interest in economically cultivating Hawthorn for consumers, leading to the establishment of Hawthorn orchards (Tunç et al., 2025). Researcher also reported that fresh Hawthorn fruit can be eaten directly without any fumigation or washing by consumers, Hawthorn fruit has also been processed into many types of products with the continued development of science and technology, such as the enzyme industry, homogenizer products and membrane technology, a new phase in the processing of Hawthorn has been introduced and its product range is becoming more and more diverse (Zhanget al., 2022). There are many products made from Hawthorn on the market, with more than 150 types of products sold. Traditional Hawthorn products in China mainly include sugar gourd, Hawthorn cakes, Hawthorn preserves, canned Hawthorn, Hawthorn chips, and Hawthorn roll (Zhanget al., 2022). Eventually, studying ecological, horticultural uses, and potential medicinal applications is essential for understanding its classification and phylogenetic context.

## 2.3 Botanical characterization of Hawthorn

### 2.3.1 Taxonomy of the genus *Crataegus* L.

The genus *Crataegus*, commonly known as Hawthorns, belongs to the family Rosaceae (the rose family). It is a large and complex genus with significant taxonomic challenges due to hybridization, polyploidy, and morphological variability. Below is a general overview of the taxonomy of the genus *Crataegus:*

Taxonomic Classification

**Kingdom: Plantae (Plants)**

**Clade: Tracheophytes (Vascular plants)**

**Clade: Angiosperms (Flowering plants)**

**Clade: Eudicots**

**Clade: Rosids**

**Order: Rosales**

**Family: Rosaceae (Rose family)**

**Subfamily: Maloideae or Amygdaloideae**

**Tribe: Maleae (formerly part of Pyreae)**

**Genus: *Crataegus* (Hawthorn)**

**Common species: *Crataegus* spp. (common Hawthorn),**

In this study, for the taxonomy of *Crataegus*, the changes made by Christensen are explained under the related species. The taxonomic treatment of the genus based on Christensen's revision and the synonyms of the taxa are not given here.

**2.3.2 Classification and phylogenetic position within the Rosaceae family.**

Classification and phylogenetic position of Hawthorn within the Rosaceae family was linked in the 1980s, that the first studies on the intrageneric classification of the subfamily Maloideae (Amygdaloideae) began Phipps, 1983, Campbell et al. 1990 and Phipps et al. 1991). which is characterized by the production of pome fruits (a type of fleshy fruit with a core containing seeds, as seen in apples and pears). The most recent comprehensive demographic study of the Asian and European *Crataegus* species was conducted by Christensen (Christensen,1992). The genus *Crataegus* L. is one of the most important genera of the Rosaceae family and it is currently considered to comprise between 150 and 1200 species depending on the species concept employed and on the insertion of many probable taxa hybrid origins (Attard and Attard, 2019). Molecular phylogenetic studies have shown that Maloideae is a monophyletic group, meaning all its members share a common ancestor.

1. Tribe Maleae:
   - Within Maloideae, *Crataegus* belongs to the tribe Maleae, which includes other pome-bearing genera such as *Malus*, *Pyrus*, *Sorbus*, and *Cydonia*.

- The tribe Maleae is defined by shared genetic and morphological traits, including the structure of the pome fruit and floral characteristics.

2. Genus *Crataegus*:
   - *Crataegus* is a large and complex genus with over 200 species, many of which are difficult to distinguish due to hybridization and polyploidy.
   - Phylogenetic studies suggest that *Crataegus* is closely related to *Mespilus* (medlar) and *Amelanchier* (serviceberry), with which it shares a recent common ancestor.
3. Evolutionary Relationships:
   - *Crataegus* is part of a clade within Maleae that includes other genera with pome fruits.
   - The evolution of *Crataegus* is marked by frequent hybridization and polyploidy, which have contributed to its high species diversity and adaptability.
   - Molecular data (e.g., from chloroplast and nuclear DNA) support the placement of *Crataegus* as a derived group within Maleae, with its closest relatives being *Mespilus* and *Amelanchier*.

- Phylogenetic Relationships: Within the family Rosaceae, *Crataegus* is part of the subfamily Maloideae (also referred to as Pyrinae) alongside genera such as *Malus* (apples), Pyrus (pears), and Sorbus (rowan). This subfamily is characterized by the presence of pome fruits and has genetic and morphological similarities among its members. Molecular phylogenetic studies have indicated that *Crataegus* is closely related to other genera within Maloideae. Genetic analyses often show that Hawthorn diverged relatively recently within this lineage, suggesting a shared evolutionary history with other fruit-bearing members of the Rosaceae family. From diversity the genus *Crataegus* is highly diverse, with hundreds of species, many of which exhibit polyploidy (multiple sets of chromosomes) and hybridization, contributing to its complex taxonomy and phylogenetic relationships.

Taxonomic Challenges of this *Crataegus* genus is that firstly, species delimitation: Difficulty in distinguishing species due to overlapping characteristics. Secondly, Hybridization: Natural hybrids blur species boundaries. Thirdly, Regional variation: Different taxonomic treatments in North America, Europe, and Asia (Ibrahimov et al., 2020).

### 2.4 Hawthorn species in Kurdistan and neighbouring regions

Kurdistan of Iraq is known as a home for many endemic plants and rich in plant diversity, as well as many interesting and wild species such as Hawthorn. Hawthorn species *(Crataegus)* grows in the mountainous side, foothills, and temperate regions of Kurdistan and its neighboring areas, which include parts of Iraq, Iran, Turkey, and Syria (Radha and Khwarahm, 2022). These regions are part of the broader Zagros and Taurus Mountain ranges, which provide suitable habitats for various Hawthorn species due to their temperate climates and diverse ecosystems. Based on the available investigations, four endemic Hawthorn species have been identified in Iraq and the KRI (Kurdistan region of Iraq), notably *Crataegus azarolus, Crataegus monogyna, Crataegus pentagyna, and Crataegus meyeri* (Shahbaz and Sadeq, 2003 and Ahmad and Salih, 2019) and the most common species among these four in these forests is *Crataegus azarolus* (Ahmad and Salih, 2021). Hawthorn is extremely valuable for a variety of reasons, including medicinal, ecological and economic benefits (Ahmadloo et al., 2015). However, the diversity of Hawthorn species in the region is still being studied, and hybridization makes species identification difficult (Radha and Khwarahm, 2022).

Hawthorns are known by various local names, such as "Zirinc" in Kurdish. Generally, Hawthorns are an integral part of the flora in Kurdistan and neighboring regions, contributing to biodiversity and local traditions. Their adaptability to mountainous environments makes them a resilient and ecologically important genus in the area. In addition, many commercially cultivated fruits, wild fruit species, which are frequently used by local peoples, are collected from their natural habitats and used for many different purposes. One of the most important wild fruit species is Hawthorn. Since Hawthorn is still widely used in traditional medicine from ancient times to the present day, both in our country and around the world, plants of this genus have significant development potential (Sumerli, 2024). Table (2.1) shows an overview of Hawthorn species in Kurdistan -Iraqi regions.

**Table 2.1 Shows an overview of Hawthorn species in Kurdistan -Iraqi region.**

| No | Common Hawthorn Species | Distribution: | Habitat | Characteristics | Uses |
|---|---|---|---|---|---|
| 1 | *Crataegus monogyna* (Common Hawthorn) | Widely distributed across Kurdistan, Turkey, Iran, and Iraq. | Found in forests, scrublands, and along riverbanks. | Small tree or shrub with white flowers, red fruits (haws), and thorns | Used medicinally and as a hedge plant. |
| 2 | *Crataegus azarolus* | Found in the western parts of Kurdistan, Turkey, and Iran | Prefers Mediterranean climates and rocky slopes. | Produces yellow or red fruits that are edible and used in traditional foods. | Fruits are consumed fresh or used in jams and beverages. |
| 3 | *Crataegus aronia* (Syn. *Crataegus azarolus* var. *aronia*) | Common in the eastern Mediterranean region, including parts of Kurdistan. | Grows in open woodlands and scrublands. | Similar to *C. azarolus* but with smaller fruits | |
| 4 | *Crataegus pentagyna* (Small-flowered Hawthorn) | Found in the Caucasus region and northern parts of Kurdistan. | Thrives in mountainous areas and forest edges. | Small white flowers and dark red fruits. | |
| 5 | *Crataegus meyeri* | Found in the Caucasus region and northern parts of Kurdistan | Found in forests, scrublands, and along riverbanks. | | |
| 6 | *Crataegus orientalis* | Found in Turkey, Iran, and the Caucasus region. | Grows in rocky and mountainous areas | Large, orange-red fruits and deeply lobed leaves. | |

### 2.4.1 Distribution of the Hawthorn species in Kurdistan Iraq

Many factors, have effect of the distribution of plant species which has un significant role on the organization, sustainable use and conservation, especially in plant habit and population of species. However, there is not enough information available on the distribution of species. due to increased human degradation, Climate changes and pests and diseases are limited.

Geographical Distribution - Habitat and widespread species in Cyprus, happening on rocky mountainsides, cereal fields, hedges and along roads, across the altitudinal zone from 0 to 1500 m. It is original also in other Mediterranean countries and eastwards towards Iran and Iraq. It is not a demanding plant and can thrive on dry, infertile soils. It is a characteristic feature of the Mesaorian plain, where it is the only indigenous tree and therefore of significant aesthetic and ecological value. Its fruits provide food to many fauna species, like the hair, the moufflon and the fruit bat. It can be grown at altitudes from 0 to 1800 m and can be easily raised from seed and cuttings.

### 2.4.2 Distribution of the Hawthorn species in Iran

In this regards, Hawthorn (*Crataegus* spp.), a member of the Rosaceae family, is a fruit bearing plant that grows naturally in Iran. Its distribution predominantly spans the temperate regions of the world, typically between 30° and 50° N latitude (Phipps, 1983). Hawthorn, characterized by its thorny nature, thrives primarily on limestone soils in wooded and sunny areas up to 1300 m above sea level in Iran. As a prominent tree of the Zagros Mountains, it is widely distributed across Iran, Syria, Palestine, Lebanon, Iraq, Anatolia, and Turkmenistan (Sagheb Talebi et al., 2014).

However, in each area of Iran, wild Hawthorn trees or shrub is possible to get, however, mountainous area of Alborz and Zagroz of Iran are a native place (*Crataegus* spp.) species with every region in semi-arid to the forests of Caspian Sea. According to Donmez (2009), Sharifnia et al., (2011) Numerous research have absorbed on the genus *Crataegus* counting the treatment of the Iranian species and also studied reported that represented 27 taxa for four sections such as (Sanguinae Zabel ex Schneider, Pentagynae Schneider, Azaroli Loud and Oxyacanthae Zabel ex Schneider) that seven taxa are endemic. In the treatment by Christensen (1992), only 13 taxa were accepted from Iran, and thus Khatamsaz revision was overlooked (Christensen and Zielinski, 2008). Many *Crataegus* (Hawthorn) species are polyploids and can reproduce both sexually and apomictically (Lo et al., 2010). In addition, polyploidy, hybridization and clinal variations are those factors are affecting *Crataegus* taxonomically which is difficult genus. Therefore, increase the available knowledge for the species in this genus is more important due to incomplete sampling and low-values, the taxonomy of the Hawthorns in Iran is not satisfying and need a more comprehensive evaluation.

One of the rich gene pools of *Crataegus* is Iran which has become famous because of they have adapted to various conditions and dispersed in various parts of the country. The largest number of *Crataegus* populations is observed in the mountainous regions of Iran (Rahmani et al., 2015 and Emami et al., 2018). Furthermore, Eghlima et al., (2025) reported that 17 species of the genus *Crataegus*, exhibits widespread distribution across Iran. Hawthorn has garnered significant attention in recent years as a prominent herbal remedy in phytotherapy and culinary applications. The ecotypes under investigation were sourced from elevations ranging from 1205 to 1681m above sea level. Recherché reported that with approximately 22 hybridized varieties identified in Iran (Alirezalu et al., 2018). In addition, researcher demonstrate that molecular ISSR markers was used to determine molecular relationships among five *Crataegus* species in Iran that selected species have common medicinal properties and accessibility in the Northwest including *C. monogyna, C. meyeri, C. pentagyna, C. pontica, and C. aronia* (Beigmohamadi et al., 2021).

### 2.4.3 Distribution of the Hawthorn species in Turkey

Turkey is one of the key genetic centers the diversity for wild-grown *Crataegus* species. Browicz (1972) described some of Turkey's *Crataegus* species in Flora of Turkey and studies also by Doenmez have contributed information about new and existing *Crataegus* species (Doenmez, 2007).

Currently, more than 20 *Crataegus* species have been identified in Turkey, including *C. monogyna Jacq., C. pentagyna* Willd., *C. azarolus* L., *C. orientalis* M. Bieb., C. rhipidophylla Gaud., and C. laevigata (Poir) DC (Table 2.2). In addition, there are some hybrid species in different ecological regions (Table 2.3) (Caliskan et al., 2016). Asia, Central America, and the Mediterranean are those place that Many Hawthorn species are grown for their edible fruit that C. *monogyna* Jacq. being the species commonly cultivated in Mediterranean countries especially, *C. azarolus* and *C. orientalis* genotypes are found to be promising in front of fruit quality characteristics in horticultural sector.

Hawthorns naturally grow in mountainous regions, shrublands, and rocky areas throughout our country, without any cultural interventions. Consequently, it is found in almost every region of Turkey, including the Aegean, Eastern Mediterranean, Southern Anatolia, Inner Anatolia, and Northeast Anatolia (Dönmez, 2014). There is not development a typical Hawthorn cultivar whereas some accessions have very good fruit quality characteristics. The most important limiting factor is propagation for the Hawthorn growing. We can say that more detail studies are needed for the widespread of Hawthorn growing in the future.

In addition, Serce et al., (2011) indicated that C. aronia var. aronia genotypes are closely related to *Crataegus aronia var. dentata* genotypes, however *C. monogyna var. azarella* genotypes are not included in same clustered with *C. aronia.*

In addition, numerous taxonomists are divided *Crataegus spp* into 40 sections that it is closest to the genus *Osteomeles* Lind according to the latest phylogenetic analysis of the genus. Turkey to Iran is good center for section *Crataegus* about the genetic diversity, that new diversity centers are proposed, according to field observations in Turkey.

**Table 2.2 Hawthorn species are found in turkey.**

| Species | Species |
|---|---|
| *C. tanacetifolia* (poir) Pers | *C. meryeri* Pojark |
| *C. orientalis* M.Bieb. FI.taur.cauucas | *C. caucasica* C.Koch |
| *C. x bornmuelleri* Zabel in Beissner | *C. ambigua* C.A.Mey. ex Backer |
| *C. azarolus* L.Sp. | *C. heterophylloiders* pojark |
| *C. pontica* C.Koch Verh | *C. longiper* Pojark. |
| *C. pentagyna* Waldst. And kit | *C. microphylla* C. Koch |
| *C. davisii* Browicz | *C. rhipidophylla* Gand. |
| *C. pseudoheterophylla* pojark | *C. monogyna* Jacq |

**Table 2.3 Hybrid Hawthorn species in the flora of turkey.**

| Species | Species |
|---|---|
| *C. orientalis* X *C. tanacetifolia* | *C.* X *bornmuelleri* |
| *C. monogyna* X *C. tanacetifolia* | *C.* X *yosgatica* |
| *C. azarolus* X *C. monogyna* | *C.* X *sinacica* |
| *C. monogyna* X *C. pentagyna* | *C.* X *rubrinervis* |
| *C. microphylla* X *C. rhipidophylla* | *C.* X *browicziana* |
| *C. monogyna* X *C. rhipidophylla* | *C.* X *kyrtostyla* |

### 2.4.4 Distribution of the Hawthorn species in Syria

According to the Shahat et al., (2002) wild plant *Crataegus* is well spread in the southern area of Syria regarding medicinal and edible fruits that this genus is considered as an extra source of income for local farmers. The data for *Crataegus* in the region of Syria and the Middle East are confused, ranging from two (Muzher, 1998) to 13 (summarized in Dmeria, (2001) *Crataegus* species described. In the Flora of Syria, Palestine and Egypt (Post, 1896), five Crataegus species are described: *Crataegus azarolus* L., *Crataegus* × *sinaica* Boiss., *Crataegus oxycantha* L., *Crataegus monogyna* Jacq. and *Crataegus orientalis* M.Bieb. This leads to the conclusion that the morphological characters are somewhat weak for identifying different species and the borders between each individual species. In the Al Arab mountains, Sweida province, Syria, the occurrence of *Crataegus* × *sinaica* and *Crataegus azarolus* is known. Many species of *Crataegus* are used and cultivated in orchards as edible fruit plant. In the Arab mountains, Sweida Province, Syria, three *Crataegus* species were recognized: *Crataegus azarolus var. aronia* L., *Crataegus* × *sinaica* Boiss. ssp. *sinaica* and *Crataegus monogyna var. monogyna* Jacq (Albarouki and Peterson, 2007).

## 2.5 Morphological characterization of Hawthorn

Morphological characterizations are the first step in describing germplasm before biochemical and molecular studies (Hoogendijk and Williams, 2002). The inventory of plants by describing morphological features helps in their preliminary study of phenotypic variation and is important for adopting conservation strategies for plant genetic resources as well as for the establishment of germplasm collections (Podgornik et al., 2010). Hawthorn comprises a semi-evergreen shrub or small tree mostly growing up to 5-15 m that considered to be recently among the global important edible wild horticultural fruits (Obideen et al., 2024). Phenological parameters are highly diverse including leaf morphology, the color of the fruits, seed numbers and seed traits. Generally, description of Hawthorn is bark typical is smooth grey at the juvenile stage, with shallow longitudinal fissures and narrow ridges in the older phenological stage. The thorns consist of sharp-tipped branches coming from other branches or the trunk, with a length of 1-3 cm., The leaves are alternate, multi-row coils, simple, lobed or serrated and have a straight or toothed edge2-6 cm long dark green depending on the species. They grow spirally positioned on long shoots and they are organized in clusters on spur shoots on the branches or twigs and roots can be shallow and fibrous, tolerant of various soil types. (Benabderrahmane *et al.,* 2021). Flowers usually appear in clusters (corymbs) during spring, attracting various pollinators, including bees and it has 1–2 cm diameter, mildly fragrant from 5 to 25 anthers, 5 sepals and 5 petals and a greenish calyx, that petals are white and pink usually, they are longer than the sepals. The fruit, known as a “haw,” is similar to a berry. It is, anatomically, a pome with between one and five pyrenes having “stones” somewhat similar to plums (Orhan 2018). Fruit can be yellowish, reddish, or blackish-purple and it is usually fleshy, 5-15 mm diameter that each fruit has between one and five hard seeds (Martinelli et al. 2021).

In addition, Researcher confirmed that characterization of morphological parameters has been performed in many species of the genus *Crataegus*, such as plant height, branches/plant, thorn number, plant canopy and berry weight and data from five randomly selected plants (in triplicate replications) were analyzed using ANOVA (Ahmad, et al., 2005). Intra- and interspecific variability was observed, allowing the conventional breeding for better varieties (Wu et al., 2021). Morphological characterization of the *Crataegus* genus has been performed in wild regions of Ukraine using vegetative characteristics (Sydora, 2018). The section Oxyacanthae (containing the common Hawthorn) was divided into five species subgroups based on their morphological vegetative features that correlated with their geographical distribution (Sydora, 2018).

## 2.6 Morphological diversity of the *Crataegus* species

Hawthorn exhibits a diverse range of morphological characteristics that enable it to thrive in various temperate habitats. Significant morphological diversity of Hawthorn plants shows, especially in their flowers, fruits and leaves. Researcher also confirmed environmental conditions, biological variability including cross-pollination via some birds which disperse the seeds in the environment and generative propagation through both natural and anthropogenic processes are those factors to largely attributed to the region's unique for diversity. This variation is thought to result from birds and some mammals that consume the fruits and act as dispersal agents (Yılmaz et al., 2010).

Techniques for morphological analysis:

- Traditional descriptive approaches.
- Quantitative traits for species differentiation.
- Hawthorn refers to several species within the genus *Crataegus* in Kurdistan Iraq which belongs to the family Rosaceae. These shrubs or small trees are widely known for their ornamental value, medicinal properties, and ecological importance. Below is a botanical characterization of Hawthorn:
- The genus *Crataegus* comprises numerous species, with notable examples including *C. azarolus, C. monogyna* (common Hawthorn), *C. laevigata* (English Hawthorn), *and C. crus-galli* (cockspur Hawthorn).

## 2.7 Phytochemicals Compositions of *Crataegus* species and their medicinal uses

Hawthorn (genus *Crataegus*) is renowned for its rich phytochemical composition, which contributes to its medicinal properties and health benefits (Eghlima, et al., 2025). Therefore, Hawthorn, as an edible and medicinal fruit, contains a large number of naturally occurring bioactive components with abundant beneficial functions that are generally safe and reliable to be used to gradually improve health in daily life. Martinelli et al., (2021) reported that Biochemical markers are properties discovered through the study of biochemical constituents of plants, including primary and secondary metabolites. Researcher also confirmed that Hawthorn fruits contain high levels of numerous valuable secondary metabolites, including flavonoid, vitamin C, glycoside, anthocyanin, saponin, tannin and antioxidant levels (Żurek et al., 2024) and phenolic compounds (Karar and Kuhnert 2015).

It is also a valuable source of dietary bioactive components for the development of functional foods or other nutraceuticals and for the prevention and management of certain chronic diseases (Ma et al., 2024). The leaves, flowers, and berries of Hawthorn contain an abundance of many chemical properties and rich in macro- and micronutrients which are thought to be responsible for its pharmacologic effect. Among them Hawthorn fruits are the main active constituents of include polyphenols, flavonoids, antioxidants, vitamins, and organic acids, and the pigments of fruits are mainly carotenoid and anthocyanin, while these chemicals have important health benefits (Alirezalu et al. 2020). (See Figure 2.1) In addition, fruit rich in organic and phenolic acids, the organic acids in Hawthorn are mainly malic, citric, succinic, ascorbic, quinic, oxalic, linolenic, and lauric acid (Cosmulescu et al., 2020). The number of organic acids varies depending on the variety of Hawthorn. Citric and malic acids are the highest in Hawthorn fruit, with malic acid averaging 1128.68 mg/100 g FW (Cosmulescu et al., 2020). In addition, apart from organic acids, Hawthorn is also rich in phenolic acids. The main phenolic acid in Hawthorn is chlorogenic acid (8410–13826.7 μg/g), accounting for more than 80% of the total phenolic acid (Sun et al., 2021).

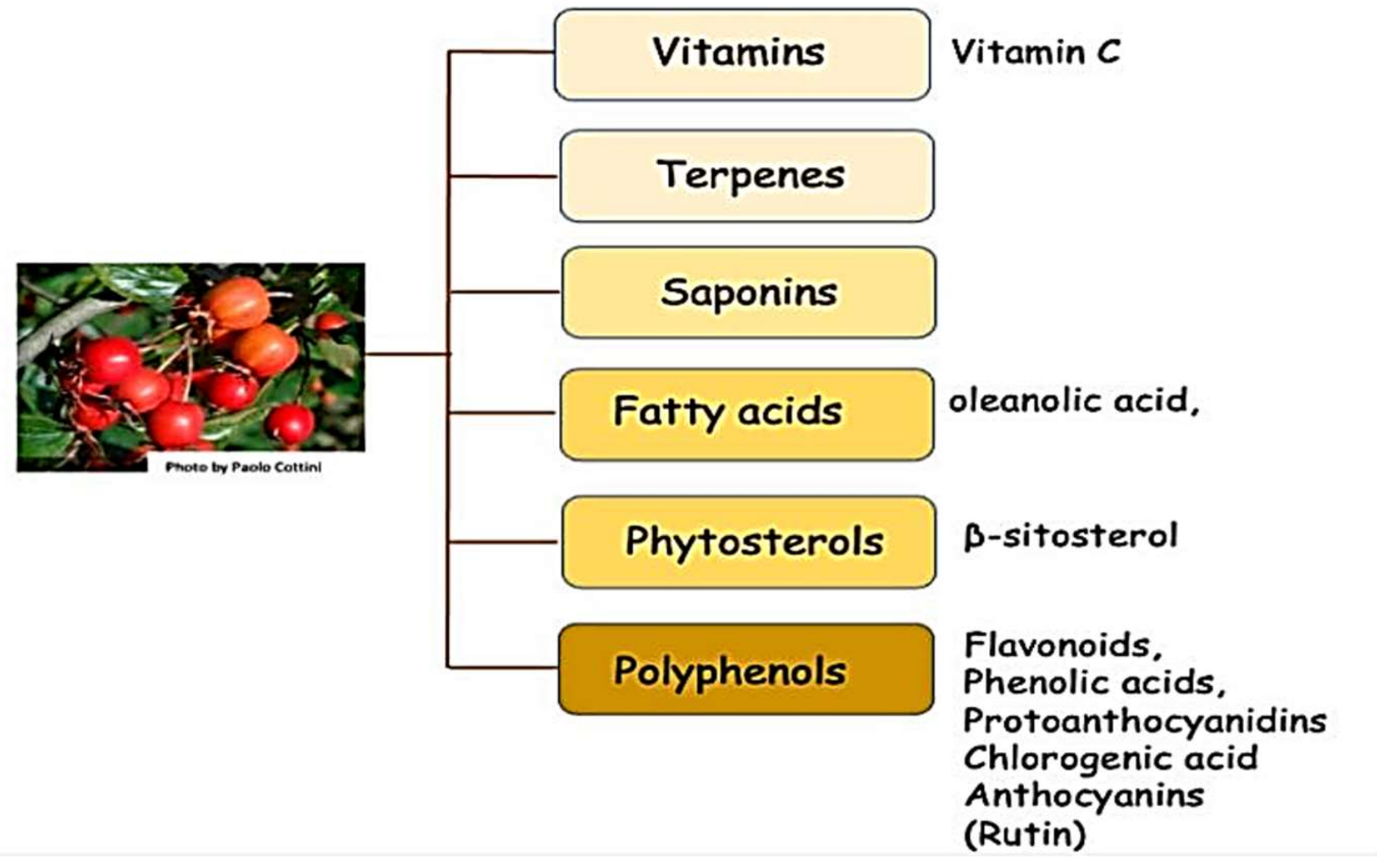

**Figure 2.1 The main phytochemical components of *C. monogyna* are divided into different principal classes.**

Acids, sugars and sugar alcohols, minerals, vitamins, and amino acid composition were among the analyzed constituents, however, overwhelming number of published articles have been focusing on polyphenolic compounds, as the most important Hawthorn bioactive phytochemicals which may provide health benefits to humans (Venskutonis, 2018). The interest in phenolic and polyphenolic compounds particularly increased in the era of functional foods (Shahidi, 2004). Besides strong antioxidant activity, phenolic compounds have demonstrated numerous protective effects against chronic diseases, which have recently been reviewed (Shahidi and Yeo, 2018). Phytochemical composition of 20 phenolic compounds quantified in *C. azarolus* and *C. monogyna* fruits were compared (Mraihi et al., 2015). Selected phytochemicals and antioxidant potential were studied in the C. monogyna ethanolic extracts from bark, leaves and berries: the highest Total phenolics content (TPC), radical scavenging potency as well as the levels of oleanolic acid, quercetin and lupeol were found in the bark extract, while the highest ursolic acid content was in the berries extract (Rezaei-Golmisheh et al., 2015). Thirty-six compounds were reported in different extracts of Hawthorn fruit, 15 of them were tentatively identified in Hawthorn fruits for the first time (Miao et al., 2016) (See figure 2.2).

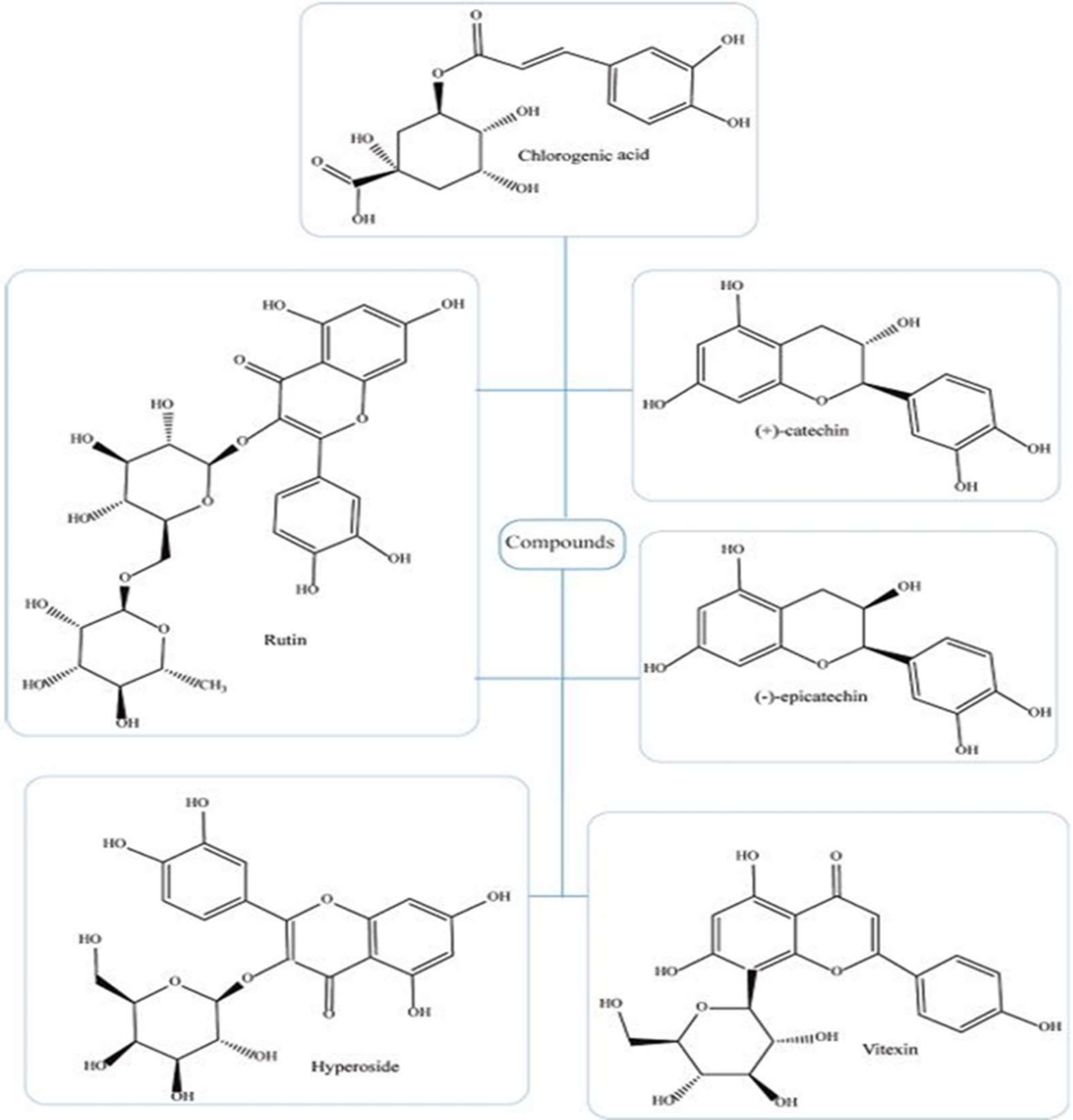

**Figure 2.2 Overview of the main compounds found in Hawthorn.**

Recent research progress indicates that some of the most critical biological activities natural

Hawthorn fruit are anti-inflammatory, antimicrobial, gut-protective, antidiabetic, cardioprotective, hepatoprotective, and anti-cancer properties that may prevent or even treat diseases, indicating an impact on the management of health problems. However, the underlying mechanisms of their relevant functions are not fully understood, such as the mechanisms and targets of anti-cancer and neuroprotective effects, requiring further research (Ma et al., 2024).

Although synthetic antioxidants and antimicrobial agents can actively be used in food processing due to high stability and efficiency and low cost, there are significant concerns related to their potential health risks and toxicological aspects (Munekata et al., 2022). Therefore, some research has been performed to evaluate the performance of natural antioxidants and antimicrobials such as essential oils and plant extracts as alternatives to synthetic antioxidants (Lorenzo, et al., 2018 and De Carvalho et al., 2019). On the other hand, limited and anti-microbials and high price as well as shortage of new sources of safe and in expensive antioxidants and antimicrobials of natural origin could be a plausible reason for the food and pharmaceutical industries to use synthetic antioxidants instead (Ramos et al., 2014).

In addition, the number of bioactive compounds such as flavonoids and phenolic acids is also affected by genetic variation among species, within the same species, and maturity of plant organs at its harvest (Wang, et al., 2009). The variation in physicochemical characteristics, phytochemicals, and antioxidant activity among different species of *Crataegus* revealed were well documented by previous studies (Mraihi et al., 2015). Eghlima et al., (2025) reported that generally, the observed morphological, biochemical and antimicrobial characteristics among different *Crataegus persica* ecotypes can be influenced by altitude which is associated to general climatic trends, including reduced temperature, increased radiation, decreased atmospheric pressure and other environmental changes.

### 2.7.1 Flavonoids

Flavonoids are an important category of polyphenols that have important biological effects. Flavonoids are usually extracted by a combination of ethanol and water in different proportions and can also be extracted by the natural deep eutectic solvents, which could solubilize moderately polar flavonoids in a low-cost and environment friendly manner (Chaves et al., 2020). It was found that (+) catechin was a main flavonoid compound in Hawthorn pulp, ranging from 9.57 to 29.9 mg/100 g based on different extraction methods (Özcan et al., 2022). In addition, apigenin, luteolin, chrysin, and quercetin were also found in the whole Hawthorn fruit (Li, Gao et al., 2022). The cultivation regions can influence the total flavonoid content (TFC) in the whole Hawthorn fruit. For example, Hou et al. (2020) evaluated the TFC in the whole Hawthorn fruits from five places in China, and they found that the TFC in samples from Henan was the highest, whereas the TFC of the Jiangsu samples was the lowest, and epicatechin was the predominant flavonoid in all samples from all regions. Flavonol-O-glycoside and flavono-C glycoside, vitexin-2-O-rhamnoside and acetyl vitexin-2-O-rhamnoside are three an essential flavonoid that present in the *Crataegus* species. Flavonoids own antimicrobials and antioxidant properties and stimulate antibody production (Kumar et al., 2013).

Flavonoids, more than other secondary metabolites, are extensively employed in chemotaxonomic studies because of their distinct quantitative and qualitative patterns (Ringl et al. 2007). Extant literature suggests that the phenolic composition of C. persica fruit differs significantly from other species and ecotypes within the genus. A comparative study of various *Crataegus* species, including *C. monogyna, C. meyeri, C. pseudoheterophylla, C. pentagyna, and C. pontica, revealed that C. meyeri, Crataegus pentagyna, and Crataegus pseudoheterophylla* contained notably high percentages of phenolic, tannic, and flavonoid compounds (Alirezalu et al., 2020). Hawthorns are an ideal source of antioxidants (Venskutonis 2018).

### 2.7.2 Antioxidants Activity

The therapeutic potential of *Crataegus* spp. is primarily attributed to their antioxidant content, which is associated with phytochemicals such as ascorbic acid, carotenoids, and flavonoid compounds (Donno et al., 2017). Furthermore, all three *Crataegus species* tested demonstrated significant antioxidant activity against 2,2-diphenyl-1-picrylhydrazyl (DPPH) and in the ferric reducing antioxidant power (FRAP) assay (Radi et al. 2023). The quality parameters of *Crataegus* fruits include fruit acids, which facilitate nutrient digestion and stimulate blood circulation. These acids are rapidly oxidized in the body and do not exert harmful effects. Moreover, their salts are alkaline-forming elements, rendering them highly important in human nutrition. Although the antioxidant activity of Hawthorn fruits has been extensively reported in the literature (Ruiz-Rodríguez et al. 2014, Ding et al. 2010 and Alirezalu et al., 2020) *C. persica* remains a prominent species in Iran, traditionally used for medicinal purposes. Given its endemic status in Iran and the paucity of comprehensive phytochemical information on its various ecotypes. Oxidative stress is a redox imbalance accompanied by increased production of reactive oxygen species (ROS) and overwhelming antioxidant defenses (He et al., 2021). The antioxidant effects of plant flavonoids have been frequently reported, and possible mechanisms include inhibiting the production or direct elimination of free radicals, or inhibiting lipid peroxidation (Huang et al., 2022). Recent studies have reported that the Hawthorn pulp and its polyphenols exhibit potent antioxidant activity in vitro. It was shown that the total antioxidant capacity of the Hawthorn pulp was 0.321.84 mmol Fe2+/g dry weight (DW) (Alirezalu et al., 2020), and it had a more stable anti-lipid peroxidation capacity than vitamin C (Feng et al., 2022). The antioxidant activity of the Hawthorn pulp was mainly attributed to its free phenolics, accounting for 35.3%–37.8% of its antioxidant activity, followed by insoluble-bound, soluble esterified bound, and soluble glycosylated–bound phenolics, accounting for 25.0%–27.0%, 23.4%–25.7%, and 9.4%–15.7% of its antioxidant activity, respectively (Feng et al., 2022).

In addition, Ma, et al., (2024) suggest that polyphenols are the main antioxidants in the whole Hawthorn fruit, and the underlying antioxidant mechanisms involve the activation of the Nrf2/HO-1 signaling pathway, enhancement of antioxidant enzyme activity, and ultimately induction of cellular antioxidant defense systems. Further studies are needed to verify whether other signaling pathways and oxidative stress molecules are involved in the invivo antioxidant effects of whole Hawthorn fruit and their polyphenols and to provide more reliable evidence for the development of natural antioxidant functional foods in the future.

### 2.7.3 Phenolic acids

Polyphenols are a large family of organic compounds that are characterized by at least one hydroxyl group in the benzene ring. Polyphenols are abundant in many plant-based foods (Lund, 2021). Usually, methanol and ethanol are common solvents for the extraction of polyphenols from plants. A study by Sultana et al. (2009) found that aqueous methanol was comparatively more efficient than ethanol and acetone for polyphenol extraction from different medicinal plant extracts. The predominant polyphenolic compounds in whole Hawthorn fruits include phenolic acids, flavonoids, and proanthocyanidins. In addition, Phenolic acids are one of the main categories of phenolic compounds in the whole Hawthorn fruit. Besides the Soxhlet extraction and ultrasonic-assisted extraction that are commonly used in phenolic acid extraction, supercritical fluid extraction and accelerate solvent extraction are also developed as alternative techniques (Arceusz et al., 2013).

Gallic acid, caffeic acid, and syringic acid were found in fresh and dehydrated whole Hawthorn fruits (Benabderrahmane et al., 2021). In addition, gallic acid, neochlorogenic acid, and cryptochlorogenic acid were revealed to be active components of the whole Hawthorn fruit for the treatment of hyperlipidemia and cardiovascular diseases (CVDs) (Sun and Zengetal., 2022). The total phenolic content (TPC), the 1,1-diphenyl,2 picrylhydrazyl (DPPH) free radical scavenging activity and metal content (Zn, Fe, Cu, Mn, Cd, Cr and Pb) have been identified in wild C. monogyna from Serbia (Velickovic et al., 2016).

As mention before, Hawthorn is rich in nutrients and beneficial bioactive compounds found in its various parts, including fruits, flowers and leaves, and also contains vitamins A, C, E, K and B group vitamins. Therefore, Hawthorn is a promising raw material for traditional treatment and has also been demonstrated to possess health benefits for humans, especially with cardiovascular protective effects. Nazhand et al., (2020) In summary, the whole Hawthorn fruit has been demonstrated to benefit the cardiovascular system, mainly through regulating blood pressure, regulating blood lipids, and alleviating heart failure symptoms. However, its other health benefits on humans require further studies for verification. For example, the beneficial effects of the whole Hawthorn fruit on human intestinal health and its anti-inflammatory effects need to be confirmed in more clinic arterials (Çakmak, 2024).

Cardiovascular health benefits of various *Crataegus* drugs and other preparations have been most widely studied. In addition, the potential role of Hawthorn in cardiovascular diseases was specifically reviewed by Chang et al. (2005). Hawthorn extracts have been used for this purpose in many countries, particularly against mild forms of chronic heart failure. Many studies have reported the anti-inflammatory effects of the whole Hawthorn fruit and its bioactive components in vitro and in vivo. The whole HFE is rich in phenolic compounds, especially chlorogenic acid and (−)-epicatechin, which are widely known for their anti-inflammatory effects (Abdel-Rahman et al., 2021and Nguyen et al., 2021). Overall, a large number of studies support the anti-inflammatory activity of the whole (Hereditary hemochromatosis) HFE and Hawthorn pulp, and its anti-inflammatory mechanism is related to the regulation of NF-κB, ERK1/2, MLCK PMLC, MAPK, and TLR4 signaling pathways, leading to the down regulation of related pro-inflammatory cytokines, such as TNF-α, IL-6, IL-8, and IL-1β. However, as mentioned earlier, the mechanism by which Hawthorn pulp alleviates inflammation- associated chronic diseases is not fully understood and requires further validation in animal experiments.

## 2.8 Molecular characterization of Hawthorn

An essential classification of germplasm for providing information about the characters of the genotype is mor important that can be utilized in conservation and breeding programs. Information about the genetic variations present within and between various plant populations and their structure and level can play a beneficial role in the efficient utilization of plants (Cole, 2003). Numerous types of agro-phenological parameters have been positively utilized to explore the diversity of genotypes that also many significant factors are used to discovery of structure and level of these variations with evolutionary background including, process of gene flow, mating system population density. During the last three decades, the world has witnessed a rapid increase in the knowledge about the plant genome sequences and the physiological and molecular role of various plant genes, which has revolutionized the molecular genetics and its efficiency in plant breeding programmers (Nadeem et al., 2018). Molecular markers play a vital role in determining the genetic relationships of entire plant populations, identification and classification of plant species, which one of the reliable markers they provide a reliable and stable method to analyze genetic diversity and variation (Benchimol-Reis, 2024). High polymorphism, co-dominancy, and reproducibility has offer significant effective tools that are of importance to characterize and quantify plant species (Goswami et al., 2022).

Molecular characterization of Hawthorn provides a comprehensive understanding of its genetic diversity, gene expression, and metabolic pathways. These insights are crucial for optimizing its medicinal use, conservation, and crop improvement. Advances in molecular biology techniques continue to enhance our knowledge of this complex and valuable genus. (Salgotra and Chauhan, 2023). From botanical point of view, the Hawthorn is a fruit. It is a diploid plant with 2n = 34 chromosomes. Genetically Hawthorn is assembled a genome size of 823.41 Mb (2n=2X=34). In addition, more than 50,000 nucleotide sequences deposited in NCBI regarding the phylogenetically related species *Crataegus pinnatifida* Bunge. The sequence of the entire genome of this species is available (Zhang et al., 2022). Limited information is available on the genetic resources of wild Hawthorn at the global level. Even though the Kurdistan region is considered to be a diversity center for *Crataegus* spp. with a wide distribution across the Irano-Turanian region and Zagros mountains, little information is available on the extent of the species genetic diversity. There is a lack of or no knowledge of the taxonomic profile and genetic diversity of Hawthorn species in Iraq. *Crataegus* species in the Kurdistan region of Iraq have been declining due to long-term land use and rapid construction; these problems may become more serious due to climate change. To certify the longevity of this species, it is necessary to take an integrated approach, which includes field surveys for genetic diversity, cultivation, and natural habitat restoration. This work represents the first comprehensive taxonomic and structural study of Iraqi *Crataegus* species using SSR, ISSR and SCoT markers, and to the best of our knowledge.

### 2.9 Application of DNA marker techniques in studying Genetic diversity of Hawthorn

The fate of plant breeding has changed during the development of molecular marker technology in the 1980s. Different types of molecular markers have been developed and advancement in sequencing technologies has geared crop improvement. in the last decades, numerous reviews have been available regarding molecular markers information. The development made genomic selection with genome editing, plant breeding and molecular genetics has donated to diversity available for plant and greatly complemented breeding stratagems which is comprehensive understanding of molecular markers and provided deeper insights for plant diversity. Altogether, the history, the types of markers, their application in plant sciences and breeding, and some recent advancements in genomic selection and genome editing are discussed (Nadeem et al., 2018) During the last decade several novel DNA-based markers have been rapidly developed for characterizing the *Crataegus* genome and to study genetic diversity within and among wild landraces of this species. Such marker techniques include randomly amplified polymorphic DNA (RAPD) (Erfani-Moghadam et al., 2016), inter-simple sequence repeats (ISSRs) (Sheng et al. 2017 and Emami et al. 2018), and simple sequence repeats (SSRs) (Khiari et al. 2015). SSR, ISSR and SCoT markers are widely used in genetic research because they are convenient codominant, and highly polymorphic (Amiteye, 2021; Sharef et al., 2026).

#### 2.9.1 PCR-based techniques

The PCR technique was developed by Cary Mullis in 1983, as a technique which could amplify a small quantity of DNA without the application of any living organ isms (Mullis et al., 1986). The vital steps involved in PCR reactions are denaturation, annealing and extension. For more information about PCR and its protocol, see the article of Joshi and Deshpande (Joshi and Deshpande, 2011).

Amplification of some discrete DNA products by PCR technique which derived from regions of DNA that flanked by regions of high homology with the primers. These regions must be close enough to one another to permit the elongation phase (Kumar *et al.*, 2009). Primers in the first category are designed arbitrarily/or semi-arbitrarily, there is no information about the flanking sequence of the region which is amplified (Agarwal *et al.*, 2008).

To control genetic relationships within plant populations with almost 100% reliability molecular markers are utilized (Güney et al., 2018). And also used for indicated plant breeding, plant systematics, and the evaluation of genetic resources therefore, to assess the genetic variation in *Crataegus* spp different molecular marker have been used including, random amplified polymorphic DNA (RAPD), sequence-related amplified polymorphism (SRAP), simple sequence repeats (SSRs), amplified fragment length polymorphism (AFLP), inter simple sequence repeat (ISSR) and single nucleotide polymorphism (SNP) (García-Martínez *et al.*, 2006; Henareh *et al.*, 2016; Mata-Nicolás *et al.*, 2020; Brake *et al.*, 2021; Caramante *et al.*, 2021 and El-Mansy *et al.*, 2021; Sharef et al., 2026), gene-targeted markers, a novel marker system, such as conserved DNA-derived polymorphism (CDDP) and start codon targeted polymorphism (SCoT) ( Collard and Mackill 2009ab and Abdeldym *et al.*, 2020).

#### 2.9.1.1 simple sequence repeats (SSRs)

Microsatellites are also called as SSRs which are tandem repeat motifs of 1–6 nucleotides that are present abundantly in the genome of various taxa. SSRs represent the lesser repetition per locus with higher polymorphism level (Kalia et al., 2011). This high polymorphism level is due to the occurrence of various numbers of repeats in microsatellite regions and can be detected with ease by PCR (Nadeem et al., 2018).

Of these molecular marker's types, SSRs markers are DNA fragments utilized to amplify specific DNA regions, primarily used in genetic mapping, diversity studies, and marker-assisted selection in plant breeding programs. These short, tandemly repeated DNA sequences are highly polymorphic, convenient and can be developed for many species, making them co-dominant and highly abundant (Hong et al., 2011). Ninety-one Hawthorn genotypes have been genetically analyzed using simple sequence repeat (SSR) markers (Güney, et al., 2018). A total of 265 alleles have been found from 32 SSR primer pairs (Güney et al., 2018), four populations, eight sub-populations and four major clusters were identified using phylogenetic analysis (Martinelli et al., 2021).

**2.9.1.2 Inter simple sequence repeat (ISSR)**

ISSR marker technique is very simple, fast, low-cost, highly discriminative, reliable, requires a small quantity of DNA sample, does not need any prior primer sequence information and non-radioactive (Bhatia *et al.*, 2009). Importantly, the same ISSR primers can potentially be used universally across plant phylogenetic diversity and also the basic technique of ISSRs is flexible and can be modified with options for implementation for a broad range of projects and budgets (Gemmill and Grierson, 2021). ISSRs markers are genomic regions flanked by microsatellite sequences, PCR amplification of these regions yields multiple products, making ISSR markers a dominant multilocus marker system for studying genetic variation in organisms (Ng et al. 2015 and Sevindik et al. 2023).

In ISSR molecular markers the target sequences are plentiful, revealing many polymorphic loci compared to other dominant markers in the eukaryotic genome. It is composed of di-, tri-, tetra-, or pentanucleotide repetitions, with or without a one-to-three nucleotide anchor targeting the microsatellite region of the genome. Additionally, ISSR markers are also widely used in genetic studies due to their ability to reveal high levels of polymorphism, making them valuable for assessing genetic diversity and population structure (Dhutmal et al., 2018).

**2.9.1.3 Start codon targeted (ScoT) marker**

Start codon targeted (SCoT) polymorphism marker, a targeted fingerprinting marker technique, has gained considerable importance in plant genetics, genomics, and molecular breeding due to its many desirable features., a highly conserved region in plant genes is flanking the start codon during SCoT marker targets. Therefore, it is a simple, novel, cost-effective, highly polymorphic, and reproducible molecular marker for which there is no need for prior sequence information that it can distinguish genetic variations in a specific gene that link to a specific trait. In the recent past, SCoT markers have been employed in many commercially important and underutilized plant species for a variety of applications, including genetic diversity analysis, (Rai 2023). SCoT markers are simple and less time-consuming and do not require prior sequence information, and generate polymorphism is linked to functional genes and their corresponding traits (Gupta et al. 2019). The arbitrary SCoT markers present within gene regions that cover genes on both plus and minus strands of DNA (Majeed et al. 2024). SCoT offers stability, reproducibility, and reliability over ISSR, AFLP, and RAPD, making it effective for population studies, genetic mapping, and marker assisted selection programs (Rai 2023). ScoT markers were usually reproducible but exceptions indicated that primer length and annealing temperature are not the sole factors determining reproducibility. ScoT marker PCR amplification profiles indicated dominant markers like RAPD markers (Collard and Mackill 2009a; Thakur *et al.,* 2021).

Recently, to diversity in Hawthorn plant SCoT marker are used, that seven *Crataegus* species were examined using five newly designed with thirteen reported SCoT markers for the diversity of gene and population structure. Results confirmed that 148 polymorphic bands with 94.87% polymorphic loci with eighteen SCoT markers were produced, and also five clades in the phylogenetic tree were constructed for Thirty-six *Crataegus* accessions by SCoT markers. *C. bretschneideri* accessions were clustered into a single clade which had a closer relationship with *C. pinnatifida* Bunge accessions. mixed gene pool in *C. bretschneideri* accessions was indicated by Population structure analyses. These results confirmed that *C. bretschneideri* may be of hybrid origin. According to Rahmani et al., (2015) that to evaluate genetic variation among and within the populations of *C. pontica* SCoT markers could be effectively utilized. Genus *Crataegus* is genetically confirmed by using Newly designed SCoT markers (Zhang et al., 2023).

### 2.10 Importance of DNA marker techniques in studying diversity

The increase in the global population, coupled with the unregulated exploitation of plant resources to satisfy human demands, as well as habitat destruction due to land expansion, urban sprawl, and industrial activities, has contributed to the decline and swift depletion of plant germplasms (Corlett, 2016). Detecting genetic variation represents the initial step toward the efficient utilization and safeguarding of genetic resources (Koornneef et al., 2004). Germplasms play a crucial role in supplying breeders with the necessary materials for the advancement of plant species (Nadeem, 2021). therefore, germplasms with wide diversity are preferred. For years, morphological, pomological, and molecular marker systems have been used to determine genetic diversity among different plants, including Hawthorn (Yildiz et al., 2023). Among these marker systems, molecular techniques provide more precise results than other methods because they are not influenced by environmental conditions (Nadeem et al., 2018) Given that certain traits contributing to plant resistance to abiotic and biotic conditions are governed by multiple genes (Yaman, 2021).

It is clear that the genic molecular markers and especially the functional markers are extremely useful source of markers in plant breeding for marker-assisted selection because these markers may represent the genes responsible for expression of target traits (Madhumati, 2014). The application of biomolecular information to know taxonomic and evolutionary relationships contributes to solving most of the classification problems by diagnosing and inferring the existence of a close relationship between living organisms. In addition, this molecular information is one of the important contributions used recently in the classification of individual plants it is a field of molecular systematic whether or not that led to the emergence of what is known as molecular taxonomy it is an essential field of study, as molecular taxonomists believe that molecular data is more reliable than Phenotypic data to know the true origin between organisms and because they reflect changes at the genetic level directly affected by environmental changes, such as those that occur with phenotypic traits (Hilu and Liang, 1997). Conservation of genetic resources and maintenance of plant diversity are critical for food security and genetic diversity (Uzun et al., 2025). Hybridization and apomictic breeding often occur in the genus *Crataegus*. Indeed, there are some hybrids in the diversity centers and propagated apodictically. These interactions have produced intermediate forms, rendering the taxonomic analysis more complex (Dönmez and Özderin, 2019). In addition, there is high variability in the genome dimension since many species of this genus are polyploid. Some species of the genus *Crataegus* are used as ornamental plants and some parts of these plant species are edible fruits and might prevent cardiovascular diseases, diabetes and anxiety disorders (Kumar et al., 2012).

## CHAPTER THREE

## MATERIALS AND METHODS

### 3.1 Plant Materials

Surveys were conducted to collect sixty-one Hawthorn accessions from April 2021 to September 2022 at different locations in the Iraqi Kurdistan region (Table 3.1 and Fig. 3.1). In these surveys, fresh leaves and fruits were collected directly from the Hawthorn plants.

**Table 3.1 Accessions codes, locations, and GPS for the Hawthorn accessions used in this study.**

| No | Acce. cod | Species | Locations | Coordinates |
|---|---|---|---|---|
| 1 | G1 | *C. azarolus* | Goizha m | 45.29.39, 35.35.02, |
| 2 | G2 | *C. azarolus* | Azmar m | 45.28.12.1 , 35.37.38 |
| 3 | G3 | *C. azarolus x C. meyeri* | Sharbazher area | |
| 4 | G4 | *C. azarolus x C. pentagyna* | Sharbazher area | 45.39.64, 35.74.20 |
| 5 | G5 | *C. azarolus* | Sharbazher area | |
| 6 | G6 | *C. azarolus* | Sharbazher area | |
| 7 | G7 | *C. azarolus* | Dar bandikhan area | 45.46.18, 34.56.37 |
| 8 | G8 | *C. azarolus* | Dar bandikhan area | |
| 9 | G9 | *C. azarolus* | Dar bandikhan area | |
| 10 | G10 | *C. azarolus* | Dar bandikhan area | |
| 11 | G11 | *C. monogyna* | Qaradagh area | 45.20.54.4 , 35.15.32.9 |
| 12 | G12 | *C. monogyna* | Qaradagh area | |
| 13 | G13 | *C. azarolus* | Qaradagh area | |
| 14 | G14 | *C. azarolus* | Qaradagh area | |
| 15 | G15 | *C. azarolus* | Choman area | 44.52.31.6 , 36.32.49.1 |
| 16 | G16 | *C. azarolus x C.* meyeri | Choman area | |
| 17 | G17 | *C. monogyna* | Choman area | |
| 18 | G18 | *C. meyeri* | Choman area | |
| 19 | G19 | *C. azarolus x C. meyeri* | Choman area | |
| 20 | G20 | *C. orientalis* | Choman area | |
| 21 | G21 | *C. azarolus* | Choman area | |
| 22 | G22 | *C. azarolus* | Choman area | |
| 23 | G23 | *C. azarolus* | Choman area | |
| 32 | G32 | *C. azarolus* | Penjwen area | 45.55.07.4, 35.34.36.1 |
| 33 | G33 | *C. meyeri* | Penjwen area | |
| 34 | G34 | *C. monogyna* | Penjwen area | |
| 35 | G35 | *C. azarolus x C. meyeri* | Penjwen area | |
| 36 | G36 | *C. azarolus* | Penjwen area | |
| 37 | G37 | *C. pentagyna* | Penjwen area | |
| 38 | G38 | *C. azarolus* | Penjwen area | |
| 39 | G39 | *C. meyeri* | Penjwen area | |
| 40 | G40 | *C. azarolus x C. meyeri* | Penjwen area | |
| 41 | G41 | *C. monogyna* | Penjwen area | |
| 42 | G42 | *C. monogyna* | Soran area | |
| 43 | G43 | *C. monogyna* | Choman area | |
| 44 | G44 | *C. monogyna* | Choman area | |
| 45 | G45 | *C. meyeri* | Azmar mountain | 45.28.12.1, 35.37.38 |
| 46 | G46 | *C. azarolus* | Halabja area | 46.02.35.2, 35.08.41.5 |
| 47 | G47 | *C. azarolus* | Khurmal area | 46.06.16.9, 35.15.53.1 |
| 48 | G48 | *C. azarolus* | Azmar mountain | |
| 49 | G49 | *C. monogyna* | Azmar mountain | |
| 50 | G50 | *C. monogyna* | Azmar mountain | |
| 51 | G51 | *C. azarolus* | Azmar mountain | |
| 52 | G52 | *C. azarolus* | Sharbazher area | 45.46.18, 34.56.37 |
| 53 | G53 | *C. azarolus* | Sharbazher area | |
| 54 | G54 | *C. azarolus* | Sharbazher area | |
| 24 | G24 | *C. meyeri* | Choman area | |
| 25 | G25 | *C. azarolus* | Choman area | |
| 26 | G26 | *C. azarolus* | Mawat area | 45.26.40.4 , 35.50.04.7 |
| 27 | G27 | *C. azarolus* | Mawat area | |
| 28 | G28 | *C. azarolus* | Mawat area | |
| 29 | G29 | *C. azarolus* | Chwarta area | |
| 30 | G30 | *C. azarolus* | Chwarta area | 45.36.07, 35.44.42 |
| 31 | G31 | *C. azarolus* | Mawat area | |
| 55 | G55 | *C. meyeri* | Sharbazher area | |
| 56 | G56 | *C. azarolus* | Sharbazher area | |
| 57 | G57 | *C. azarolus* | Mawat area | |
| 58 | G58 | *C. azarolus* | Mawat area | |
| 59 | G59 | *C. azarolus* | Mawat area | |
| 60 | G60 | *C. pentagyna* | Mawat area | 45.26.40.4, 35.50.04.7 |
| 61 | G61 | *C. meyeri* | Mawat area | |
| | | | | |

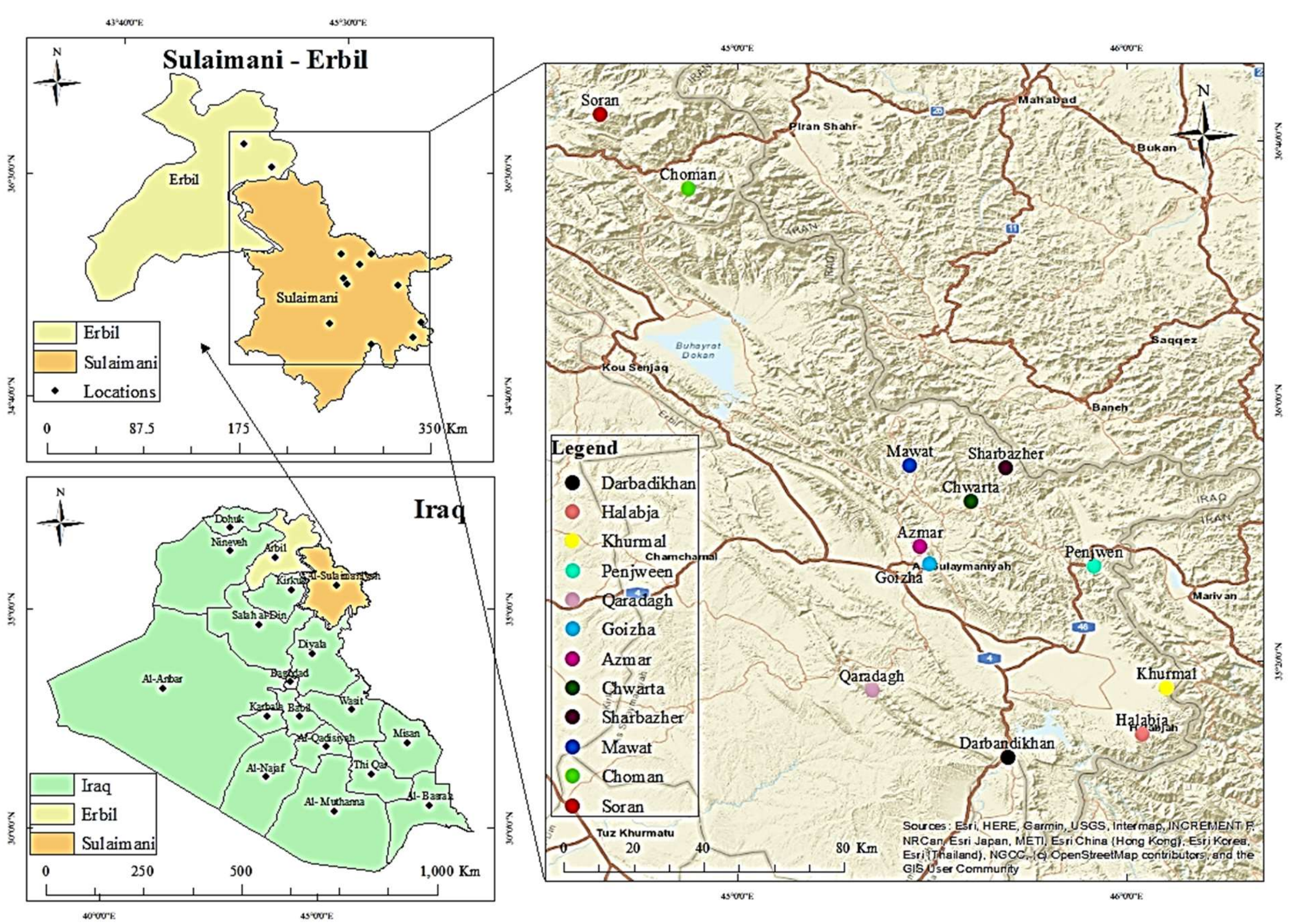

**Figure 3.1 A map of geographic distribution for collected wild Hawthorn in the Iraqi Kurdistan regionA colored circle presents the location: black for Darbandikhan area, pink for Halabja area, yellow for Khurmal area, light green for Penjwen area, dirty pink for Qaradagh area, light blue for Goizha mountain, violet for Azmar mountain, dark green for Chwarta area, purple for Sharbazher area, blue for Mawat area, green for Choman area, and red for Soran area.**

## 3.2 Phenological characteristic

Homogeneous samples from Hawthorn parts for all the 61 accessions were collected at different stages, including the harvest time from the determined naturally grown plants and then transported to the laboratory in cool box, and kept in fridge for floristic study, morphological and fruit quality and physiochemical measurements:

### 3.3 Plant materials collection

To investigate branches, and flowers characterization of the genotypes, Randomely,15 leaf and flower samples from each genotype and for each (three replicate) was collected at the time of full blooming (April and May in 2022). Leaf and flower samples were taken from different regions (locations) of the genotype to which these samples belong. Leaf type, shape, margin, arrangement, attachment, Persistence, stipule presence, blade margin, number of lobs, leaf width, leaf length and petiole length were measured in mm with the help of a digital caliper (Valkyrie brand) sensitive to 0.01. Regarding flowers about 30 parameters were taken including: Flowering period, inflorescence type, flower color, flower attachment, pedicel if present length, sepal length, sepal width, sepal shape, sepals color, indumentum, petal length, petal width, petal shape, petal color, pistil shape, stigma shape, style number, stamen number, shape, color, indumentum, filament length, anther length, anther shape, color and anther arrangement.

In addition, 15 fruits and seeds per a replicate as a plant material were randomly indicated from each genotype and many morphological parameters were documented including; fruiting period, fruiting stalk, fruit shape, fruit color, fruit surface, fruit length, fruit width, seed Shape, seed color, seed surface, seeds length, seeds width and number of seeds per fruit (Pyrene).

### 3.4 Determination of physio-pomological characteristics of fruits

For physio-biochemical analyses, approximately 0.5 kg fully matured, fresh Hawthorn fruits were harvested manually from sixty-one Hawthorn genotypes and transferred to laboratory for analysis. A total of 45 fruits (3 replicates) were randomly selected per replicate and desirable physio biochemical characteristics were measured for fruit weight (WF-g), fruit size (FS-$cm^3$)by (VS: Volume solution) , fruit length (FL mm), fruit width (FW mm), seed weight (SW-g), Seed length (SL.mm), seed width (S w- mm), number of seeds per 5 fruits ( NSF), Weight of fresh fruit (WOF- g). WS: weight of seeds (WS-g) and Fruit external color was determined by visible means. To determine flesh/seed ratio, first the fruits were cut and flesh separated from seeds. Both parts weighted separately then divide each component to determine the ratio of flesh/seed ratio.

### 3.5 Phytochemical analysis

Phytochemical properties of Hawthorn genotypes were determined with 3 replicates and 15 fruits per a replicate. The stones of the fruits were removed, and the extractions were carried out by making them homogeneous with a hand blender to make juice. Juice from the Hawthorn fruit were used to determine total phenolic content (TPC), antioxidant activity (AA) total flavonoid content (TFC), carotenoid content (CAC), soluble solid content (SSC, %), Potential of hydrogen (pH), total soluble solids (TSS), Titratable acidity (TA%) and Moisture content (MC-%). A standard procedure for fruit biochemical assays was followed.

To prepare extraction 5 g of fruits was taken and make a juice and mixed with 10 mL of 80% methanol. The samples were centrifuged at 6000 g for 5 min at 4 °C. Supernatant was collected and used in total phenolic, anthocyanin, flavonoid, and antioxidant assay procedures (Tahir, et al., 2023).

### 3.5.1 Total phenolics content (TPC)

To determine the total phenolic content of the samples, Ahmad et al. (2020)'s Folin-Ciocalteu method was used with minor modifications. According to (Lateef *et al.*, 2021) total phenolic content (TPC) was evaluated for this purpose, 200 μL of the supernatant mixed with 1800 μL of 1: 9 Folin–Ciocalteu reagent: water (v/v) after 7 min added 850 μL 10% $Na_2CO_3$ and incubated in dark for 30 minutes. After reaction, the colour of mixture solution was changed to light blue and read at 750 nm against the blank (150 μL $dH_2O$ mixed with 1050 μL 1: 9 Folin–Ciocalteu reagent: water (v/v) and 850 μL 10% $Na_2CO_3$), a UV-visible spectrophotometer (UVM6100, MAANLAB AB, Sweden) was used at a wavelength of 760 nm. The results obtained were calculated in gallic acid and expressed as mg/100 g (fresh weight). Gallic acid (GAE) was employed as a standard, the standard solution was prepared by dissolving 9 mg of gallic acid in 9 mL of methanol to attain a final concentration of 1 mg/mL. A sequence of dilutions of gallic acid (0, 50, 100, 150, 200, 250, 300 μg/mL) had been used to produce a standard curve and linear association between the absorbance values at 750 nm and the gallic acid content was observed. The total phenolic content in each sample was determined using the standard curve (Figure 3.2). The following equation was used to calculate the TPC:

$$\text{TPC } (\mu\text{g GAE/gm FW}) = \frac{\mathbf{V}}{\mathbf{W}} \mathbf{\ x\ C}$$

Where V is the volume of extract (mL), W is the fresh weight of the sample (g), and C is the concentration of gallic acid collected from the standard curve.

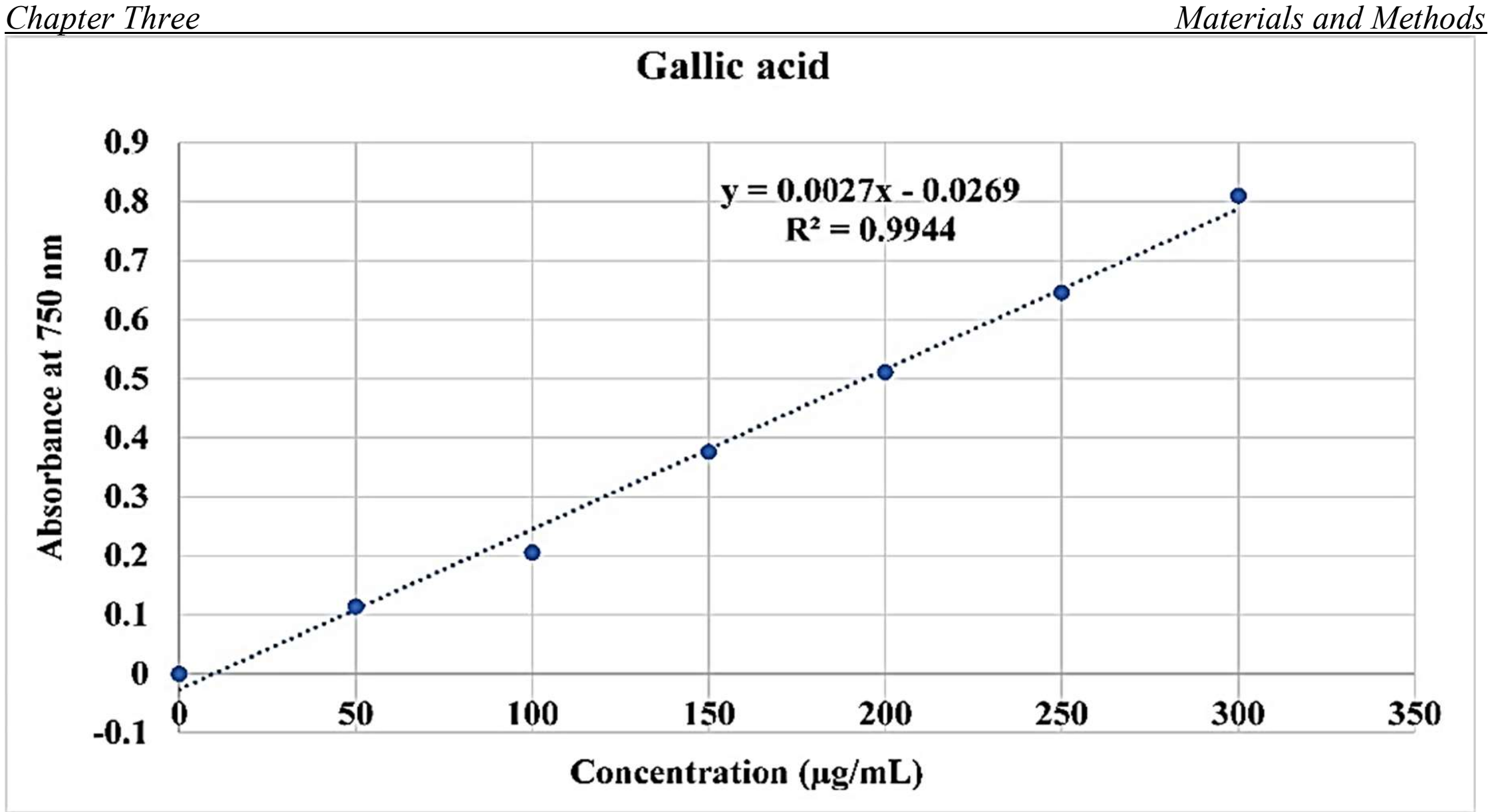

**Figure 3.2 Standard curve of gallic acid.**

### 3.5.2 (DPPH) antioxidant activity (Free radical scavenging activity) (AA)

The method of Brand-Williams et al. (1995) was used in the determination of antioxidant activity. 100 µL of extract was mixed with 1.9 mL of 1-diphenyl-2-picrylhydrazyl (DPPH) solution (0.01g DPPH dissolved in 260 mL of %95 methanol). The sample mixtures were incubated in dark for 30 minutes at room temperature, absorbed the samples at 517 nm against the blank (95% methanol) using a UV-visible spectrophotometer (UVM6100, MAANLAB AB, Sweden) was used (Lateef *et al.*, 2021).

The standard compound, 6-hydroxy-2,5,7,8-tetramethylchroman-2-carboxylic acid (Trolox), was used to build the calibration curve. Trolox (12 mg) was combined with 12 mL of 75% ethanol (v/v) solvent and diluted to achieve concentrations of (0.00, 0.33, 0.66, 1.320, 2.00, 2.7, and 3.4 µg/mL) (Figure 3.3). Linear regression was found between the absorbance values at 517 nm and the varied Trolox concentrations. The following equation was used to estimate the antioxidant capacity:

$$\text{Antioxidant capacity by DPPH } (\mu\text{g Trolox/g FW}) = \frac{\mathbf{V}}{\mathbf{W}} \mathbf{x\ C}$$

Where V is the volume of extract (mL), W is the fresh weight of the sample (g), and C is the concentration of Trolox determined from the standard curve.

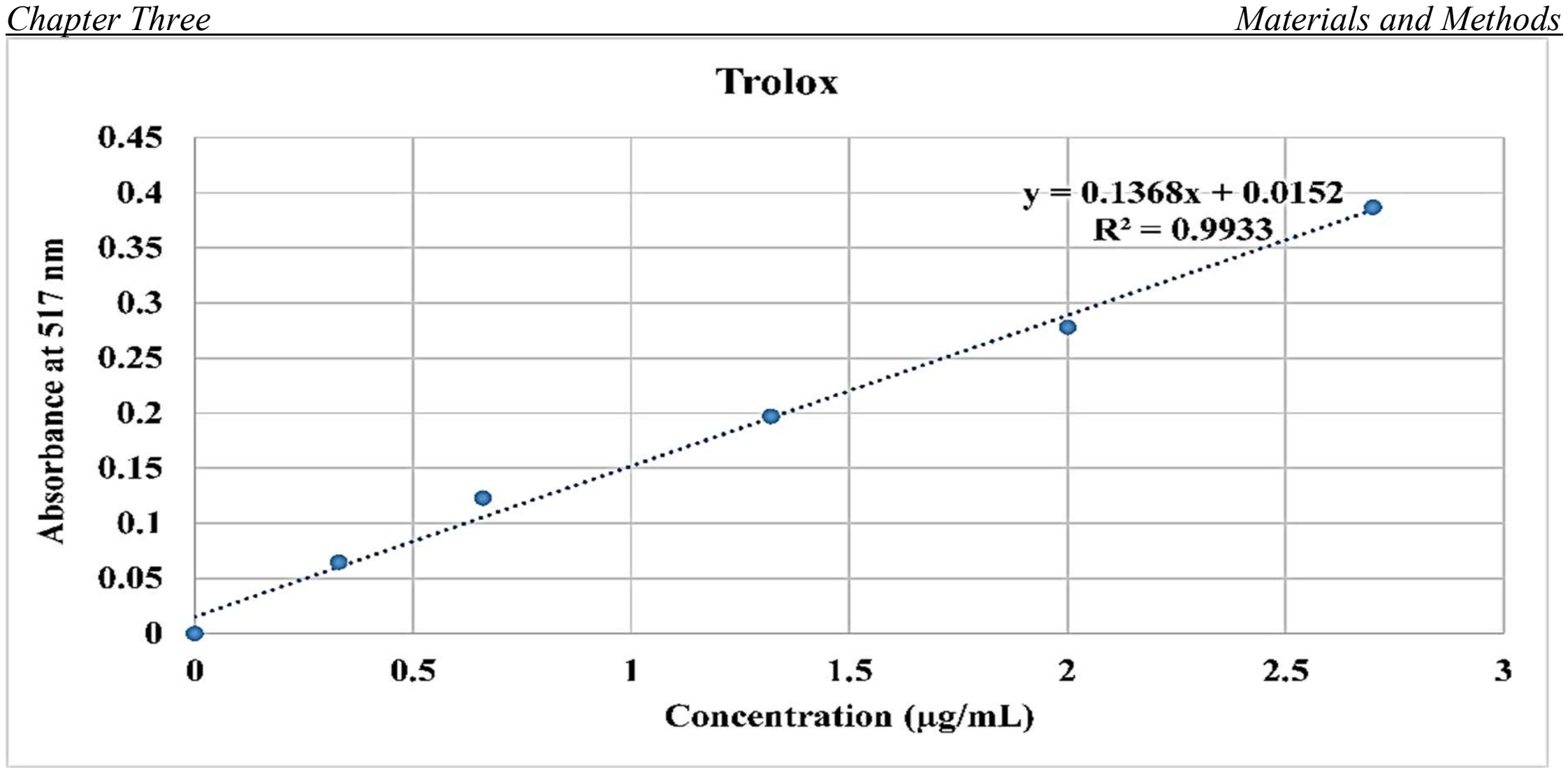

**Figure 3.3 Trolox standard calibration curve.**

### 3.5.3 Total flavonoids content (TFC)

Flavonoid contents were determined according to method reported by (Lateef, Mustafa and Tahir, 2021; Tahir et al., 2023a) with some modifications. A concentration of 1 mg/mL was obtained by dissolving 11 mg of quercetin in 11 mL of deionized water to create the stock solution. Standard curve for quercetin and linear regression between absorption values at 415 nm and quercetin concentrations have been created (Figure 3.4). for analysis of flavonoid 500 µL of supernatant was taken and mixed with combined as 300 µL of aluminum chloride (2% w/v), 80 µL of potassium acetate (1 M), and 1.7 mL of deionized water. At 415 nm, the absorbance of the solution was recorded after 30 min of incubation at 28 °C using a UV-visible spectrophotometer (UVM6100, MAANLAB AB, Sweden). Three replications were used to create each genotype's mean value. Each juice's total flavonoid concentration was reported as µg quercetin (QE) per gram of fresh flesh matter using the following formula: TFC (µg QE g-1 FW) = Volume of juice (mL) Fresh weight of flesh (g) x Concentration from standard curve of quercetin (µg/mL).

Concentration was reported as µg quercetin (QE) per gram of fresh flesh matter using the following formula:

formula: $\text{TFC } (\mu\text{g QE g}^{-1}\text{ FW}) =$

$$\frac{\text{Volume of juice (mL)}}{\text{Fresh weight of flesh (g)}} \times \text{Concentration from standard curve of quercetin } (\mu\text{g/mL})$$

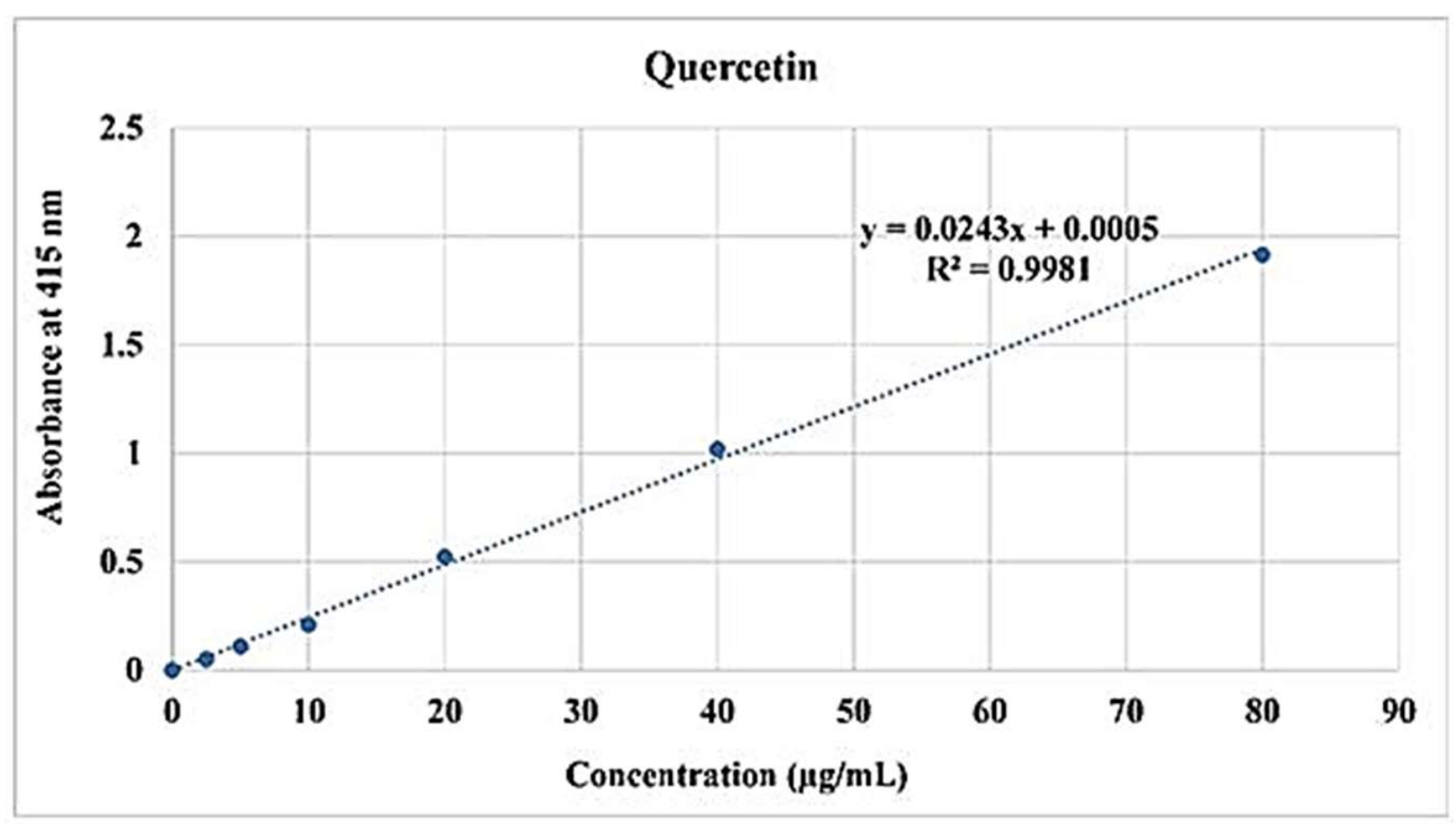

**Figure 3.4 Standard calibration curves of quercetin.**

### *3.5.4 Quantification of carotenoid content (CAC)*

According to Ferrante et al. (2008) fresh CAC was measured by Methanol (99.9%) as the solvent for extracting carotenoids. One gram of crushed flesh was combined with 1.5 mL of 99.9% methanol and stored at 5°C for overnight in the dark. The mixture was centrifuged at 5500 rpm for 15 min. The supernatant was collected from the centrifuged mixture. A total of 1 ml of supernatant was combined with 1.5 mL of methanol (99.9%). The spectrophotometer was used to take readings at 470 nm, and the carotenoid concentrations were expressed as µg per gram of fresh flesh weight and estimated by this formula: CAC (µg g-1) = Absorbance reading x Total volume of juice (mL) x 10000 Carotene extension coefficient in methanol x Fresh weight of flesh (g) Three replications have been used to calculate the average value of each genotype.

### 3.5.5 Soluble sugar content (SSC) measurement

According, Tahir et al. (2022) Soluble sugar content was determined. Glucose stock solution was created by added to 12 mg of glucose with 12 mL of deionized water to achieve a final concentration of 1 mg/mL. To construct the standard curve, a series of dilutions (0, 6, 12, 24, 36, 48, 60, 72, 144, 288, and 576 µg) were performed. The relationship between the 620 nm absorbance values and the glucose concentrations was found to be linear (Figure 3.5). Deionized water was used to immerse 0.1 g of freshly ground flesh in 1000 µL. The solution was brought to a boil at 99 °C for 33 min. After cooling, it was centrifuged at 4000 rpm for 15 min. Anthrone reagent (1997 µL of a solution: 0.15 g anthrone in 84 mL of sulphuric acid and 16 mL of deionized water) was mixed with 25 µL of a sample or glucose. This concoction was brought to a rolling boil and maintained that temperature for 4 min. A UV-visible spectrophotometer was used to estimate the soluble sugar concentration at 620 nm.

Three replications have been used to determine the average value of each genotype. The soluble sugar concentration was given as µg g-1 of fresh flesh weight using the following formula: SSC (µg glucose g-1 FW) = Volume of juice (mL) Fresh weight of flesh (g) x Concentration from standard curve of glucose (µg/mL).

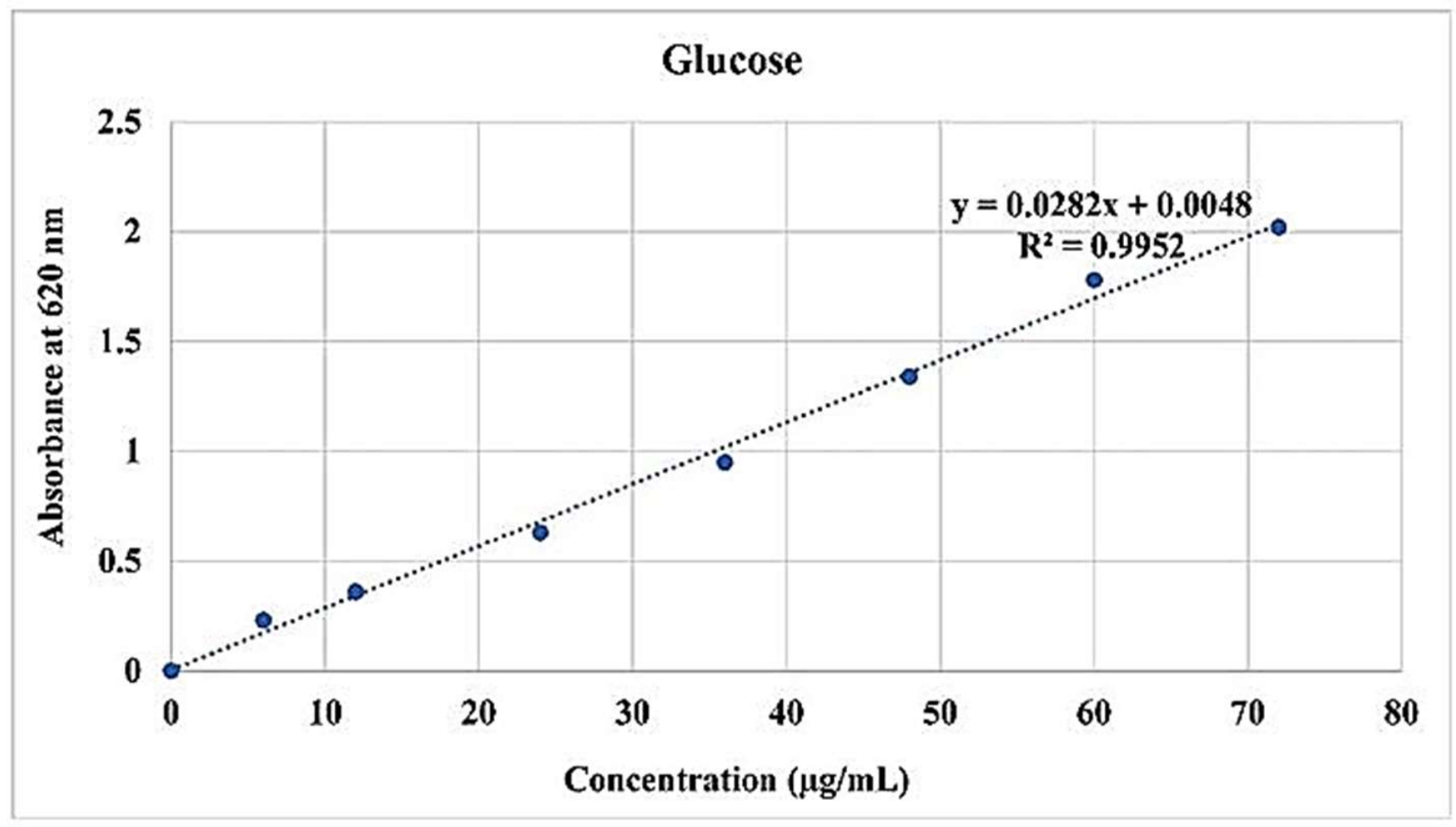

**Figure 3.5 Standard calibration curves of glucose.**

### 3.5.6 pH measurement

Fresh juice was collected and homogenized from known weights of flesh Hawthorn and a known volume of deionized water was added to each juice sample. The pH was obtained after calibrating the pH meter (Hanna Instruments, Romania) for pH 4 and 7 with standard solutions. Furthermore, three readings were taken from each juice genotype, and such values from triplicate samples were used to calculate the mean.

### 3.5.7 Measurement of total soluble solids (TSS)

TSS was determined using a standard procedure. Fruit juice was recovered by pulping and crushing the fruits. A handheld refractometer (ATAGO Pocket PAL-2, Japan) was used to measure the TSS in the juice. Following the cleaning and calibration of the refractometer, a known volume of juice (drop) was put on top of the refractometer at the designated spot. The Brix unit was used to express TSS results (Rasul et al. 2022). Three replications have been used to calculate the average value of each genotype.

### 3.5.8 Assessment of titratable acidity (TA)

The titratable acidity of Hawthorn genotypes was evaluated using the method described previously by Ranganna (1986). To extract the juice, ten grams of fruit flesh were compressed. The juice was centrifuged for 10 min at 5500 rpm, and the clear supernatant was collected. In a conical flask, a 5 mL aliquot was taken. Two drops of phenolphthalein indicator were added to the mixture, which was then titrated against 0.1 N sodium hydroxide (NaOH). The appearance of pink color was noted as the titration's endpoint. The titratable acidity was then estimated by the formula of Nielsen (2017) and expressed as percentage unit. Three replications have been used to calculate the average value of each genotype.

$$\mathbf{TA\%} = \frac{\mathbf{V} * \mathrm{N} * \mathrm{Eq.\,wt.}}{\mathbf{S.v}} \mathbf{X\ 100}$$

**Where TA, Titratable acidity v, volume of NaOH, N acid side from Hawthorn, Sv. Sample volume.**

### 3.5.9 Fruits moisture content (MC-%)

The moisture contents of the samples were determined using oven drying method. Before drying, the weight of the fruit samples and empty glass petri-dishes were recorded, as was the weight of the glass petri-dishes and samples after 72 hours of oven drying at 70 °C. The following formula was used to calculate the moisture contents of the fruits: (AOAC 1990).

$$\mathbf{MC\%} = \frac{\mathbf{FW - DW}}{\mathbf{FW}} \mathbf{X\ 100}$$

Where FW and DW are the weights of fresh and dried fruits, respectively.

## 3.6 Molecular Marker Assay

### 3.6.1 DNA Isolation

Hawthorn fresh leaves crushed with liquid nitrogen were used for the DNA extraction for all the 61 accessions according the procedure described by Majeed et al. (2024). To lyse powdered leaf material, 1 mL of lysis buffer [0.50% (w/v) SDS, 8.00% (w/v) PVP, 0.25 M NaCl, 0.025 M EDTA, and 0.2 M Tris-base] and 9 µL RNase (10 mg/mL) was added and incubated for 74 min at 64 °C and the tubes were inverted 15 times during this period. After cooling for 7 min at room temperature, 300 µL of 5 M potassium acetate (pH 6.5) was added, combined, and refrigerated for at least 9 min. The samples were then centrifuged at 16200 rpm for 18 min. The upper phase was gathered and transferred to a new Eppendorf tube (2 mL). GTE buffer (2M guanidine thiocyanate dissolved in 75% ethanol) was gently added to the upper layer. A volume of the mixture (750 µL) was deposited into a spin column and allowed to stand for 7 min. It was centrifuged for 6 min at 1000 rpm and the flow solution was thrown away. At this time, a volume of 550 µL of NTEH washing buffer (10 mM NaCl, 10 mM Tris-base pH 6.5, and 80% ethanol) was added to the spin column. The column was centrifuged for 6 min at 8200 rpm resulted in the removal of the flow solution from the column. A second wash of NTE buffer was applied to the spin column. After centrifuging, the flow solution was removed. The spin column was centrifuged at 11000 rpm for 6 min to dry. The spin column was placed in a new 1.5 mL Eppendorf tube, which was then filled with 108 µL of elution buffer and incubated for 5 min at room temperature. The eluted DNA was collected by centrifuging for 5 min at 9500 rpm and storing it at -20 °C. a 1.1% agarose gel and a nanodrop spectrophotometer (NanoPLUS-MAANLAB AB, Sweden) were used to check the quality and amount of the isolated DNA.

### 3.6.2. ITS- PCR amplification

For alignment of sequences and phylogenetic reconstruction of the *Crataegus* spp. amplification of the ITS region, ITS1 (5′ TCCGTAGGTGAACCTGCGG 3′) and ITS4 (5′ TCCTCCGCTTATTGATATGC 3′) were employed (Hsiao et al., 1995; Karim et al., 2025). In brief, the PCR amplification was carried out in a 26 µL volume that contained 5.0 µL of genomic DNA, 10 µL of master mix (AddStart Taq Master, AddBio, Daejeon, Republic of Korea), 2 µL of each ITS primer (10 µM), and 7 µL of deionized water. The first step of the PCR program was to separate the strands at 94 ◦C for 10 min, followed by 35 cycles of denaturation at 94 ◦C for 1 min, annealing at 55 ◦C for 1 min, elongation at 72 ◦C for 2 min, and a final extension at 72 ◦C for 10 min. A negative control (no DNA template) was utilizedto monitor each set of reactions. Following PCR amplification, the product was separated on 1.5% agarose gels with 1X TBE buffer, stained with ethidium bromide, and observed under UV light. The ITS bands were cut, and the DNA amplicon was extracted from the agarose using a Gel Extraction Kit (AddBio, Daejeon, Republic of Korea). The DNA was then sequenced at Macrogen Company in South Korea. The results were compared to DNA sequences previously recorded in the National Center for Biotechnology Information dataset (NCBI).

### 3.6.3 PCR amplification of the markers

Genetic diversity of the Hawthorn accessions was assessed using 31 markers of three assays to include eleven SSR primers (Tahir and Maeruf 2016; Tahir et al. 2021; Ahmad et al. 2022), ten ISSR primer (Faraj 2023; Khal et al. 2023), and ten SCoT primer (Ahmad 2024; Rasul et al. 2022). The detail of primers' sequence and annealing temperature are present in (Table 3.2). These marker types were chosen based on the quality work of our previous studies and availabilities in our lab as well. The mixture of PCR contained 8 µL of master mix buffer (Sinaclon, Iran), 4 µL of isolated DNA, 4 µL of deionized water, and 4 µL of primer (Addbio, Korea), and completed finally to 20 µL of mixture were produced. A conventional PCR machine (Prime UK) was used for the PCR amplification. The PCR machine was run with initial denaturation for 10 min at 94 °C in one cycle, then one minute at 94 °C for denaturation, depending on the primer, annealing temperature was set for one minute, and two minutes for extension at 72 °C, these three stages were 36 times repeated, and final extension for 10 min at 72 °C. By using a 1.4% agarose gel dissolved in 1X TBE buffer (see appendix 1) and mixed with 0.13 µg/mL ethidium bromide, the PCR products were electrophoresed for 85 min at a voltage of 85 V. UV transilluminator digital imaging equipment was utilized in the gel visualization.

**Table 3.2 Simple sequence repeat (SSR), Inter simple sequence repetition (ISSR), and Conserved DNA derived polymorphism (SCoT) primer, their sequences and used annealing temperature.**

| SSR markers | Sequence 5' to 3' | Annealing temperature °C |
|---|---|---|
| SSR09-F | GACCACCTCCTTGTCTTCCA | 56.50 |
| SSR09-R | GTGGGAAAACCAAACCTGAA | |
| SSR17-F | TGTATCAGCCTCGACGACAG | 59.00 |
| SSR17-R | TTCCCCTCCTCCACTTTACC | |
| SSR19-F | CTTGGAAGGTGAGGACGGTA | 56.50 |
| SSR19-R | CCGGAGGGAAACTAAAACAA | |
| SSR35-F | TGTTCTTGGTCCACTGTTGG | 59.00 |
| SSR35-R | AAATTCGCCTTTGGTCAAAAT | |
| SSR38-F | TTTTCCACCGTTAGGAGTCG | 59.00 |
| SSR38-R | TTAATGGACCGCCATAGAGC | |
| SSR40-F | CTTTCCCAAAAATCAGCGAA | 58.00 |
| SSR40-R | GGAAGAATTTCGGACGTCAA | |
| SSR46-F | CATCTGCCAACCTTGTTTCA | 58.00 |
| SSR46-R | AACATCCACACTTGACAGACAAA | |
| SSR56-F | GGTCAGGAGAGAGGATGCTG | 58.00 |
| SSR56-R | CGCTGAGAGAGAGTGCTTCTT | |
| SSR58-F | AGAAGAAGTGGCGACAGCAT | 58.00 |
| SSR58-R | AACTCTCTCGAACCGACGAA | |
| SSR61-F | CGCTAGACGCGGTAGAAAAA | 58.00 |
| SSR58-R | GCAGGGTTTTAGTGGGGACT | |
| SSR99-F | TCCACACCACCTTCTGATGA | 58.00 |
| SSR99-R | CTGTCATTTCAATTTCCGTCA | |
| ISSR markers | | |
| UBC-808 | AGAGAG AGAGAGAGAGC | 50.00 |
| UBC-810 | GAGAGAGAGAGAGAGAT | 50.00 |
| UBC-812 | GAGAGAGAGAGAGAGAA | 50.40 |
| UBC-818 | CACACACACACACACAG | 52.80 |
| UBC-822 | TCTCTCTCTCTCTCTCA | 50.00 |
| UBC-823 | TCTCTCTCTCTCTCTCC | 50.00 |
| UBC-826 | ACACACACACACACACC | 50.00 |
| UBC-846 | CACACACACACACACAAT | 50.00 |
| UBC-880 | GGAGAGGAGAGGAGA | 48.00 |
| UBC-891 | ACTACTACTTGTGTGTGTGTGTG | 52.00 |
| SCoT markers | | |
| SCoT 6 | CAACAATGGCTACCACGC | 52.05 |
| SCoT 10 | CAACAATGGCTACCACGG | 51.27 |
| SCoT 12 | CAACAATGGCTACCAGCC | 51.20 |
| SCoT 14 | ACGACATGGCGACCACGC | 58.60 |
| SCoT 16 | ACGACATGGCGACCGCGA | 59.90 |
| SCoT 22 | ACCATGGCTACCACCGAC | 54.05 |
| SCoT 23 | CACCATGGCTACCACCAG | 52.43 |
| SCoT 33 | CCATGGCTACCACCGCAG | 55.60 |
| SCoT 34 | ACCATGGCTACCACCGCA | 56.30 |
| SCoT 35 | CATGGCTACCACCGGCCC | 57.90 |

### 3.7 Statistical analysis and data scoring

Analysis of variance for all the biochemical traits of Hawthorn fruits were analyzed using XLSTAT version 2021.1 (Tahir et al., 2024). One-way analysis of variance (one-way ANOVA) and Duncan's multiple range tests were performed with three replications under completely randomized design (CRD). The dendrogram was created by JMP Pro 18, using binary data of 1 for the presence and 0 for absence of alleles, the visualized bands were manually scored. The PIC was assessed by utilizing Power Marker 3.25 software. The genetic distance and accession groups were detected using the UPGMA in XLSTAT version 2021.1. Diversity indication and molecular variance of the studied access within and among populations were measured by Gen AlEX software, (Peakall and Smouse, 2006). version 6.5. STRUCTURE program version 2.3.4 was carried out to generate the structure of population (Pritchard et al. 2000).

# CHAPTER FOUR
# RESULTS AND DISCUSSION

## 4.1 Morphological Study of *Crataegus* L.

A critical initial step in the development of appropriate varieties is selection of ecotype, therefore, wild species of edible fruits exhibit considerable diversity in morphology, fruit quality, performance, and nutrient content compared to their domesticated counterparts. These traits can be significantly influenced by environmental conditions and ecotype variations (Gundogdu et al., 2014). It is frequently stated that phenotypic markers have proven to be extremely useful in studies of genetic diversity in Hawthorn genotypes (Ansari et al., 2020 and Tira-umphon, and Ketudat-Cairns, 2018). Therefore, the results of the present morphological study showed that there are five species with two hybrids were founded including, *C. azarolus, C. meyrei, C. monogyna, C. orientalists, C. pentagyna, C. azarolus x C. meyrei and C. azarolus x C. pentagyna*. They were significant variations among different taxa in terms of plant, fruit morphological and functional characteristics were observed, description and classification of species and hybrids were documented as following.

## 4.2 Crataegus L.

Common English name: Hawthorn, common Kurdish name:

Deciduous trees or shrubs, Leaves pale green, petiolate lobed or serrate, obovate, cuneate, triangular, broadly ovate oblong or flabellate, pinnately lobed, apex acute or truncate. Inflorescence corymb (or rarely solitary), 3-12 flowered Flowers white or pinkish; sepal ovate, ovate-oblong, triangular, pubescent; petal five, orbicular, sub orbicular, broadly obovate, triangular; stamens 5-20, white, milky or light purple; Styles as many as carpels carpels 1-5 free at apex, pale green, united on the inner margin at least in basal parts. Fruit fleshy, globose or ovoid, or globose downy, reddish orange yellowish, red tinged ripe or yellowish pink, glabrous or thinly hairy, especially around and just below persistent calyx, usually crowned with the persistent sepals. Pyrene (seeds) pale brown or light brown, sub globose, outer surface furrowed or slightly furrowed, inner surface keeled or almost flat.

Style 4-5 1

Style 1-3 2

1a Fruit blackish when ripe, calyx, receptacle and pedicels conspicuously villose 5.*C. pentagyna*

1b. Fruit reddish orange when ripe, receptacle and pedicels conspicuously tomentosa 4. *C. orientalis*

2a. Young shoots densely grey or whitish pubescent; fruit as large as a cherry, yellow or tinged reddish 1. *C. azarolus*

2b. Young shoots glabrous or thinly hairy; fruit distinctly smaller than a cherry, red or blackish 3

3a. Style 1 (rarely 2); fruit usually less than 1.2 cm in diam. 3. *C. monogyna*

3b. Styles 2 (rarely 3); fruit usually more than 1.2 cm in diam. Long *2. C. meyeri*

#### 4.2.1 ***Crataegus azarolus*** **L.**

Deciduous trees, 3-5 m tall, young shoots densely gray or whitish pubescent, twigs grayish-brown, Bark type finely fissured, minimum to maximum length of spine was 1.0 -1.19 cm that spine average was 0.78 cm long. Leaves pale green, petiolate, obovate, cuneate, 3.12-5.06 × 2.12-4.84 cm; pinnately lobed, 3-7, apex acute or truncate. Inflorescences compact corymbs, 3-12 flowered. Flowers white, pedicels 0.5-1.6 cm long; sepal ovate, ovate-oblong, triangular, 1.90-3.23 × 1.81-3.67 mm, pubescent; petals white, orbicular, sub orbicular, broadly obovate, 5.17- 8.13 × 3.30-4.98 mm; stamens 15-20, white, milky, 4-6 mm long, anthers milky, 1.87- 3.09 mm long, styles pale green, 1 or 2 (occasionally 3), stigmas capitate. Fruit globose, yellowish or red tinged ripe, 8.76 - 13.28 mm in diam., glabrous or thinly hairy, especially around and just below persistent calyx, pyrene (seeds) pale brown, sub globose, outer surface furrowed, inner surface keeled or almost flat, 6.41- 9.81 × 4.87- 7.77 mm (see figure 4.1 and appendix 2 and 3).

**Habitats:** foothills, mountainsides, oak woodlands, rocky slope, stony soil; elevation 1243 m.

**Flowering:** Mar–Jun.

**Occurrence:** common throughout of the mixed forest in Kurdistan region of Iraq.

**Distribution:** Kurdistan Iraq, Iran, Turkey, Syria, Mediterranean region, Cyprus, Lebanon, and Palestine, Europe, and North Africa.

**Collections:**

MSU: Karzan, Goizha 21(1), Azmar (21)2, Qaiwan (21)5, Qaiwan (21)6, Sartaki bamo (21)7, Sartaki bamo (21)8, Darbandikhan- qashty (21)9, Darbandikhan- qashty (21)10, Qaradakh-Smaela (21)13, QaraDag (21)14, Mro: Karzan, Saman, Nariman, Qasery-Kanibasta (21)15, Qasery-Walasha (21)19, Sakran-Shewakaly (21)21, Choman-Gomi Felaw (21) 22, Choman-Gomi Felaw (21)23, Choman-Khoshkan (21)25, MSU, Sharbazher-kany bewaka (21)26, Sharbazher -Zenal (21)27, Sharbasher-kele (21)28, Sharbazher- Chwarta (21)29, Sharbazher- Sarsir (21)30, Sharbazher-Amaden (21)31, penjwen-nalpareaz (21)32, Penjwen- Hangazhall (21)36, Penjwen- Qalandarawa (21)38, Halabja-Klazhdr (21)46, Khurmal-Bakhakone (21)47, Azmar (21)48, Azmar (21)51, Azmar (21)52, Sharbazher-Bazagir (21)53, Sharbazher- Gwezila (21)54, Sharbazher-Qshlax (21)56, Sharbazher-Saraw (21)57, Sharbazher-Saraw (21)58, Sharbazher-Saraw (21)59.

MSU: Karzan, Goizha 22(62), Azmar (22)63, Qaiwan (22)66, Qaiwan (22) 67, Sartaki bamo (22)68, Sartaki bamo (22)69, Darbandikhan-qashty (22)70, Darbandikhan-qashty (22)71, Qaradakh-Smaela (22)74, QaraDag (22)75, Mro: Karzan, Saman, Nariman, Qasery-Kanibasta (22)76, Qasery-Walasha(22)80, Sakran-Shewakaly (22)82, Choman-Gomi Felaw (22)83, Choman-Gomi Felaw (22)84, Choman- Khoshkan (22)86, MSU, Sharbazher-kany bewaka (22)87, Sharbazher -Zenal (22)88, Sharbasher-kele (22)89, Sharbazher-Chwarta (22)90, Sharbazher-Sarsir (22)91, Sharbazher-Amaden (22)92, penjwen-nalpareaz (22)93,Penjwen- Hangazhall (22)97, Penjwen- Qalandarawa (22)99, Halabja-Klazhdr (22)107, Khurmal-Bakhakone (22)108, Azmar (22)109, Azmar (22)112, Azmar (21)113, Sharbazher-Bazagir (22)114, Sharbazher- Gwezila (21)115, Sharbazher-Qshlax (22)117, Sharbazher-Saraw (22)118, Sharbazher-Saraw (21)119, Sharbazher-Saraw (21)120.

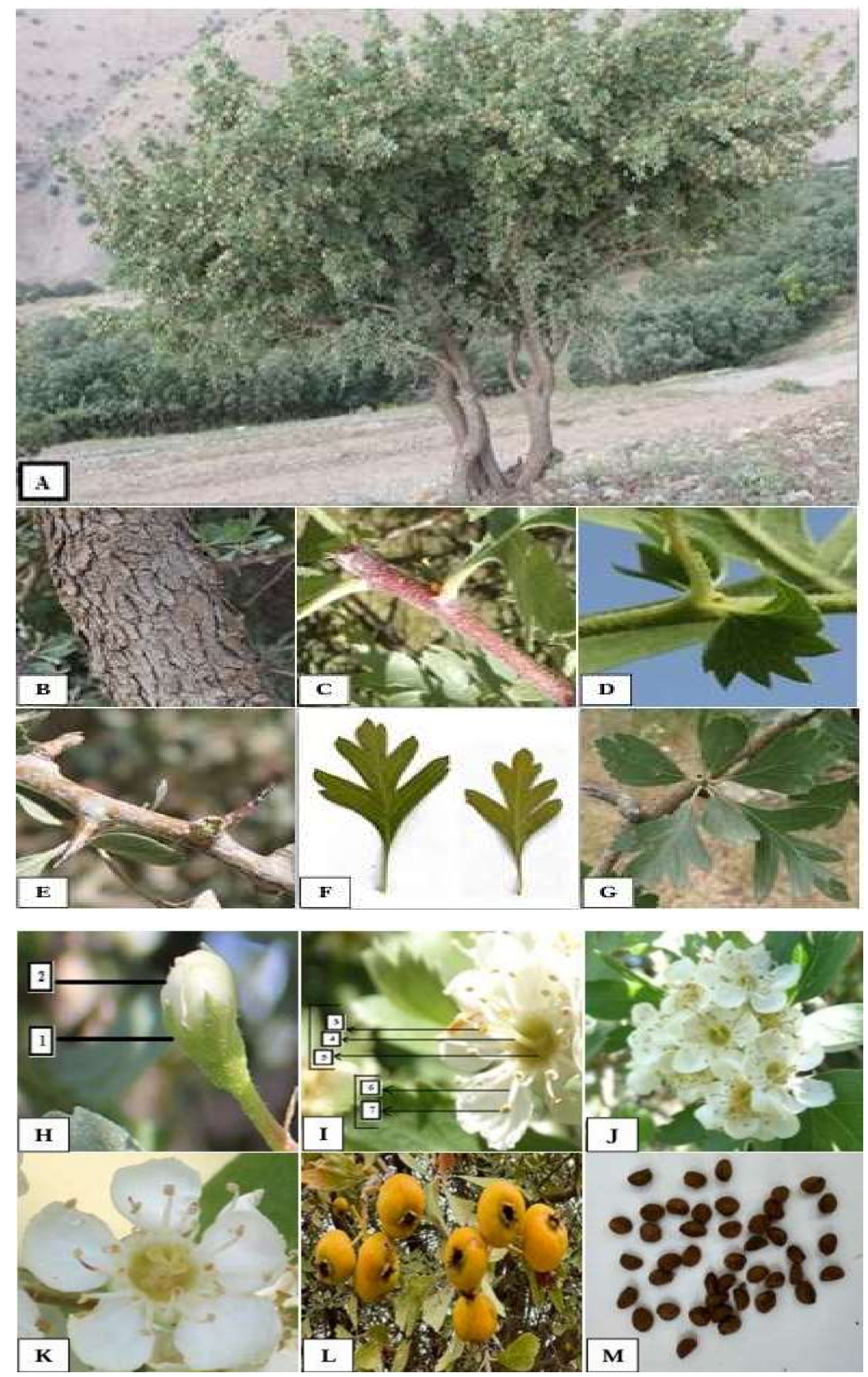

**Figure 4.1** ***Crataegus azarolus*** **var. , A. Plant habit; B. Bark; C. Young shoot; D. Stipules; E. Spine; F & G Leaves; H. Bud 1- Sepal. 2- Petals; I. Disect flower. 3- Stigmas. 4- Style. 5- Ovary. 6- Filament. 7- Anthers; J. Inflorescences; K. Single flower: L. Fruits; M. Seeds.**

### 4.2.2 *Crataegus meryeri* Pojark

Deciduous shrub or small tree, 3.5 - 4.5 m tall, young shoots densely grayish or whitish pubescent, twigs grayish-brown, Bark type rough scaly, minimum to maximum length of spine was 0.79 -1.105 cm spine average 0.88 cm long. Leaves pale green, petiolate, obovate, 3.79 - 5.42 × 3.46 - 5.05 cm; pinnately lobed, 3-5, apex acute or truncate. Inflorescences compact corymbs, 3-12 flowered. Flowers white; pedicels 1.23 -1.67 cm long; sepal ovate, ovate-oblong, triangular, 1.96 - 2.76 × 1.31- 1.95 mm, pubescent; petals white, orbicular, sub orbicular, broadly obovate, 3.99 - 6.31 × 3.29 - 5.00 mm; stamens 16-20, white, milky, 4.36 - 5.81 mm long, light purple, 1.79 - 2.94 mm long, styles pale green, (3-4), stigmas capitate. Fruit globose, red ovoid, 10.33 - 14.37 mm in diam, glabrous or thinly hairy, especially around and just below persistent calyx; pyrene (seeds) pale brown, sub globose, outer surface furrowed, inner surface keeled or almost flat, 7.18 - 8.58 × 5.93- 6.97 mm (see figure 4.2 and appendix 2 and 3).

**Habitats:** foothills, mountainsides, oak woodlands, rocky slope, stony soil; elevation 1243 m.

**Flowering:** March-June.

**Occurrence:** Frequent the mixed forest in Kurdistan region of Iraq.

**Distribution**: Kurdistan Iraq, Iran, Turkey, Syria, Mediterranean region, Cyprus, Lebanon, and Palestine, Europe, and north Africa.

**Collections:** Mor: Karzan, Saman, Nariman, Qasery-Kanibasta (21)18, Choman-Khoshkan (21)24, Msu: Penjwen-Nalpareaz (21)33, Penjwen-Nzara (21)39, Azmar (21)45, Sharbazher-Qshlax (21)55, Sharbazher-Saraw (21)61.

Mor: Karzan, Saman, Nariman, Qasery-Kanibasta (22)79, Choman-Khoshkan (22)85, Msu: Penjwen-Nalpareaz (22)94, Penjwen-Nzara (22)100, Azmar (22)46, Sharbazher-Qshlax (22)116, Sharbazher-Saraw (22)122.

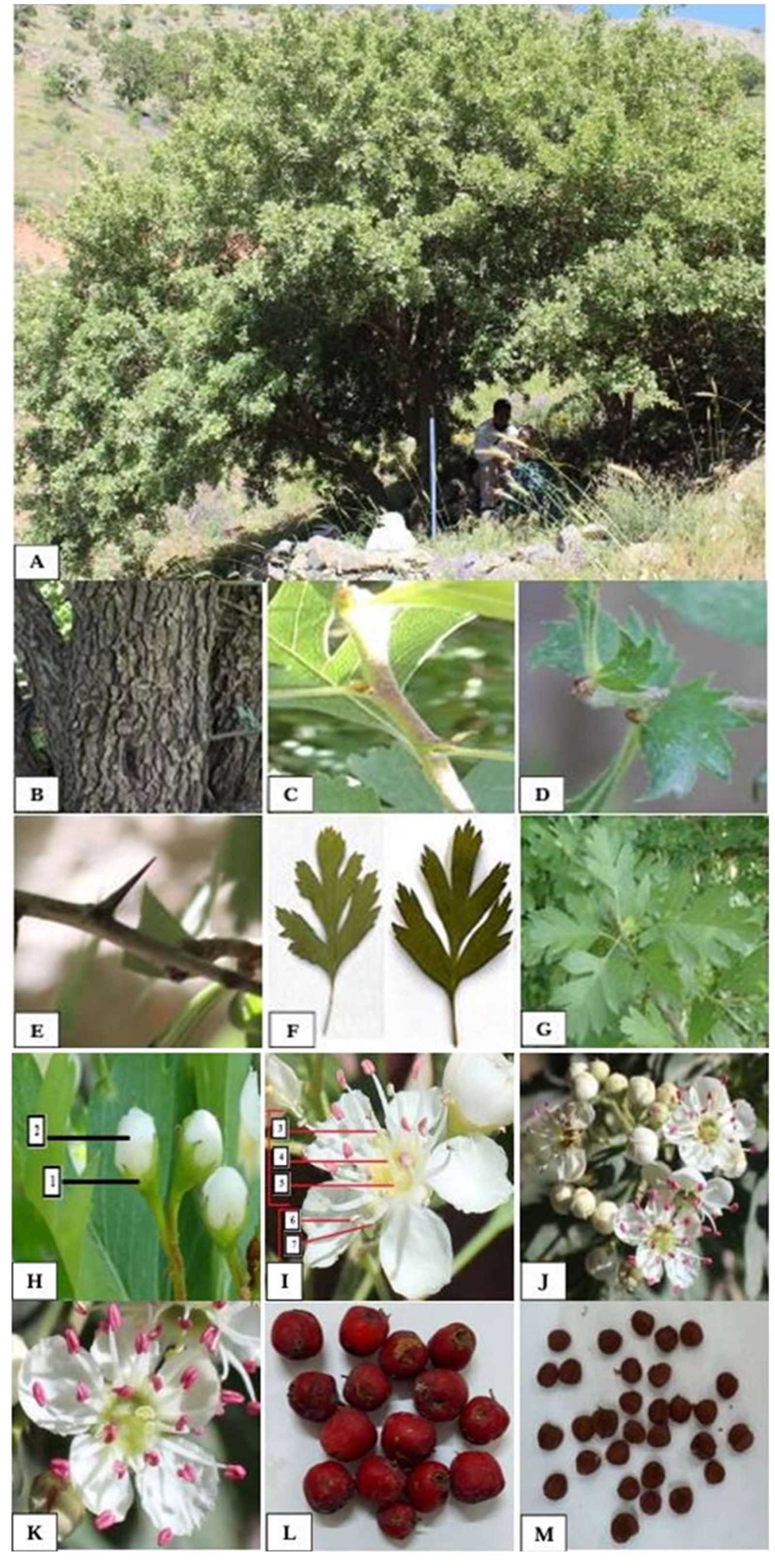

**Figure 4.2** ***Crataegus meyrei*****, A. Plant habit; B. Bark; C. Young shoot; D. Stipules; E. Spine; F & G Leaves; H. Bud 1- Sepal. 2- Petals; I. Disect flower. 3- Stigmas. 4- Style. 5- Ovary. 6- Filament. 7- Anthers; J. Inflorescences; K. Single flower: L. Fruits; M. Seeds.**

### 4.2.3 *Crataegus monogyna* Jacq

Deciduous small tree, 4 - 4.5 m tall, young shoots glabrous or grayish or dark reddish-brown sympathy or slightly fissured twigs grayish or dark reddish brown, Bark type scaly roughie, minimum to maximum length of spine was 0.9 -1.41 cm, spine average 0.71 cm long. Leaves pale green, petiolate, obovate, 2.94 - 5.05 × 2.21- 4.00 cm; pinnately lobed, 3-5, apex acute or truncate. Inflorescences compact corymbs, 2-15 flowered. Flowers white; pedicels 1.30 -1.85 cm long; sepal ovate, ovate-oblong, triangular, 1.30 - 2.01 × 1.11- 1.87 mm, pubescent; petals white, orbicular, sub orbicular, broadly obovate, 3.35 - 7.05 × 3.07 - 6.14 mm; stamens 18-20, white, milky, 3.47- 4.73 mm long, light purple, 1.21- 3.02 mm long, styles pale green, (1), stigmas capitate. Fruit globose, red ovoid glabrous, 9.05 - 13.62 mm in diam, glabrous or thinly hairy, especially around and just below persistent calyx Pyrene (seeds) light brown, sub globose, outer surface slightly furrowed, inner surface keeled or almost flat, 7.36 - 9.15× 56.07- 7.25 mm (see figure 4.3 and appendix 2 and 3).

**Habitats:** foothills, mountainsides, oak woodlands, rocky slope, stony soil; elevation 1243 m.

**Flowering:** March-June.

**Occurrence:** Frequent in the mixed forest in Kurdistan region of Iraq.

**Distribution:** Kurdistan Iraq, Iran, Turkey, Syria, Mediterranean region, Cyprus, Lebanon, and Palestine, Europe, and north Africa.

**Collections:** Msu: Karzan, Nareman, Qara Dag (21)11, Qara Dag (21)12, Nalparea-Milakawa (21)34, Penjwen- Blkian (21)41, Azmar (21)49, Azmar (21)50, Mor: Qasery-Kani basta (21)17, Soran -Hasan bag (21)42, Choman-Gomi Felaw (21)43, Khoshkan, (21)44.

Msu: Karzan, Nareman, Qara Dag (22)72, Qara Dag (22)73, Nalparea-Milakawa (22)95, Penjwen-Blkian (22)102, Azmar (22)50, Azmar (22)111, Mor: Qasery-Kani basta (22)78, Soran-Hasan bag (22)103, Choman-Gomi Felaw (22)104, Choman-Gomi Felaw, (22)45.

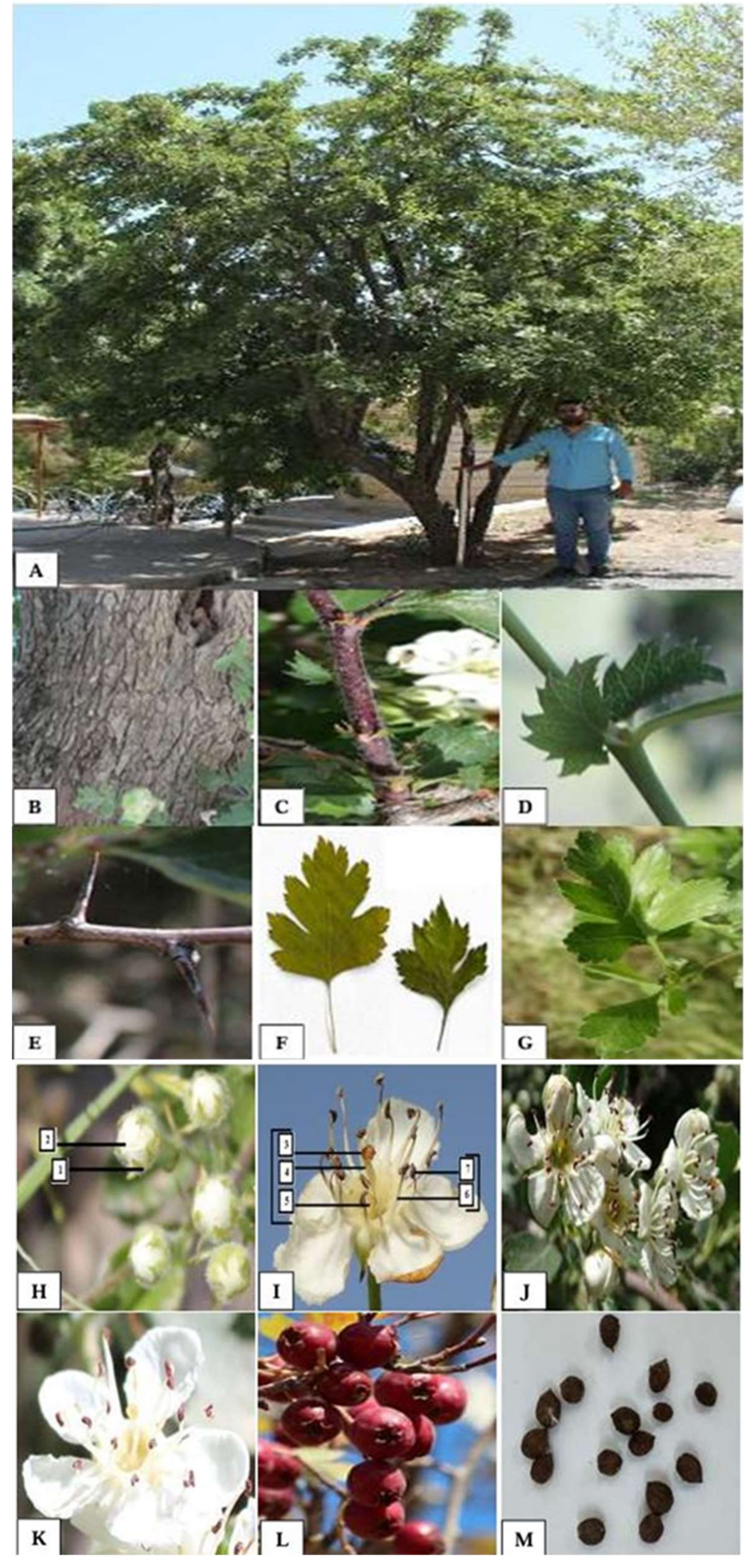

**Figure 4.3** ***Crataegus monogyna*****. A. Plant habit; B. Bark; C. Young shoot; D. Stipules; E. Spine; F & G Leaves; H. Bud 1- Sepal. 2- Petals; I. Disect flower. 3- Stigmas. 4- Style. 5- Ovary. 6- Filament. 7- Anthers; J. Inflorescences; K. Single flower: L. Fruits; M. Seeds.**

#### 4.2.4 *Crataegus orientalis* M. Bieb

Deciduous tree or shrubs, 5.5 - 6 m tall, young shoots whitish pubescent twigs grayish brown, bark with whitish hairs covered, minimum to maximum length of spine was 0.8 -1.45 cm, spine average 0.85 cm long. Leaves pale green, petiolate, triangular, 4.13 - 5.98 × 3.80 - 5.60 cm; pinnately lobed, 3-7, apex acute or truncate. Inflorescences compact corymbs, 3-15 flowered. Flowers white; pedicels 1.39 - 1.75 cm long; sepal ovate, ovate-oblong, triangular, 2.96 - 3.10 × 1.11- 1.23 mm, pubescent; petals white, orbicular, sub orbicular, broadly obovate, 3.65 - 5.85 × 3.07 - 5.20 mm; stamens 20, white, milky, 3.67 - 4.73 mm long, light purple, 1.12 - 1.33 mm long, styles pale green, (2-3), stigmas capitate. Fruit globose downy, reddish orange, 13.05 - 12.99 mm in diam., glabrous or thinly hairy, especially around and just below persistent calyx; pyrene (seeds) light brown, sub globose, outer surface slightly furrowed, inner surface keeled or almost flat, 7.56 - 9.45 × 6.17- 7.35 mm (see figure 4.4 and appendix 2 and 3).

**Habitats:** foothills, mountainsides, oak woodlands, rocky slope, stony soil; elevation 1243 m.

**Flowering:** March-June.

**Occurrence:** Occasional in the mixed forest in Kurdistan region of Iraq.

**Distribution:** Kurdistan Iraq, Iran, Turkey, Syria, Mediterranean region, Cyprus, Lebanon, and Palestine, Europe, and north Africa.

**Collections:** Mor: Karzan, Qasery-Walasha (21)20.

Mor: Karzan, Qasery-Walasha (22)81.

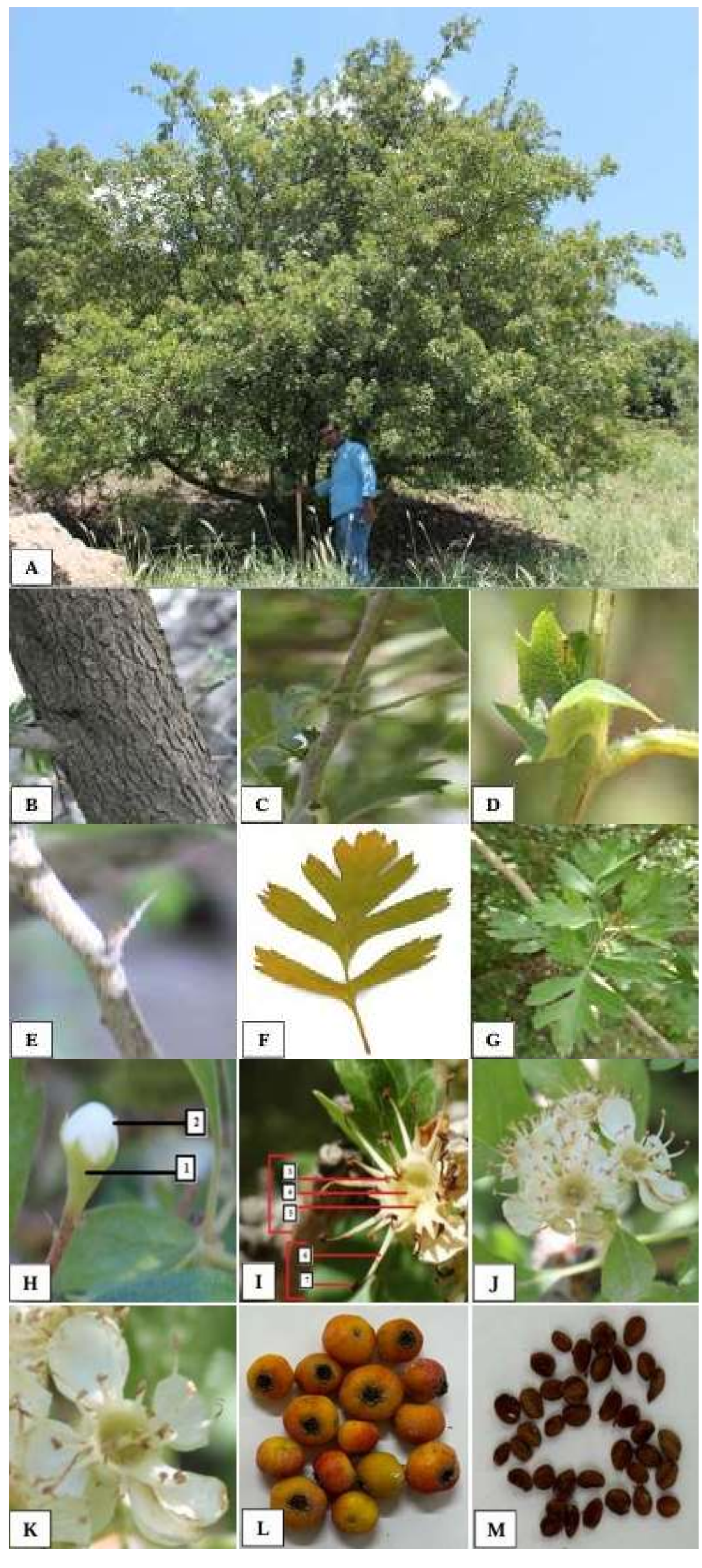

**Figure 4.4** ***Crataegus orentalies*** **. A. Plant habit; B. Bark; C. Young shoot; D. Stipules; E. Spine; F & G Leaves; H. Bud 1- Sepal. 2- Petals; I. Disect flower. 3- Stigmas. 4- Style. 5- Ovary. 6- Filament. 7- Anthers; J. Inflorescences; K. Single flower: L. Fruits; M. Seeds.**

### 4.2.5 *Crataegus pentagyna* Waldst. and Kit.exWilld

Deciduous small tree, 4 - 4.5 m tall, young shoots smooth or irregularly fissured twigs dark grayish reddish brown, Bark type dark grayish, minimum to maximum length of spine was 0.5 - 1.13 cm, spine average 0.56 cm long. Leaves pale green, petiolate, broadly ovate oblong or flabellate, 4.23 - 4.78 × 3.84 - 3.90 cm; pinnately lobed, 3 - 8, apex acute or truncate. Inflorescences compact corymbs, 3 - 15 flowered. Flowers white; pedicels 1.32 - 1.43 cm long; sepal ovate, ovate-oblong, triangular, 1.37 - 1.51 × 1.20 - 1.47 mm, pubescent; petals white, orbicular, sub orbicular, broadly obovate, 4.02 - 4.57 × 3.13 - 3.28 mm; stamens ca. 18 - 20, white, milky, 4.26 - 4.85 mm long, light purple, 1.32 - 1.43 mm long, styles pale green, (4 - 5), stigmas capitate. Fruit globose downy, reddish orange, 11.35 - 12.29 mm in diam., glabrous or thinly hairy, especially around and just below persistent calyx; pyrene (seeds) light brown, sub globose, outer surface slightly furrowed, inner surface keeled or almost flat, 5.68 - 5.73 × 3.19 - 4.45 mm (see figure 4.5 and appendix 2 and 3).

**Habitats:** foothills, mountainsides, oak woodlands, rocky slope, stony soil; elevation 1243 m.

**Flowering:** March-June.

**Occurrence:** Frequent in the mixed forest in Kurdistan region of Iraq.

**Distribution:** Kurdistan Iraq, Iran, Turkey, Syria, Mediterranean region, Cyprus, Lebanon, and Palestine, Europe, and north Africa.

**Collections:** Msu: Karzan, Saman, Nariman, Penjwen-Tatan (21)37, Sharbazher-Saraw (21)60.

Msu: Karzan, Saman, Nariman, Penjwen-Tatan (22)83, Sharbazher-Saraw (22)121.

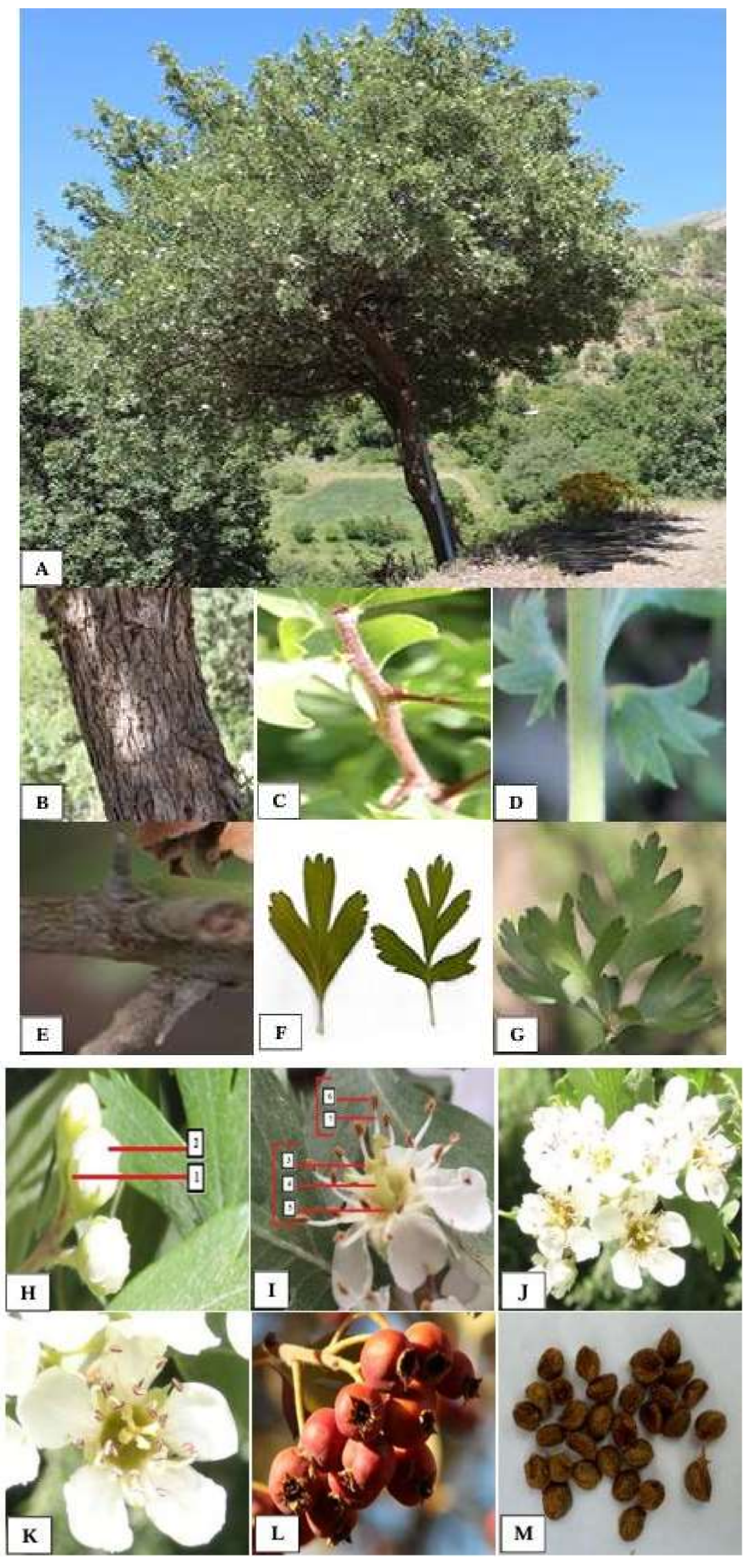

**Figure 4.5** ***Crataegus pentagyna*** **. A. Plant habit; B. Bark; C. Young shoot; D. Stipules; E. Spine; F & G Leaves; H. Bud 1- Sepal. 2- Petals; I. Disect flower. 3- Stigmas. 4- Style. 5- Ovary. 6- Filament. 7- Anthers; J. Inflorescences; K. Single flower: L. Fruits; M. Seeds.**

#### 4.2.6 *Crataegus azarolus* L. x *Crataegus meyrei* Pojark

Deciduous shrubs or small trees, 3.5 - 6 m tall, young shoots densely grayish or whitish pubescent, twigs grayish-brown, Bark type rough scaly bark, minimum to maximum length of spine was 0.50 - 0.95 cm, spine average 0.52 cm long. Leaves pale green, petiolate, obovate, 4.33 - 4.82 × 3.80 - 4.80 cm; pinnately lobed, 3 - 6, apex acute or truncate. Inflorescences compact corymbs, 3 - 15 flowered. Flowers white; pedicels 1.07 - 1.53 cm long; sepal ovate, ovate-oblong, triangular, 1.11 - 2.30 × 2.19 - 3.67 mm, pubescent; petals white, orbicular, sub orbicular, broadly obovate, 4.97 - 8.53 × 3.3 - 5.7 mm; stamens 15-20, white, milky, 4.97 - 8.53 mm long, light purple, 3.3 - 5.7 mm long, styles pale green, 1 or 2 (occasionally 3), stigmas capitate. Fruit globose, yellowish pink, 8.76 - 13.28 mm in diam., glabrous or thinly hairy, especially around and just below persistent calyx. Pyrene (seeds) pale brown, sub globose, outer surface furrowed, inner surface keeled or almost flat, 6.41 - 9.81 × 11.24 - 12.81 mm (see figure 4.6 and appendix 2 and 3).

**Habitats:** foothills, mountainsides, oak woodlands, rocky slope, stony soil; elevation 1243 m.

**Flowering:** March-June.

**Occurrence:** common in the foothills and mixed forest in Kurdistan region of Iraq.

**Distribution:** Kurdistan Iraq, Iran, Turkey, Syria, Mediterranean region, Cyprus, Lebanon, and Palestine, Europe, and north Africa.

**Collections:** Msu: Karzan, Saman, Nariman, Sharbazher-Khamza (21)3, Mlakawa (21)35, Penjwen-Nzara (21)40. Mor: Qasery-Kanibasta (21)16, Qasery-Walasha (21)19.

Msu: Karzan, Saman, Nariman, Sharbazher-Khamza (22)64, Penjwen-Mlakawa (22)96, Penjwen-Nzara (22)101. Mor: Qasery-Kanibasta (22)77, Qasery-Walasha (22)80.

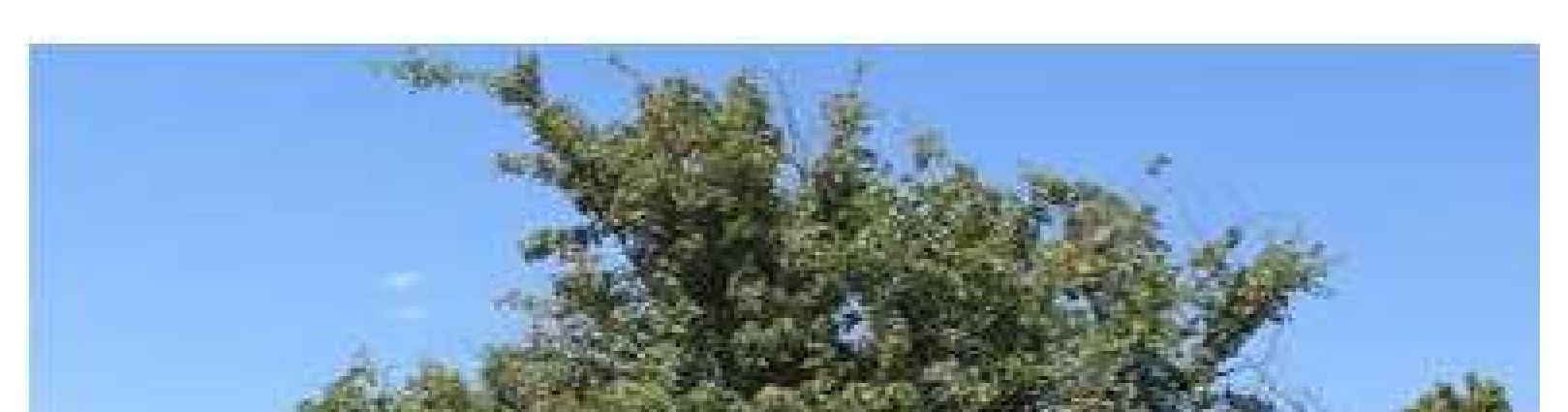

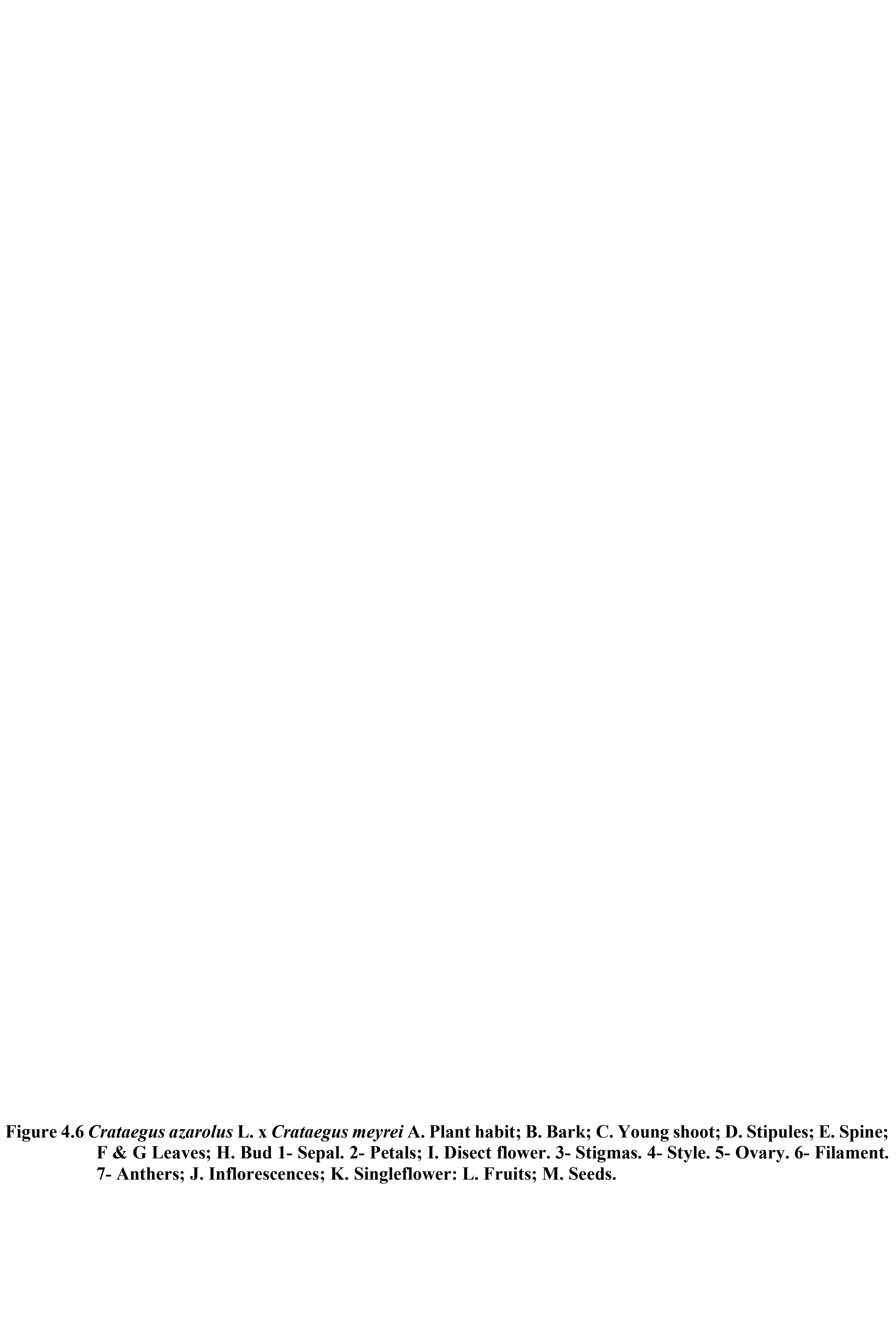

**Figure 4.6 *Crataegus azarolus* L. x *Crataegus meyrei* A. Plant habit; B. Bark; C. Young shoot; D. Stipules; E. Spine; F & G Leaves; H. Bud 1- Sepal. 2- Petals; I. Disect flower. 3- Stigmas. 4- Style. 5- Ovary. 6- Filament. 7- Anthers; J. Inflorescences; K. Singleflower: L. Fruits; M. Seeds.**

### 4.2.7 *Crataegus azarolus x Crataegus pentagyna* Waldst. and Kit.exWilld

Deciduous small tree, 3 - 4 m tall, young shoots densely grayish or whitish pubescent, twigs grayish-brown, Bark type finely fissured, minimum to maximum length of spine was 1.3 - 1.6 cm, spine 1.4 cm long. Leaves pale green, petiolate, obovate, 4.33 - 4.82 × 3.50 - 4.10 cm; pinnately lobed, 3 - 7, apex acute or truncate. Inflorescences compact corymbs, 3 - 12 flowered. Flowers white; pedicels 1.09 - 1.93 cm long; sepal ovate, ovate-oblong, triangular, 1.70 - 3.13 × 1.11 - 1.77 mm, pubescent; petals white, orbicular, sub orbicular, broadly obovate, 3.6 - 7.7 × 3.4 - 4.5 mm; stamens 18 - 20, white, milky, 4.97 - 8.53 mm long, light purple, 1.11 - 3.02 mm long, styles pale green, (3 - 4), stigmas capitate. Fruit globose, yellowish pink, 10.43 - 14.57 mm in diam., glabrous or thinly hairy, especially around and just below persistent calyx. Pyrene (seeds) pale brown, sub globose, outer surface furrowed, inner surface keeled or almost flat, 7.12 - 9.05 × 4.93 - 4.82 mm (see figure 4.7 and appendix 2 and 3).

**Habitats:** foothills, mountainsides, oak woodlands, rocky slope, stony soil; elevation 1243 m.

**Flowering:** March-June.

**Occurrence:** very common throughout of the mixed forest in Kurdistan region of Iraq.

Distribution: Kurdistan Iraq, Iran, Turkey, Syria, Mediterranean region, Cyprus, Lebanon, and Palestine, Europe, and north Africa.

**Collections:** Msu: Karzan, Qaiwan (21)4.

Msu: Karzan, Qaiwan (22)65.

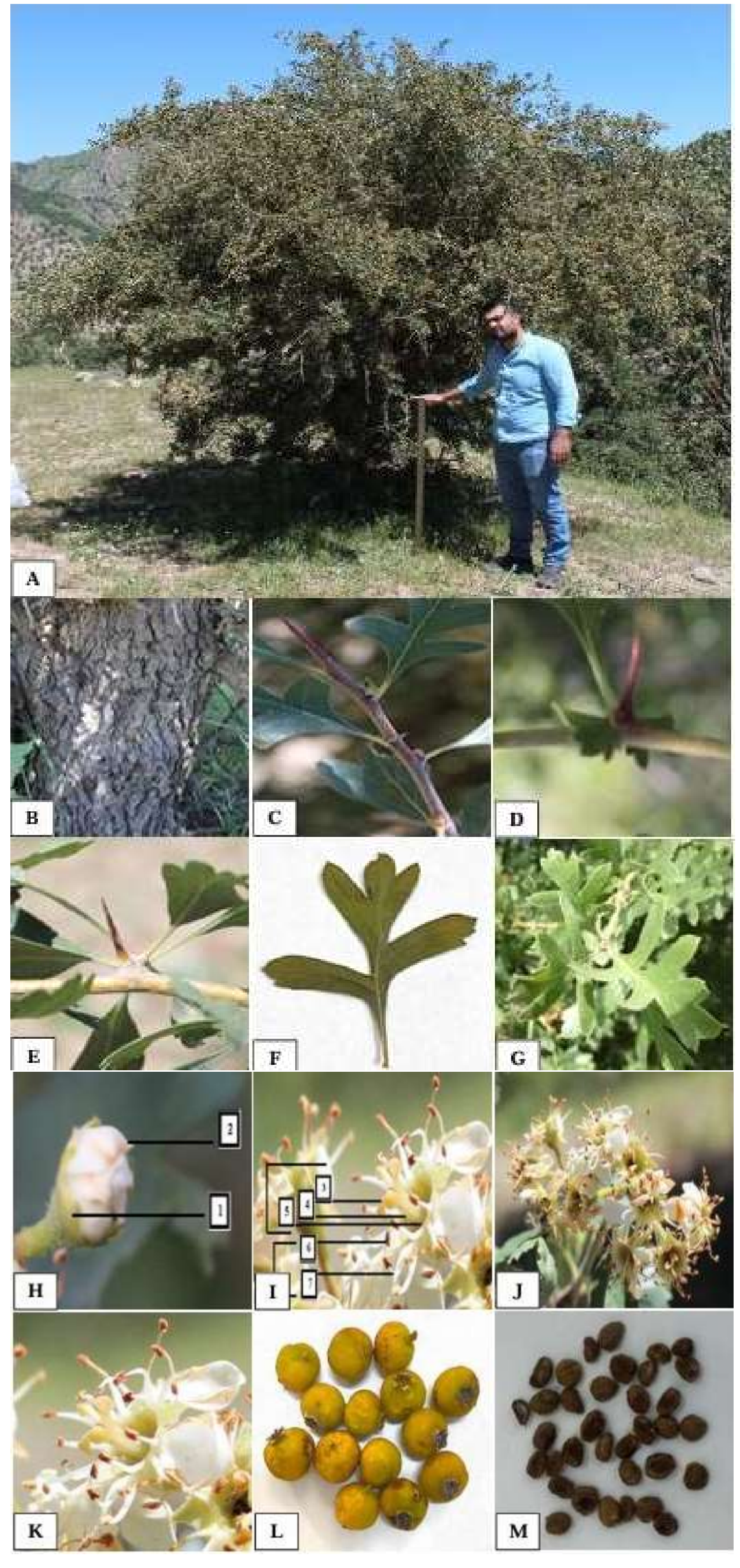

**Figure 4.7** ***Crataegus azarolus*** **x** ***Crataegus pentagyna*****. A. Plant habit; B. Bark; C. Young shoot; D. Stipules; E. Spine; F & G Leaves; H. Bud 1- Sepal. 2- Petals; I. Disect flower. 3- Stigmas. 4- Style. 5- Ovary. 6- Filament. 7- Anthers; J. Inflorescences; K. Singleflower: L. Fruits; M. Seeds.**

Through studying the morphological characteristics of *Crataegus* Spp. in detail in Kurdistan region of Iraq and comparing the order of *Crataegus* from the flora of Iraq, it became clear that there is similarity and overlap with relative differences between individuals of the same species. Thus, it was possible to isolate them taxonomically, which reinforced the necessity of studying these species. Therefore, in this study there are high levels of growth tree, fruit phenotypic, and physio-pomological diversity for 61 *Crataegus* species. It was deliberated that morpho-physio diversity is avital morphological marker alongside with genetic diversity to indicate a new wild Hawthorn species around different locations from Kurdistan region of Iraq. However, taxonomist observed that morphological marker of wild *Crataegus* species was difficult issue to indicate new variety because of, difficulty in distinguishing species due to overlapping characteristics, natural hybridization and also ecological factors (Ibrahimov et al., 2020). In addition, the study of the external characteristics of a part tender or reproductive plants are of great importance in plant taxonomy. According to Dönmez (2004) that the most significant taxonomic characters of *Crataegus* are indumentum, teeths and shape of leaves. Additionally, members of this genus exhibit yellow-red fruits and greenish calyx, therefore, the taxonomy of *Crataegus* is a difficult issue because of the polymorphic characters, hybridisation, inadequate collections and deficiency of field observations. Despite the progress achieved in the field of modern taxonomic sciences the general morphology remains the basis of taxonomic studies, because it represents the visual characteristics that give the species appearance and it is clearly visible to the researcher. We identified the critical components that mostly contribute the diversity of previously uncharacterized Iraqi *Crataegus* genotypes based on these features.

### 4.3 Physio-pomological descriptions of Hawthorn species

Data in (Table 4.1 and Appendix 4) demonstrate significant differences among 61 accessions for 11 variables investigated. it can be seen the maximum and the minimum fruit weight was revealed in G24 (11.26 g) and G1, G 12 and G49 (2.32, 2.33 and 2.41 g), respectively. Highest value of fruit length and fruit width which were G24 and G42 (13.76 and 13.75), G59 (14.79) while lowest values had the G10 (8.46), G11, G12 and G49 (7.52, 7.17 and 7.86) respectively. The highest and the lowest value of seed length and width were recorded as G38 and G47 (9.67 and 9.63), G10, G21, G22 and G53, (6.40, 6.35, 6.40 and 6.34) and G18 (7.92), G1 (4.77). The number of seeds per 5 fruits was observed in G56 and G59 (15.33) as a maximum while in G33, G34 and G41 (7.33, 7.0 and 7.33) as a minimum. In addition, G18 (13.50) had the highest rate of volume per 5 fruits while the lowest rate was notified in G22 (2.40). Fresh weight of fruit was indicated, as G24 (8.67) while in G12 (0.63). weight of seeds were measureed that G8 (4.20) had the highst value while G53 had (1.13). The pH value was observed in G42 and G49 had the maximum value as (4) while in G4 (2.47) had minimum value. G26 (87.20 %) and G3 (60.23) had the highest and lowest percentages of moisture contents, respectively.

**Table 4.1 Descriptive statistics of fruit characters in Hawthorn accessions.**

| Traits | Minimum | Maximum | Mean | Std. deviation | F | Pr > F |
|---|---|---|---|---|---|---|
| FW | 2.32 | 11.26 | 6.27 | 2.36 | 54.07 | <0.0001 |
| FL | 8.46 | 13.76 | 11.56 | 1.40 | 16.79 | <0.0001 |
| FW | 7.17 | 14.79 | 11.70 | 1.80 | 32.23 | <0.0001 |
| SL | 6.34 | 9.67 | 7.91 | 0.83 | 13.29 | <0.0001 |
| SW | 4.77 | 7.92 | 6.21 | 0.75 | 10.00 | <0.0001 |
| NSF | 7.00 | 15.33 | 11.70 | 2.17 | 35.39 | <0.0001 |
| VS | 2.40 | 13.50 | 7.52 | 2.61 | 654.84 | <0.0001 |
| WOF | 0.63 | 8.67 | 3.82 | 1.76 | 1158.54 | <0.0001 |
| WS | 1.13 | 4.20 | 2.46 | 0.76 | 2295.47 | <0.0001 |
| PH | 2.43 | 4.00 | 3.08 | 0.44 | 43.48 | <0.0001 |
| MC | 60.23 | 87.20 | 69.34 | 5.44 | 7.11 | <0.0001 |

FW: fruit weight (g), FL: fruit length (mm), FW. Fruit width (mm), SL: Seed length (mm), SW: Seed width(mm). NSF: Number of seed per fruit, VS: Volume solution, WOF. Weight of fresh fruit (g). WS: weight of seeds (g). PH: Potential of hydrogen. MC: moisture content (%). Min: minimum, Max: maximum, SD: standard deviation.

## 4.4 Multivariate analysis of fruit traits in *Crataegus* spp.

Principal component analysis (PCA) and Agglomerative hierarchical clustering (AHC) are one of the most popular clustering and multivariate statistical methodology used for evaluating and understanding the complex and huge datasets. The pattern of variability in *Crataegus* accessions was analyzed using PCA and AHC based on the correlation between the traits and extracted clusters to assess the variety of the genotypes and their relationship with the observed traits. Results from (Figure 4.8 A and B) categorized all accessions into six distinct groups. The most populous group (C4) encompassed 14 accessions, whereas groups C3 and C6 contained the fewest genotypes, with six accessions each. Group one (C1, depicted in red) was characterized by the highest values of SW and MC traits. The second group (C2, shown in green) stood out for its high WOF, FL, WF, and SL traits. The third cluster (C3, represented in blue) was notable for its high NSF, FW, and WS. In contrast, Groups 6 and 7 had the lowest values for all studied traits.

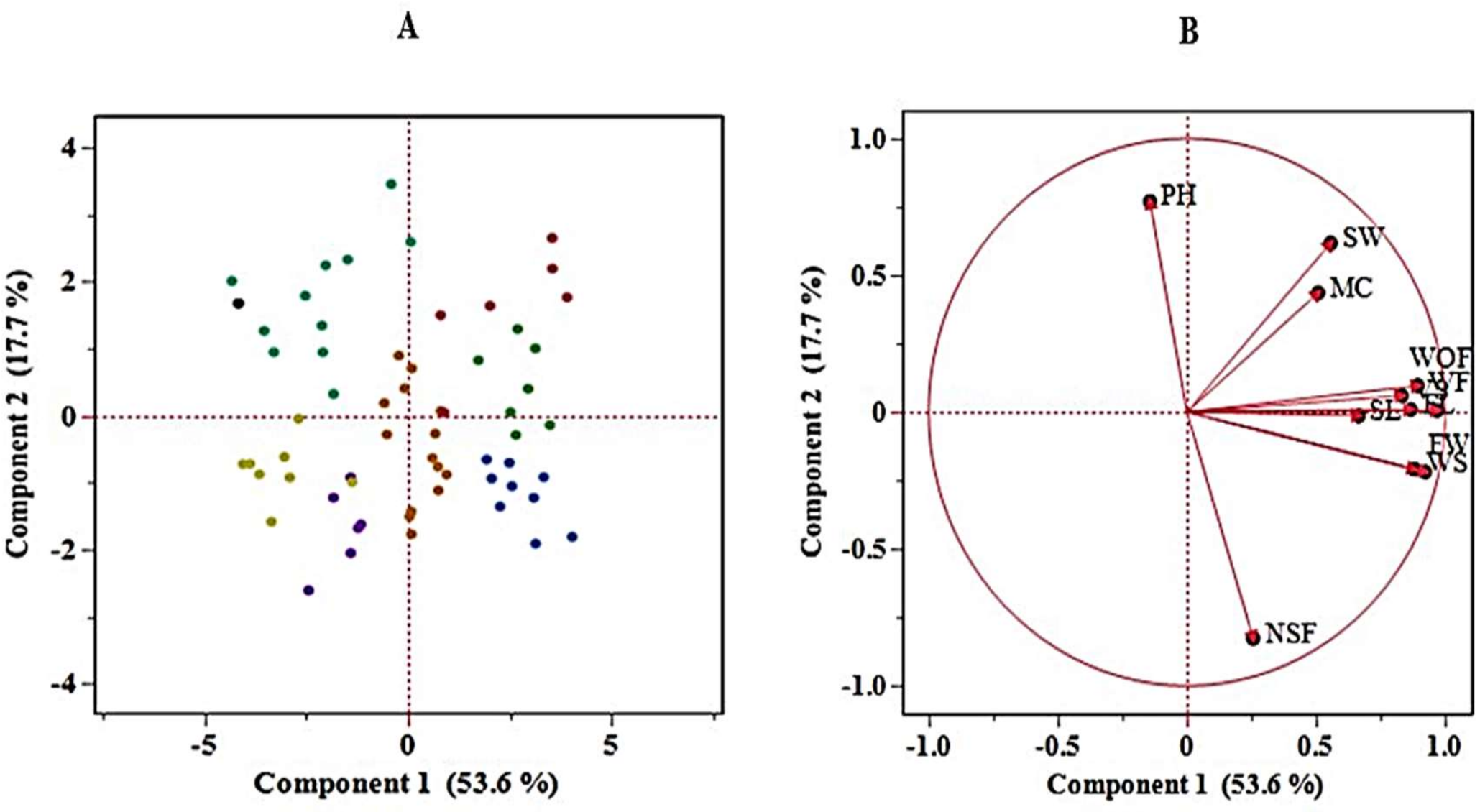

**Figure 4. 8 principal component analysis scatter plot of the studied accessions of Crataegus spp. based on morpho-pomological characteristics A: Principal component analysis (PCA) and Agglomerative hierarchical clustering (AHC).**

In addition, seven clusters were formed in the case of hierarchical clustering of fruit characteristics data from all *Crataegus* genotypes accessions (Figure 4.9). The first cluster (in red) had six genotypes: G42, G24, G18, G44, G61, and G39) while the second cluster (in green) had seven: G29, G20, G26, G48, G15, G23, and 25). However, (G31, G60, G38, G8, G47, G35, G51, G59, and G32) were the nine genotypes in Cluster 3 (in blue). G27, G36, G14, G52, G56, G54, G46, G40, G19, G16, G58, G57, G37, and G28 comprised the fourth cluster (in brown). G43, G34, G33, G17, G55, G45, G50, G41, G49, G12, and G11 were parts of cluster five (in Cyan). In the six cluster (purple) had six genotypes such as (G1, G2, G4, G3, G30, and G6), while the final cluster G21, G22, G7, G9, G53, G10, G13, and G5 were parts of cluster seven (in rust gold).

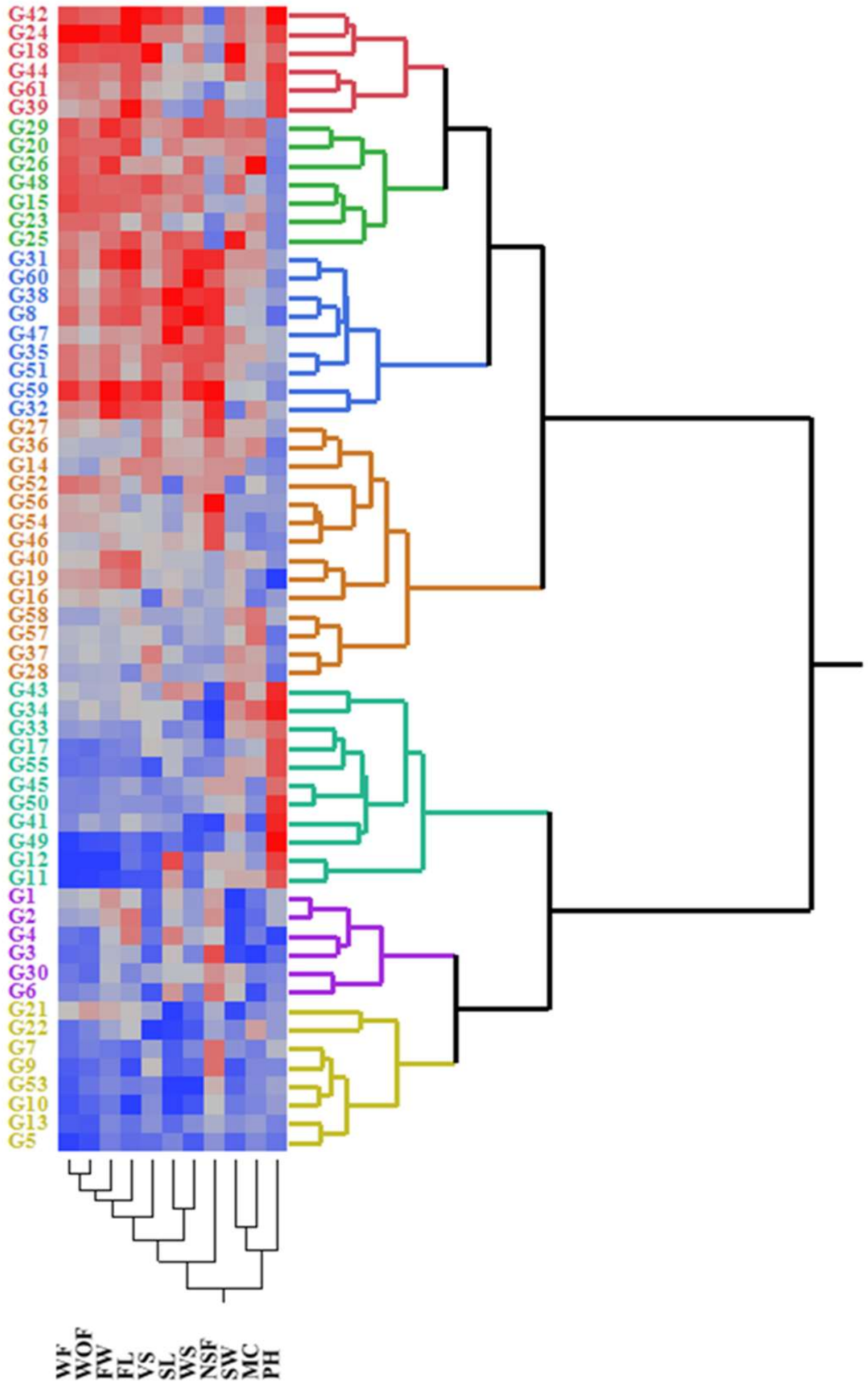

**Figure 4. 9 UPGMA dendrogram based on the physio morphological studied of 61 accessions of Crataegus spp.**

Furthermore, the 11 variables fruit physio-pomological characteristics of Hawthorn species of this study were quite effective to differentiate the among 61 accessions and to disclose the underpinning phenotypic variability. FW: fruit weight (g), FL: fruit length (mm), FW. Fruit width (mm), SL: Seed length (mm), SW: Seed width (mm). NSF: Number of seed per fruit, VS volume solution (VS), WOF. Fleshy weight of fruit (g). WS: weight of seeds (g). pH: potential of hydrogen. MC: moisture content (%) were the significant demonstrate of genetic diversity in the genotypes investigated.

Morphological markers are visible plant traits controlled by Mendelian genes which congregate with genes determining the expression of the trait of interest to allow selection for suitable individuals from a population (Akbari1, et al., 2014). Plant genotypes or species can be distinguished using morphological marker systems for breeding, mapping, or genetic variation surveys by utilizing a number of visible phenotypic characteristics. Morphological markers can be influenced by other markers because of their pleiotropic gene action (López-Caamal and Tovar-Sánchez, 2014). Many environmental factors can influence the number and appearance of plant morphological markers.

The study revealed that there were differences in terms of fruit characteristics among Hawthorn species, and thus, better quality Hawthorn genotypes can be selected within the species. Since the edible organ of this plant is the fruit, considering the high variation in fruit weight and pericarp weight, superior ecotypes in terms of these traits can be used for future breeding programs and achieve higher yielding ecotypes. The results showed that there is a significant diversity in terms of morphological and functional traits. Previous studies have shown that there is a significant genetic diversity among wild fruits, which can be caused by the repeated reproduction of wild fruit seeds in natural habitats, which ultimately increases genetic diversity (Guo et al. 2020).

### 4.5 Assessment of biochemical traits in Hawthorn fruits

The high nutritional quality of the studied fruit demonstrated the importance of the mentioned fruit in human health as a bioactive compounds source. In this study, the genetic diversity of wild Hawthorn was observed through the utilization of biochemical markers such as TSS, TPC, TFC, AA, and CAC. The diversification of fruit components has referred to genetic structure and environmental factors. In our results, a considerable variation was revealed in the biochemical traits of Hawthorn species.

The results of ANOVA analysis (Appendix 5) showed significant differences in the biochemical traits (TSS, TPC, TFC, AA and CAC) of the accessions examined ($P \leq 0.001$). Accession G21 recorded the highest TSS value at (29.70) Brix, while G6 presented the minimum value of (10.25) Brix. Both G5 and G49 revealed the highest values of TPC and TFC, with 2.58 µg/g FW and 0.78 µg/g FW, respectively, whereas G35 showed the lowest values of TPC and TFC with 0.28 µg/g FW and 0.07 µg/g FW, respectively. The highest values of AA and CAC were scored 1.40 µg/g FW and 25.26 µg/g FW of the accession G8, and the lowest values were G11 with 0.27 µg/g FW and 4.87 µg/g FW, respectively. The mean values of TSS, TPC, TFC, AA, and CAC were identified as 19.75 Brix, 1.06 µg/g FW, 0.27 µg/g FW. 1.01 µg/g FW, and 18.31 µg/g FW, respectively (Table 4.2).

**Table 4.2 Descriptive statistics of biochemical traits in the accessions of Hawthorn fruits**

| Traits | Min. value | Max. value | Mean value | SD | *F* | *P*-value |
|---|---|---|---|---|---|---|
| TSS | 10.25 (G6) | 29.70 (G21) | 19.75 | 5.97 | 427.62 | < 0.0001 |
| TPC | 0.28 (G35) | 2.58 (G5) | 1.06 | 0.53 | 47.28 | < 0.0001 |
| TFC | 0.07 (G35) | 0.78 (G49) | 0.27 | 0.17 | 45.87 | < 0.0001 |
| AA | 0.27 (G11) | 1.40 (G8) | 1.01 | 0.26 | 26.01 | < 0.0001 |
| CAC | 4.87 (G11) | 25.26 (G8) | 18.31 | 4.54 | 221.34 | < 0.0001 |

*TSS* total soluble solids (Brix), *TPC* total phenolic content (µg/g FW), *TFC* total flavonoid content (µg/g FW), *AA* antioxidant activity (µg/g FW), *CAC* carotenoid content (µg/g FW).

### 4.6 Hierarchical cluster analysis based on biochemical traits

A hierarchical clustering was used to assess the Hawthorn fruit biochemical traits (TSS, TPC, TFC, AA and CAC). Hawthorn accessions were divided into seven clusters derived from these fruit biochemical data (Figure 4.10). The first cluster (red colour) included four accessions (G1, G2, G6, and G30). The second cluster (green colour) involves fifteen accessions (G7, G8, G14, G22, G40, G46, G47, G51, G53, G54, G56, G57, G58, G59, and G60). The greatest cluster is the third cluster, which is comprised of 25 accessions. G15, G23, G28, G32, G35, and G52 were the six accessions in the fourth cluster (in brown), while the same number of accessions, G3, G10, G12, 16, G44, and G49, formed the fifth cluster (in teal). The sixth cluster (in purple) was produced from the three accessions, which were G4, G5, and G11, and only two accessions, G17 and G45, generated the seventh cluster (in olive).

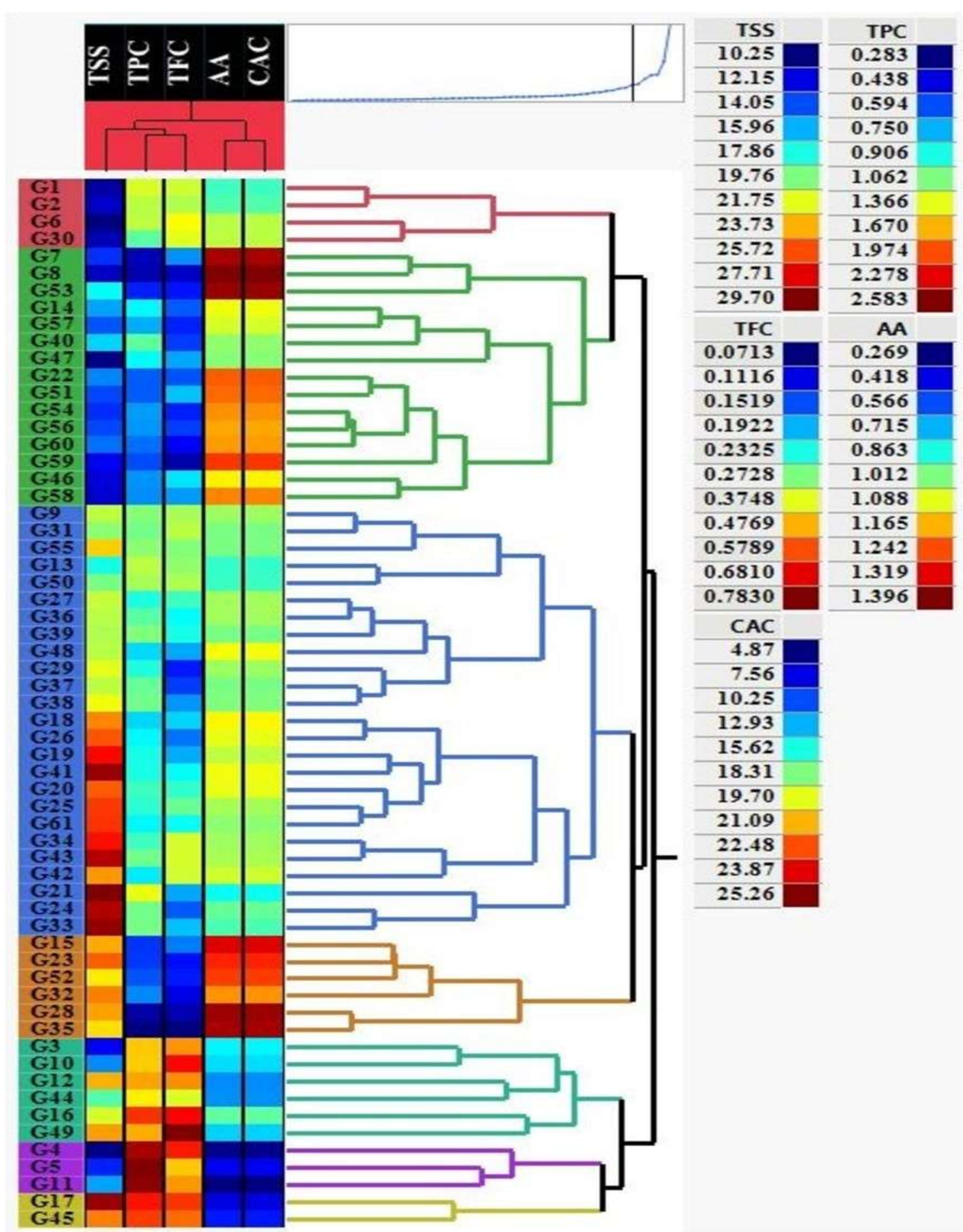

**Figure 4.10 A hierarchical cluster formed based on the bio- chemicals of 61 Hawthorn fruit accessions. Total soluble sol- ids (TSS), total phenolic content (TPC), carotenoid content (CAC), total flavonoid content (TFC), antioxidant activity (AA).**

In our study, biochemical traits were utilized to know the variation between the Hawthorn accessions, and the highest differences were found between the accessions in the contents of TSS, TPC, TFC, AA, and CAC (Appendix 5). In point of fact, seven cluster have resulted in a high distance between the accessions in different clusters. These results may be due to various factors, including the variation in TSS and TFC traits of Hawthorn fruits are influenced by Hawthorn species and sites of sample collection (Li et al. 2015). As mention before, Kurdistan region of Iraq is known as one of the primary centers of genetic diversity of *Crataegus*; however, few studies have been carried out on the content of total phenolic compounds, antioxidant activity, and antimicrobial e effects of this genus. Furthermore, environmental conditions, postharvest handling, and processing are among the factors that might have an influence on the physical characteristics, chemical composition of phenolic compounds, and their antioxidant activity (Lorenzo and Munekata, 2016). Researcher also confirmed that, environmental conditions such as temperature, light, and altitude affect the polyphenol contents in the fruits of Hawthorn plants, particularly in phenylpropanoid metabolism (Alirezalu et al. 2020). In the case of antioxidant activity in the Hawthorn fruits, phenolic compounds are detected as a main contributor to the activity of antioxidant and enhanced by fruit species variations. In addition, the flavonoid content, phenolic compound concentrations, fruit pigments such as anthocyanins and carotenoids content in Hawthorn fruits are influenced by the high temperature and the level of $CO_2$ (Chang et al. 2013).

### 4.7 Alignment of Sequences and Phylogenetic Reconstruction of the *Crataegus* spp.

Free common BLAST database was utilized to compare the result sequences of ITS rDNA PCR product amplified. All taxa's ITS sequences were checked, and new accessions were added to NCBI GenBank. According to a molecular analysis of 61 taxa of the *Crataegus* accessions Universal primer pairs ITS1 and ITS4 was used which is sequence of the tested species ranged from 605-705 bp indicated by using ITS sequencing (Figure 4.11). The sequences contained 18S rRNA, the entire ITS1, and 5.8S rRNA regions. To generates a neighbor-joining tree through a series of nucleotide sequence alignments. *Crataegus* total genome set species was used that all 61 trees divided into two clades (Figure 4.12). First group consists of one clade which only genotype (G37). On the other hand, the rest genotypes which is belonged the second group are 60 accessions that are consent of nine sub-clads. These results revealed that there were discernible divisions in the molecular phylogeny of the various taxonomic forms of *Crataegus*. An examination of the DNA sequence derived from both ITS regions revealed a significant amount of genetic diversity among the samples that were studied.

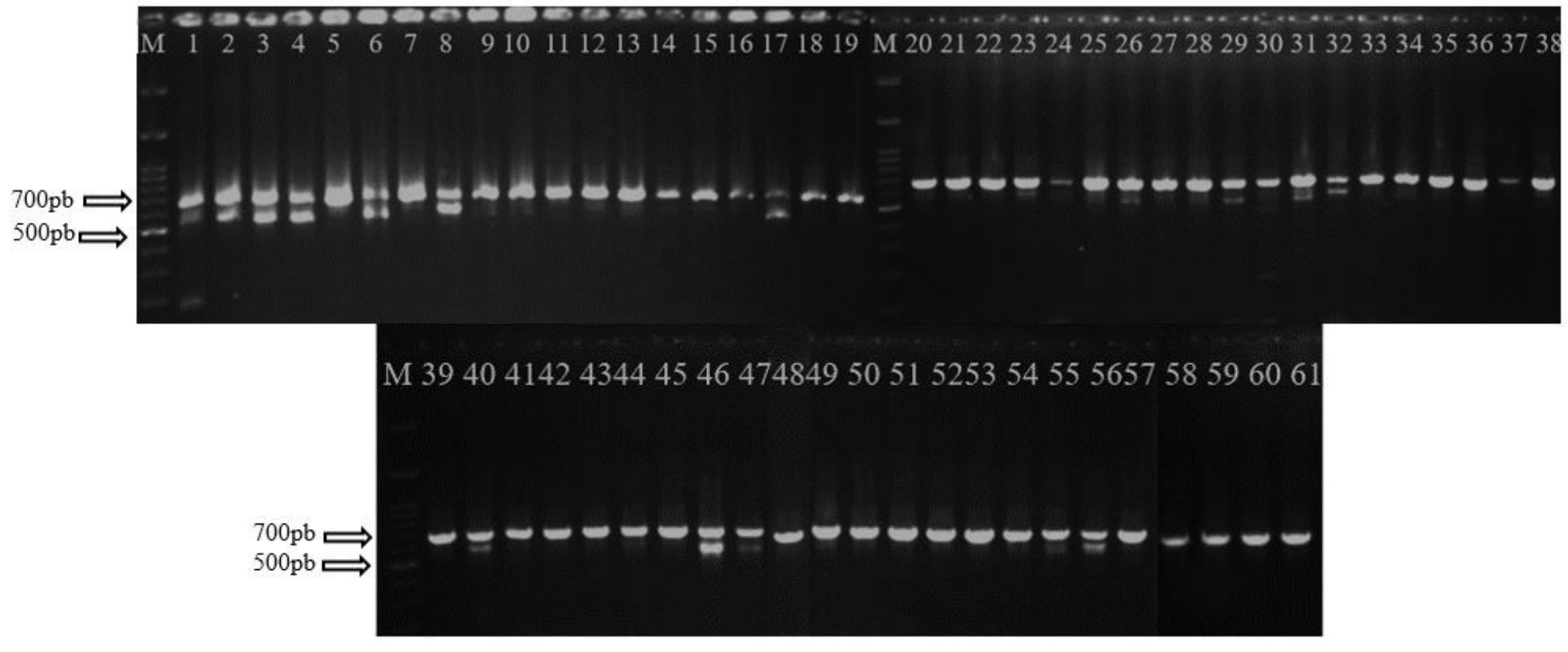

**Figure 4. 11 ITS-PCR product of investigated taxa used sequencing.**

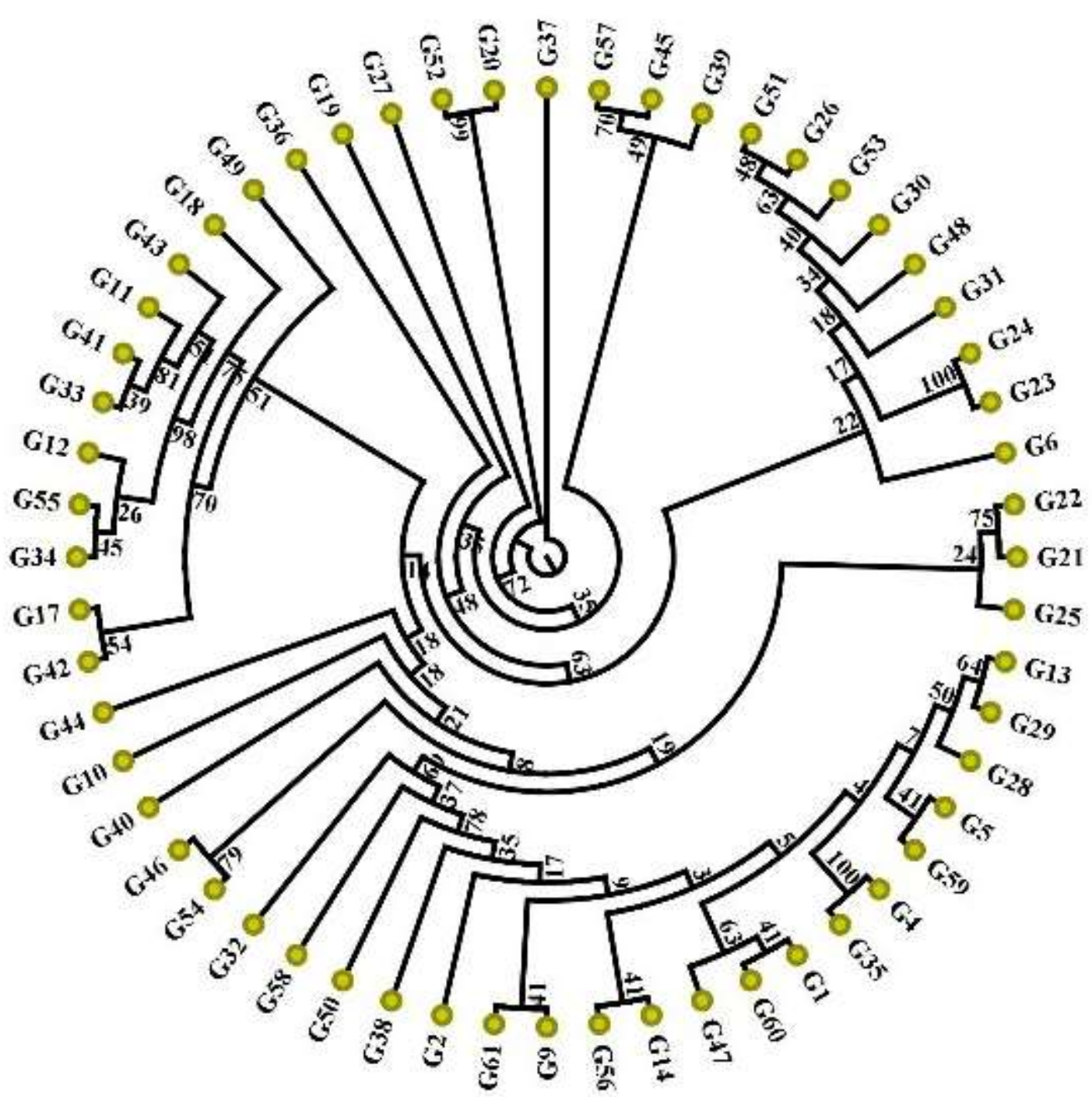

**Figure 4. 12 Clustering of 61 Crataegus genotypes based on ITS gene sequencing data. The numbers on the node indicate the bootstrap percentage.**

The illogicality in *Crataegus* taxonomy at the species has been documented by the internal transcribed spacer (ITS) region of nuclear ribosomal DNA, about ITS1 and ITS4. The current study involves the PCR product of Hawthorn, measuring roughly 605-705 base pairs. Following PCR amplification using ITS1 and ITS4 primers. This study attempted to resolve the phylogenetic relationship of the 61 genotypes invested in the region of Iraqi Kurdistan using the ITS rDNA. Five species with two hybrids were discovered by this study from the collected samples of wild *Crataegus* species. highly homogenous within the group was documented according to the ITS region of the *Crataegus* species in the same clade group, which is the reason for putting them grouped together. Moreover, the bootstrap in this investigation was 3 and 100%. The BLASTn analysis of the ITS data showed that the dominant strains in all the identified genotypes samples corresponded to which indicates that the species adapt to different environments. In the context of a species, genetic diversity refers to the degree of genetic variability among individuals belonging to a variety or population within that species (Salgotra and Chauhan 2023). All sequencing results were documented (see Appendix 6).

### 4.8 SSR, ISSR, and SCoT marker polymorphisms and discriminating features

Thirty-one (11 SSR, 10 ISSR, and 10 SCoT) markers were used to evaluate the genetic diversity among 61 accessions of Hawthorn. A total of 237 polymorphic bands were generated from the total primers used. SSR, ISSR, and SCOT markers yielded 57 (5.18 per primer), 46 (4.6 per primer), and 134 (13.4 per primer), respectively (Table 4.3 and appendix 7). According to the SSR markers, SSR 09 and SSR 19 generated only 1.00 polymorphic bands and SSR 58 recorded the maximum polymorphic bands, 13.00. The TPb of ISSR primers oscillated from 2.00 (UBC- 891) to 6.00 (UBC-810, UBC-818, UBC-822, and UBC-826). SCoT markers ranged from 10.00 (SCoT 10) to 17.00 (SCoT 6). SSR 40 among the SSR primers indicated the highest values of Na, Ne, I, He, uHe, and PIC, which were 1.42, 1.35, 0.30, 0.20, 0.22, and 0.36, respectively. Whereas the lowest values of these indices expected PIC were presented by SSR 17, with the values of 0.42, 1.01, 0.02, 0.01, and 0.01, respectively, and SSR 46 showed the minimum PIC value, which was 0.05. According to the data of ISSR, the maximum values of Na, Ne I, He, and uHe were detected by the UBC-818 primer with 1.38, 1.37, 0.31, 0.21, 0.24, respectively. The highest PIC value indicated by UBC-823 with value of 0.36. The minimum values of these indices were revealed for the UBC-810 marker, with values of 0.73, 1.11, 0.11, 0.07, 0.07, and 0.15, respectively. In case of SCoT markers, SCoT 12 recorded the highest value of 1.34, 1.37, 0.33, 0.22, 0.25, and 0.41 for the Na, Ne, I, He, uHe, and PIC, respectively. Although SCoT 10 indicated the lowest value of these indices by 0.77, 1.18, 0.16, 0.10, 0.11, and 0.2, respectively. The average values of Na, Ne, I, He, uHe, and PIC for the SSR primers were 0.98, 1.18, 0.17, 0.11, 0.12, and 0.19. Whereas the mean values of these indices for the ISSR markers were 1.12, 1.23, 0.20, 0.13, 0.15, and 0.23, and for the SCoT markers were 1.13, 1.28, 0.25, 0.17, 0.18, and 0.28, respectively.

**Table 4.3 SSR,ISSR and ScoT primers with their values of the indices of TPB, Na, Ne, I, He, uHe, and PIC.**

| | TPB | Na | Ne | I | He | uHe | PIC |
|---|---|---|---|---|---|---|---|

| *SSR markers* | | | | | | | |
|---|---|---|---|---|---|---|---|
| SSR09 | 1.00 | 1.30 | 1.21 | 0.17 | 0.12 | 0.13 | 0.17 |
| SSR17 | 3.00 | 0.42 | 1.01 | 0.02 | 0.01 | 0.01 | 0.07 |
| SSR19 | 1.00 | 1.22 | 1.20 | 0.17 | 0.11 | 0.13 | 0.21 |
| SSR35 | 4.00 | 0.92 | 1.19 | 0.18 | 0.12 | 0.12 | 0.19 |
| SSR38 | 7.00 | 0.69 | 1.12 | 0.12 | 0.08 | 0.08 | 0.16 |
| SSR40 | 6.00 | 1.42 | 1.35 | 0.30 | 0.20 | 0.22 | 0.36 |
| SSR46 | 3.00 | 0.76 | 1.05 | 0.06 | 0.03 | 0.04 | 0.05 |
| SSR56 | 9.00 | 1.24 | 1.28 | 0.25 | 0.17 | 0.18 | 0.24 |
| SSR58 | 13.00 | 1.24 | 1.26 | 0.24 | 0.16 | 0.17 | 0.24 |
| SSR61 | 8.00 | 1.00 | 1.21 | 0.20 | 0.13 | 0.15 | 0.25 |
| SSR99 | 2.00 | 0.56 | 1.09 | 0.10 | 0.06 | 0.07 | 0.14 |
| Mean | 5.18 | 0.98 | 1.18 | 0.17 | 0.11 | 0.12 | 0.19 |
| Total *ISSR markers* | 57.00 | | | | | | |
| UBC-808 | 5.00 | 1.22 | 1.28 | 0.22 | 0.15 | 0.16 | 0.21 |
| UBC-810 | 6.00 | 0.73 | 1.11 | 0.11 | 0.07 | 0.07 | 0.15 |
| UBC-812 | 5.00 | 1.20 | 1.22 | 0.19 | 0.13 | 0.14 | 0.19 |
| UBC-818 | 6.00 | 1.38 | 1.37 | 0.31 | 0.21 | 0.24 | 0.31 |
| UBC-822 | 6.00 | 0.93 | 1.27 | 0.23 | 0.15 | 0.16 | 0.33 |
| UBC-823 | 4.00 | 1.14 | 1.24 | 0.24 | 0.15 | 0.17 | 0.36 |
| UBC-826 | 6.00 | 1.26 | 1.29 | 0.23 | 0.16 | 0.17 | 0.15 |
| UBC-846 | 3.00 | 1.16 | 1.15 | 0.14 | 0.09 | 0.10 | 0.22 |
| UBC-880 | 3.00 | 1.08 | 1.20 | 0.18 | 0.12 | 0.13 | 0.20 |
| UBC-891 | 2.00 | 1.14 | 1.16 | 0.14 | 0.10 | 0.10 | 0.23 |
| Mean | 4.60 | 1.12 | 1.23 | 0.20 | 0.13 | 0.15 | 0.23 |
| Total *SCoT markers* | 46.00 | | | | | | |
| SCoT 6 | 17.00 | 1.00 | 1.23 | 0.21 | 0.14 | 0.15 | 0.22 |
| SCoT 10 | 10.00 | 0.77 | 1.18 | 0.16 | 0.10 | 0.11 | 0.21 |
| SCoT 12 | 12.00 | 1.34 | 1.37 | 0.33 | 0.22 | 0.25 | 0.41 |
| SCoT 14 | 11.00 | 1.31 | 1.34 | 0.29 | 0.20 | 0.21 | 0.32 |
| SCoT 16 | 13.00 | 1.15 | 1.29 | 0.26 | 0.17 | 0.18 | 0.27 |
| SCoT 22 | 13.00 | 1.32 | 1.31 | 0.28 | 0.19 | 0.21 | 0.29 |
| SCoT 23 | 16.00 | 1.04 | 1.26 | 0.22 | 0.15 | 0.16 | 0.21 |
| SCoT 33 | 14.00 | 1.04 | 1.27 | 0.24 | 0.16 | 0.17 | 0.27 |
| SCoT 34 | 15.00 | 1.02 | 1.22 | 0.20 | 0.13 | 0.14 | 0.29 |
| SCoT 35 | 13.00 | 1.26 | 1.35 | 0.29 | 0.20 | 0.22 | 0.27 |
| Mean | 13.40 | 1.13 | 1.28 | 0.25 | 0.17 | 0.18 | 0.28 |
| Total | 134.00 | | | | | | |

## 4.9 Cluster analysis via SSR, ISSR, SCoT data

A dendrogram was formed using Unweighted pair group method with arithmetic mean (UPGMA). The SSR markers categorized all Hawthorn accessions into seven main groups (Figure 4.13A). The first (in red) and the fifth (in light blue) clusters contained only one accession, which was G1 and G42, respectively. While two accessions (G4 and G47) produced the second cluster (green colour), and two accessions (G8 and G14) generated the third cluster (in blue). The fourth cluster (in brown) comprised 25 accessions. The fifth cluster in cyane as 1. 21 accessions generated the sixth cluster (in purple). The last cluster (in yellow) involves nine accessions. UPGMA was used to cluster the hawthorn accessions based on the ISSR data; seven main clusters were divided into (Figure 4.13B). The first cluster (in red) comprised seven accessions (G1, G2, G5, G6, G8, G9, and G10) and also another seven accessions (G3, G4, G32, G47, G52, G55, and G56) generated the second cluster (in green).

Whereas the third cluster (in blue) included only one accession (G48). The fourth cluster (in brown) included 19 accessions and 10 accessions produced the fifth cluster (in light blue). The sixth cluster (in purple) comprised 13 accessions. The seven cluster (in yellow) generated by four accessions (G58, G59, G60, and G61). Based on the SCoT marker data (Figure 4.13C), seven clusters were split into. The first cluster (in red) included only one accession (G1) and one accession (G3) generated the second cluster (in green). The third cluster (in blue) included six accessions (G54, G57, G58, G59, G60, and G61), while 12 accessions produced the fourth cluster (in brown). The fifth cluster (in light blue) comprised three accessions (G7, G8, and G9). The sixth cluster (in purple) produced 27 accessions and 11 accessions generated cluster the seven cluster (in yellow).

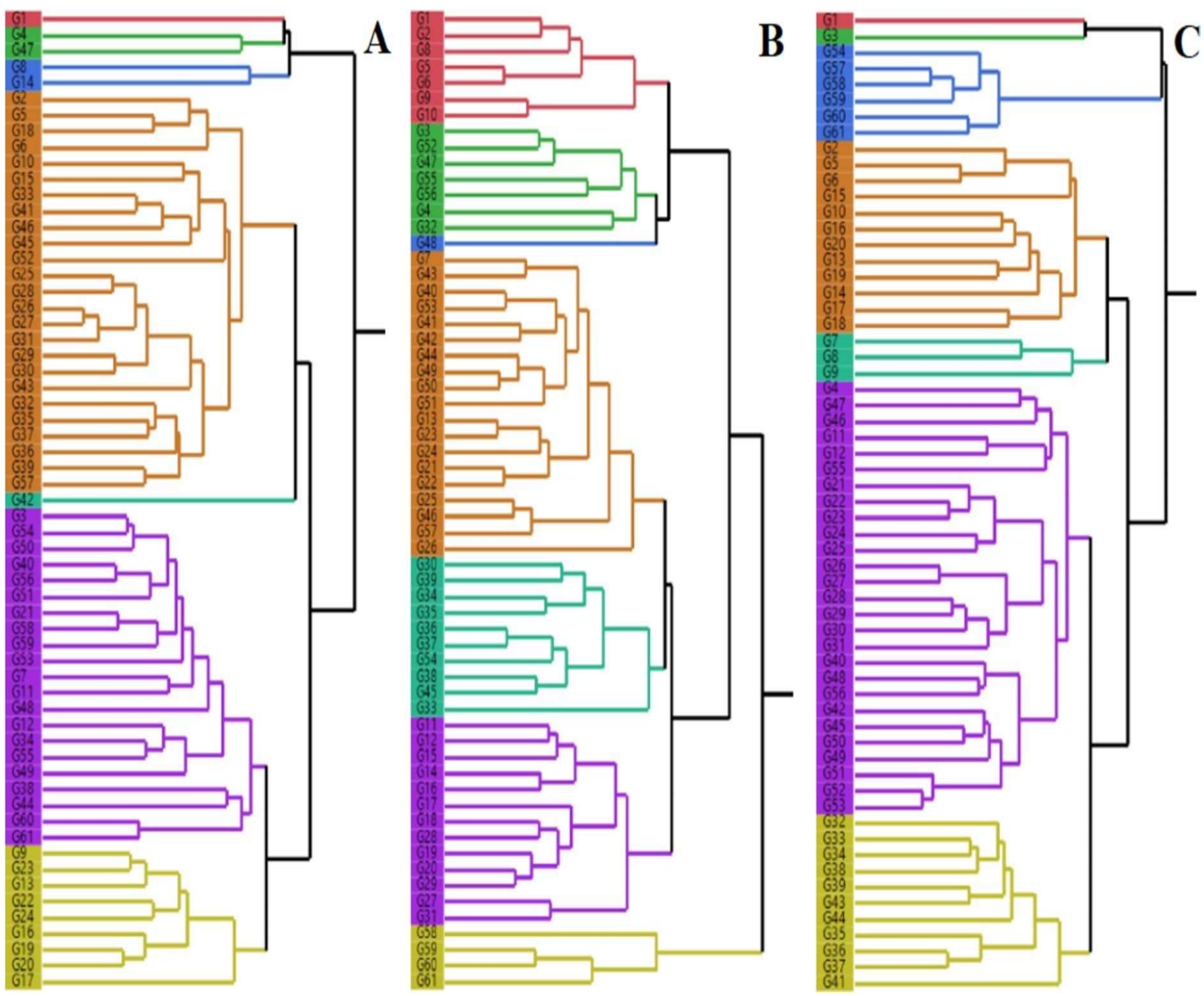

**Figure 4.13 Unweighted pair group method with arithmetic mean (UPGMA) dendrograms of the sixty-one Hawthorn acces- sions (G1–G61) based on simple sequence repeat (SSR) markers (A), inter simple sequence repeat (ISSR) markers (B), and Standard Codon Targeted (SCoT).**

## 4.10 Structural analysis of the *Crataegus* spp. accessions based on marker datasets

The structural analysis was performed to investigate the population stratification of Hawthorn accessions. Results of SSR markers, the number of accession clusters was designed using the K value. The value of optimal K was detected by graphing the cluster number versus K. The largest peak occurred at K = 2 (Figure 4.14 A). The value of optimal K indicated that two populations were created (population 1 (red) and population 2 (green) (Fig.  4.14 B). The threshold of membership probability was 0.80. The accessions subgroups were detected separately, and fractions less than 0.20 indicated admixture. The first population and second population involved twenty-one Hawthorn accessions each. Seventeen accessions were defined as admixed amongst the two populations, presenting that this accession is not pure.

The value of ideal K was measured using structure-harvester analysis. The number of clusters (K) was plotted against K, with a supreme peak occur- ring at K = 2, based on the ISSR data (Fig. 4.14C). The sixty-one accessions of Hawthorn were separated into two genetic populations (Fig. 4.14D). The first group (red) contained thirteen accessions. The second group comprised seventeen accessions (green). The remaining thirty-one were from an admixed population. Structure-harvester analysis was used to measure the optimal K based on the SCoT markers data, and the number of clusters (K) was plotted compared with K, with a determined peak at K = 3. Evano method suggests that K = 3 is the best number of clusters (Fig. 4.14E). Twenty-six accessions involved the population 1 (in red), six accessions included the population 2 (green), and the population 3 (blue) were made by the 12 accessions, which determined as pure. The remaining seventeen accessions were labeled admixed (Fig. 4F).

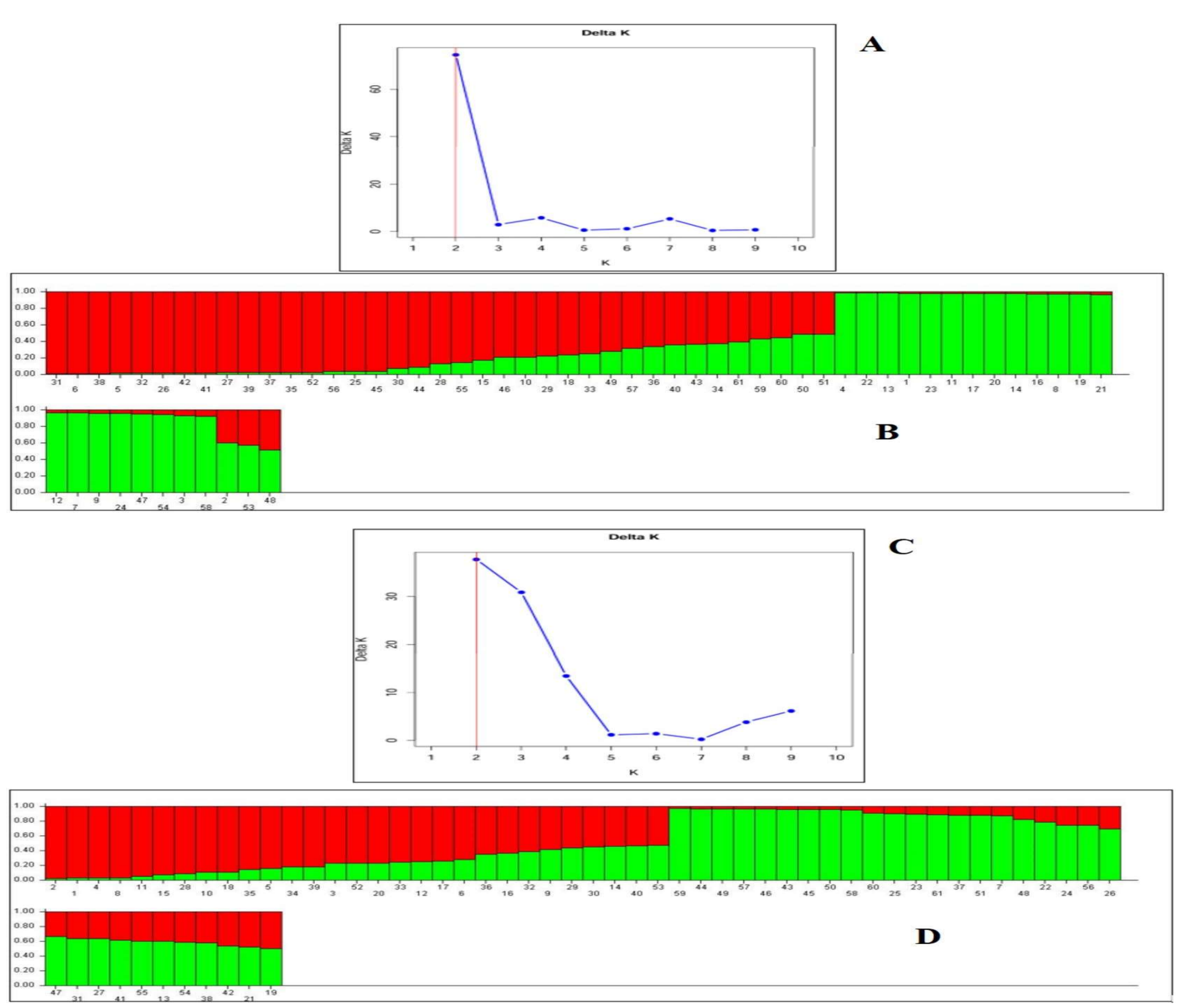

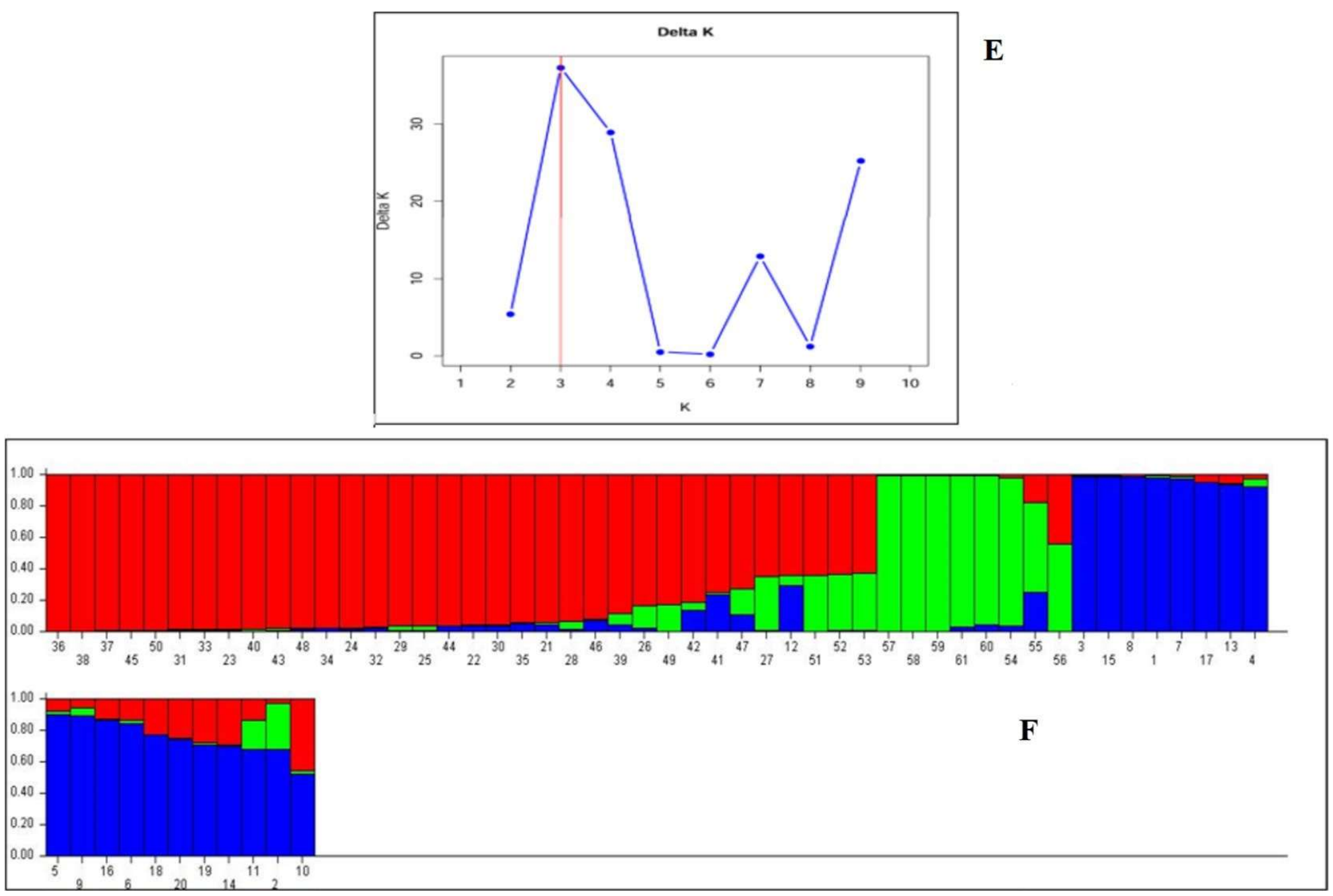

**Figure 4. 14 A Population numbers (K) creating by delta K from simple sequence repeat (SSR) data; B expected popula tion structure of the 61 hawthorn accessions (G1–G61) based on K = 2 resulting from the SSR dataset; C delta K for diverse population numbers (K) derived from the inter simple sequence repeat ISSR dataset D: estimating population structure of the 61 hawthorn accessions (G1–G61) on K = 2, which revealed by ISSR data; E based on the SCoT marker data a delta K for diverse population numbers (K) was generated which K = 3; F the SCoT marker data generating three popula tion structure of the 61 hawthorn accessions (G1–G61).**

## 4.11 Diversity indices analysis and AMOVA of Hawthorn populations

The genetic variation between and within the Hawthorn population was measured by molecular variance analysis (AMOVA) through SSR, ISSR, and SCoT marker datasets (Table 4.4). The substantial PhiPT values of 0.10, with a p-value of 0.001; 0.19, with a p-value of 0.001; and 0.17, with a p-value of 0.001, were detected for the SSR, ISSR, and SCoT markers, respectively. The variation amongst populations, described as 10.00%, 19.00%, and 17.00%, were presented for the SSR, ISSR, and SCoT marker data, respectively, of the total variance. Nevertheless, with the values of 90.00, 81.00, and 83.00%, for SSR, ISSR, and SCoT markers, respectively, the highest variation occurred within the population.

**Table 4.4 AMOVA analysis between and within Hawthorn accessions derived by three markers used.**

| Source | Df | SS | MS | Est. Var | Var (%) |
|---|---|---|---|---|---|
| *SSR markers* | | | | | |
| Among Pops | 8 | 97.34 | 12.17 | 0.79 | 10 |
| Within Pops | 52 | 365.61 | 7.03 | 7.03 | 90 |
| Total | 60 | 462.95 | | 7.82 | 100 |
| PhiPT | 0.10** | | | | |
| *p*-value *SSR markers* | 0.001 | | | | |
| Among Pops | 8 | 108.54 | 13.57 | 1.24 | 19 |
| Within Pops | 52 | 284.78 | 5.48 | 5.48 | 81 |
| Total | 60 | 393.31 | | 6.72 | 100 |
| PhiPT | 0.19** | | | | |
| *p*-value ISSR markers | 0.001 | | | | |
| Among Pops | 8 | 321.47 | 40.18 | 3.53 | 17 |
| Within Pops | 52 | 895.18 | 17.22 | 17.22 | 83 |
| Total | 60 | 1216.66 | | 20.75 | 100 |
| PhiPT | 0.17** | | | | |
| *p*-value *SCoT marker* | 0.001 | | | | |

*Df* degrees of freedom, *SS* sum of squares, *MS* mean of squares, *Est. Var* estimated variance, *Var* variance, *PhiPT* the percentage of overall genetic differences among accessions entire a population, *p*-value: probability value **statistically significant at the 0.01 level.

Molecular marker discussion observed that significant repercussions for the evolution and preservation of species are associated with genetic polymorphism, which varies between species and within genomes (Ellegren and Galtier 2016). Allelic abundance of accessions predicts genetic diversity enhancement, commonly utilized by informative molecular markers that identify populations for conservation, selection, and breeding (Civan et al. 2021). Allelic richness demonstrates the degree to which populations have been enriched with genetic diversity. As the marker numbers and the genome quantity coverage rise, the dependability of the data may likewise expand (Kushanov et al. 2021). Our result revealed 237 polymorphic bands by three diverse types of molecular markers. SCoT markers distinguished an average of 13.40 polymorphic bands based on different markers, greater than SSR primers (5.18) and ISSR markers (4.60). The polymorphism in the various regions of the Hawthorn genome is the probable cause of these differences. Single primer typically anneals to ATG initiation codon flanking regions on two strands of DNA generating SCoT markers (Col- lard and Mackill 2009). It is possible to distribute the amplicons that were created inside gene regions. As a consequence of this, it is possible to identify genetic variations in a particular gene that are associated with a specific characteristic (Rai 2023). Despite the fact ISSR markers are abundant in microsatellite-flanked genomic regions (Rasul et al. 2024) and SSR markers are sequences in which one or few bases are tandemly repeated, ranging from 1–6 base pair (bp) long units (Victoria et al. 2011).

The index of PIC is important it reveals the effectiveness of molecular markers in predicting the genetic variation of organisms. A PIC is a low-efficiency marker if it is less than 0.25 and an identical powerful marker if it is 0.5 or higher (Aziz and Tahir 2023). According to our results, PIC values for SCoT markers were above the minimum (0.25), 0.28. A higher PIC value above 0.50 in average was identified for SCoT when applied to other species (Ahmed et al. 2022). The SCoT markers used in this study revealed genetic variation among Hawthorn accessions more efficiently than SSR and ISSR markers, having PIC values of 0.19 and 0.23, respectively. This might relate to the nature of the species study to give the low rate of the PIC, as a much higher rate of PIC value was indicated in other investigations applying SSR markers (Ahmad et al 2022; Tahir et al 2023b).

In population structure, diversity indices are statistical measures of the diversity of the population, with each individual belonging to a different group. Indices with lower values indicate a lower level of diversity, whereas indices with higher values provide evidence of a higher level of diversity (Majeed et al. 2024). In our study, we found that SCoT markers had moderate average expected heterozygosity or gene diversity values (0.17), indicating that this marker type is more effective than ISSR (0.13) and SSR (0.11) markers for genetic diversity analysis in wild Hawthorn and aiding in the selection of the best SCoT markers in the genetic divergence analysis. The Shannon's information index (I) of 0.25, 0.20 and 0.17 for the SCoT, ISSR and SSR markers, respectively, demonstrates their effectiveness in detecting the level of diversity amongst the accessions. All Hawthorn accessions were divided into seven main clades with some subclades according to the UPGMA. These groups cannot differentiate all accessions according to location, based on the three types of marker data, whereas the accessions of these groups of each marker diverge as well. It is possible that the nature of any marker, which exists in various sites on the Hawthorn genome, as well as the level of polymorphisms attained, are responsible for these discrepancies between the dendrograms. All the accessions of Hawthorn were classified into two main genetic populations for the SSR and ISSR, and three genetic groups for the SCoT markers based on structural analysis, along with allele introgression amongst various wild Hawthorns, therefore the majority of the accessions are pure.

AMOVA statistics were performed to reveal a significant amount of genetic differentiating both among and within the populations. Furthermore, the differences within the populations were greater than the variation between the populations, which refers to presence of a high rate of gene exchange between the studied populations (Khdir et al. 2023). These findings may be related to gene drift and gene flow that influence Hawthorn populations. It has been demonstrated that populations with high genetic variation are due to clonal propagation, natural crossing, or long-lived species. Populations with low genetic variation are due to harsh environments (Wroblewska et al. 2003). In general, the findings regarding genetic diversity are essential for the conservation of germplasm because they not only support the development of agricultural systems that are resilient, productive, and sustainable, but they also preserve the genetic heritage of plant species simultaneously. Findings regarding genetic diversity contribute to the identification and preservation of endangered plant species that would otherwise be considered extinct. The preservation of the genetic heritage of our planet and the maintenance of the equilibrium of ecosystems are both important reasons for this (Priyanka et al. 2021).

## Conclusions

The major conclusions from this study include:

1. To the best of our knowledge, this is the first attempt to use morph- chemical characteristics and molecular markers to assess the diversity of Iraqi Kurdistan region cartages accessions.
2. The important Hawthorn genotypes revealed a wide range of variability in morpho-physicochemical markers, and based on molecular markers, that this wide range of variance will help germplasm management, classification, and preservation.
3. Fruit characterization of Hawthorn genotypes observed valuable nutritional qualities, which may enhance human health. Generally, all genotypes were rated as the highest performing genotype based on fruit morpho- physicochemical features.
4. High and significant genetic variation was found among the 61 accessions tested at both the fruit and molecular levels.
5. The results confirmed the efficacy of SSR, ISSR and SCoT, markers as useful approaches for assessing Hawthorn diversity. SCoT markers showed more effectiveness in studying the genetic diversity of Hawthorn species than SSR and ISSR markers. These markers can be relied in future genetic studies of Hawthorn.
6. The utilization of these marker techniques will be of great assistance in the management of Hawthorn germplasm as well as breeding programs for new Hawthorn cultivars.

## Recommendations

- In this regard, more investigations are needed to confirm the distinction in the gene pool between our collection and other Hawthorn accessions from other parts of the world.
- Additional molecular markers, such as single nucleotide polymorphism (SNP), should be utilized to have more precise identification of variation among the studied accessions.
- It is recommended that other morphological and anatomical characterization should be utilized in Hawthorn characterization including electronical microscopy for scanning pollen gran.
- Other techniques of gas chromatography-mass spectroscopy GCMS and HPLC could be followed to indicate phytochemical components in Hawthorn accessions in the Kurdistan of Iraq.
- Further research is recommended to explores additional locations within Kurdistan Forest, as many areas remain understand regarding *Crataegus* accessions.
- Expanding the geographical scope in future studies in order to ceftibuten to a more comprehends for understanding the specie's genetic diversity ecological adaptations and potential uses.